# CONCEPTUAL DESIGN REPORT:
# A RING-BASED ELECTRON COOLING SYSTEM FOR THE EIC

V. Lebedev[1], S. Nagaitsev, A. Burov, V. Yakovlev, I. Gonin,

I. Terechkine, A. Saini, and N. Solyak

*Fermilab, Batavia, IL 60510*

**Abstract:** This report describes a concept of an EIC cooling system, based on a proven induction-linac technology with a DC electron beam. The system would operate in a full energy range of proton beams (100 – 270 GeV) and would provide 50-100 A electron beams, circulating in a cooler ring for 5 ms. Every 5 ms a new electron pulse would be injected into the cooler ring to provide continuous cooling at collisions. Operations with a 10-ms cycle is possible but it will reduce the cooling rates by ~30 %. The system is capable of delivering the required performance in the entire EIC energy range with emittance cooling times of less than 1-2 hours.

---

[1] val@fnal.gov

# *Table of Contents*

# 1. Introduction

The Electron-Ion Collider is a proposed discovery machine, designed to unlock the secrets of the "glue" that binds the building blocks of visible matter in the universe [1]. Its realization is one of the highest priorities in the modern nuclear physics. The EIC will be an unprecedented collider that will need to maintain high luminosity ($10^{33-34}$ cm$^{-2}$s$^{-1}$) over a very wide range of Center-of-Mass energies (20 GeV to 100 GeV, upgradable to ~140 GeV), while accommodating highly polarized beams and many different ion species [2]. Addressing the challenges of this machine requires R&D in areas such as crab cavities, strong hadron beam cooling, and high-field magnets for the interaction points. This report addresses one of these challenges, proton (light ion) cooling at the collision energy.

Electron cooling was proposed [3], developed and experimentally tested at Budker INP, Novosibirsk in the 1970s [4,5]. Since that time, many low-energy machines employed electron cooling. All of them used the same principle: a single-pass scheme with a DC electron beam acceleration, immersed in the accompanying solenoidal magnetic field, similar to the first cooler, developed in Novosibirsk. The first deviation from this principle was introduced in 2005 in the FNAL cooler, which did not use the accompanying magnetic field [6]. Up to now it remains the highest energy cooler. It was capable of cooling 8.9 GeV/c antiprotons. Recently, the next step in the electron cooling development was achieved by replacing DC acceleration with RF acceleration [7]. Since its inception, a number of schemes for high-energy electron cooling were proposed. For example, an electron storage ring was proposed [8] to be employed to cool hadron beams, while the electrons in the ring were continuously cooled by the synchrotron radiation. Later publications [9-11] considered different implementations of this idea. Another popular idea is to use RF acceleration of the electron beam [12-14]. In this paper we consider a detailed proposal for a ring-based electron cooler with beam acceleration in an induction linac.

The dominant proton-beam heating mechanism in the EIC is intra-beam scattering (IBS). To counteract it, the beam cooling system must provide emittance cooling times of about 1-2 hours for the longitudinal and transverse degrees of freedom at collision energies (100 – 270 GeV). Electron cooling has been successfully employed at Fermilab at antiproton kinetic energies of up to 8 GeV with cooling times of the order of 0.5 hour [6, 15]. The electron cooling time has a very unfavorable beam energy scaling (~$\gamma^{2.5}$). This scaling dependence can be mitigated by increasing two cooling system parameters: (1) the cooling section length and (2) the electron beam current. In simplest electron cooling models, the cooling time is proportional to both of these two parameters. In more sophisticated cooling models, this dependence needs to be considered (and optimized) more carefully because other beam parameters, relevant to cooling rates, are also affected by the length and the beam current. This conceptual design report considers several such effects. Typical electron beam currents in DC electron coolers (including the one at Fermilab [1]) are ~1 A (DC) and it is very difficult to increase it to the desired level (~100 A) for the required EIC energy range (100-270 GeV/u). In our concept, we propose to employ a pulsed (up to 830-ns long) DC-like electron beam of 50-100 A at 55 – 147 MeV in a storage ring, where it circulates for about ~10,000 turns, while cooling the EIC proton beam. Every 5 ms, the electron beam is extracted from the ring into the beam dump and a new pulse is reinjected. This greatly reduces the total electron beam power, required to support the cooling process. The electron beam has the DC (coasting) structure, which simplifies the overall cooling scheme. At 147 MeV, the total electron beam power to dump is about 2.5 MW. In what follows, we will focus mostly on the design of a 100-GeV cooling system. Appendix A describes the modifications required for a 270-GeV cooling system. Figure 1 shows the proposed concept.

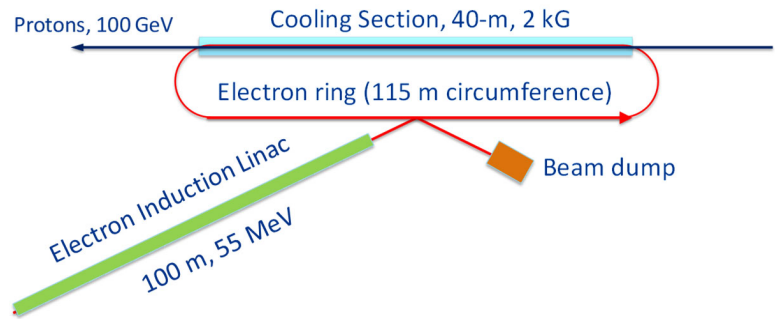

Figure 1: Schematic of a ring-based electron cooler. Note that the cooling section length may be extended in the final design.

## 2. System Parameters

The main parameters of the cooler are shown in Table 1. The electron beam is produced in a pulsed gun with a thermionic cathode, then it is accelerated in the induction linac (see Section 10) and injected into the cooling ring where it circulates for many turns as a continuous beam. The number of circulating turns is determined by the IBS rate in the electron beam. The beam optics in the electron gun, the induction linac, the transport line and the ring are designed to preserve the beam emittance. Thus, the beam transverse emittance at injection into the ring is mostly determined by the beam emittance at the cathode. The corresponding rms normalized emittance is:

$$\varepsilon_n = \frac{r_c}{2}\sqrt{\frac{T_c}{m_e c^2}} \tag{1}$$

where $m_e$ is the electron mass, $c$ is the light speed, $r_c$ is the cathode radius, and $T_c$ its temperature. For the cathode radius of 1.8 cm and its temperature of 0.12 eV we obtain $\varepsilon_n$ = 4.36 µm. Parameters presented in Table 1 correspond to this thermal emittance. Electron cooling is carried out in the cooling section with a longitudinal magnetic field. It improves the beam stability in the cooling section and keeps the electron beam radius constant. To compensate for the rotational contribution to the particle angles coming from the cooling solenoid entrance, the electron gun cathode is also immersed into the longitudinal magnetic field introducing required angular momentum to the beam.

The magnetic field of the solenoid couples horizontal and vertical beam motion. In this case the transverse phase space is characterized by two emittances called the mode emittances [16]. Their values are determined by the magnetic field in the cooling solenoid and the initial thermal emittance at the cathode. The normalized emittances for the modes 1 and 2 are:

$$\varepsilon_{1n,2n} = \frac{\varepsilon_n}{\sqrt{1+\Phi_r^2 \beta_0^2} \pm \Phi_r \beta_0}, \tag{2}$$

where $\beta_0 = a_e^2/(\varepsilon_n/\beta\gamma)$ is the effective beta-function, $a_e$ is the electron beam radius in the cooling section, $\beta$ and $\gamma$ are the relativistic factors, $B_0$ is the magnetic field in the cooling solenoid, and $\Phi_r = eB_0/(2\gamma\beta m_e c^2)$ is its focusing strength. The magnetic fields at the cathode, $B_c$, and in the cooling section are related as: $r_c^2 B_c = a_e^2 B_0$. For a fixed $r_c$ and $a_e$ an increase of the magnetic field on the cathode and, consequently, in the cooling section increases the difference between mode emittances.

**Table 1: Main parameters of the ring-based electron cooler**

| | |
|---|---|
| Proton beam energy | 100 GeV |
| Peak current in a proton bunch | < 10 A |
| Proton ring circumference (it is used for computation of cooling rates only) | 3000 m |
| Relativistic factor, $\gamma$ | 107.58 |
| Normalized rms proton beam emittance | 1 μm |
| Proton beam rms momentum spread | $<3 \cdot 10^{-3}$ |
| Proton beam rms angular spread in the cooling section | 15 μrad |
| β-functions of proton beam in cooling section center | 40 m |
| Electron beam energy | 54.48 MeV |
| Electron ring circumference | 114.2 m |
| Cooling length section | 40 m |
| Electron beam current | 50 A |
| Longitudinal magnetic field in cooling section, $B_0$ | 1.848 kG |
| Electron beam rms momentum spread, initial/final | $(1.06/1.64) \cdot 10^{-3}$ |
| Rms electron angles in cooling section | 27 μrad |
| Rms electron beam size in cooling section | 1.47 mm |
| Electron beam rms norm. mode emittances at the cycle beginning, $\varepsilon_{1n}/\varepsilon_{2n}$, μm | 233/0.081 |
| Number of cooling turns in the electron storage ring | 13,000 |
| Longitudinal cooling time (emittance)[*] | 0.5 hour |
| Transverse cooling time (emittance)[*] | 1 hour |

[*] These values account for a reduction of cooling rates due to imperfections (see details in Section 6).

## 3. Cooling Ring Optics

The longitudinal magnetic field in the cooling section strongly couples the vertical and horizontal motions in the ring. Therefore, the entire optics analysis is carried out using a formalism of coupled beta-functions [16] with usage of OptiMX software [17].

**Table 2: Main parameters of the cooling ring optics**

| | |
|---|---|
| Betatron tunes, $\nu_1/\nu_2$ | 13.8764 / 5.8548 |
| Natural chromaticities | -14.09 / -5.53 |
| Corrected chromaticities | -6.99 / -8.05 |
| Slip-factor, $\alpha$ | 0.0444 |
| Damping parameters, $g_x / g_y / g_s$ | 0.57693 / 1 / 2.42307 |
| Equilibrium rms horizontal emittance set by SR | 2.9 nm |
| Equilibrium rms momentum spread set by SR | $3.1 \cdot 10^{-5}$ |
| Amplitude damping decrements per turn, $\lambda_x / \lambda_y / \lambda_s$, [$10^{-9}$] | 2.24 / 3.88 / 9.41 |
| Energy loss due to SR | 0.42 eV/turn |
| Transverse acceptance* | 150 μm |
| Longitudinal acceptance (maximum momentum spread)* | 0.011 |

\* That implies the round vacuum paper with 1" radius, and the beam acceptance and maximum momentum spread limited by 22 mm leaving 3 mm for closed orbit distortion.

Table 2 presents the main parameters of the cooling ring optics. The ring consists of two straight sections connected by two 180° arcs. Figure 2 presents the 4D-beta-functions and dispersions for the entire ring. A minimization of the transverse electron beam temperature in the solenoid requires the beam phase space being presented by two ideal circular betatron modes at the solenoid entrance. It is achieved by equaling all four 4D-beta-functions in the solenoid and setting the 90° phase difference between the horizontal and vertical parts of the eigen-vectors. The beta-function values $\beta_{1x} = \beta_{2x} = \beta_{1y} = \beta_{2y} = 0.992$ m are matched to the solenoid focusing and stay constant at the entire length of the solenoid with magnetic field of $B_0$=1.848 kG. This field determines the magnetic field at the electron gun cathode of 12.26 G so that the rotation introduced by the entrance into solenoid would be cancelled by the beam angular momentum inherited at the cathode. At both ends the solenoid is interfaced with Derbenev's adapters [18-20] making the transformation from the circular betatron modes to the uncoupled betatron motion. Following Ref. [20] we will call them below as planar-to-circular and circular-to-planar adapters or rotators. The transformation is performed so that the betatron mode with larger emittance is placed into the horizontal plane in the arcs. Further we will call this mode the mode 1 while the mode with vertical betatron motion in the arcs is called the mode 2.

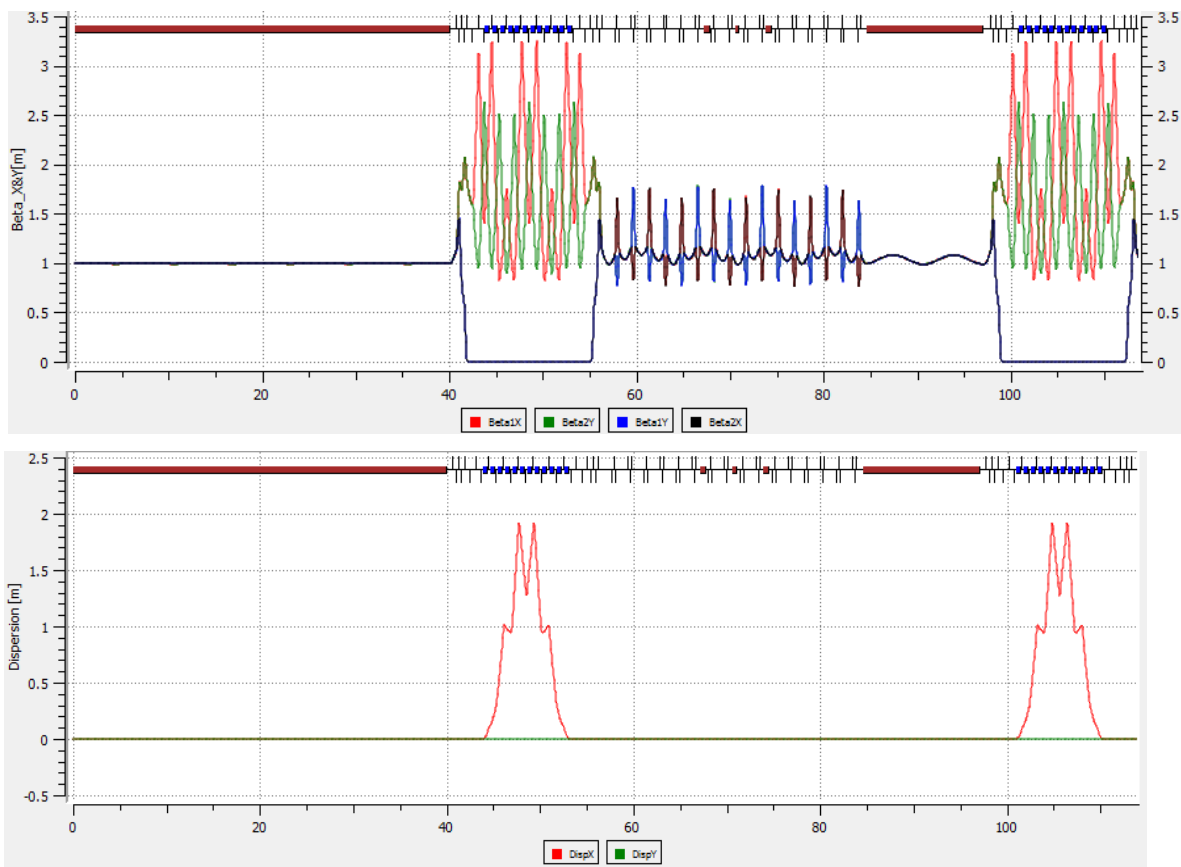

Figure 2: 4D-beta-functions (top) and dispersions (bottom) for the entire ring starting from the entrance to the cooling solenoid. Locations and lengths of elements are shown by the legends located at the top of each picture: brown – cooling (40 m) and beam stabilization (12.45 m) solenoids, injection/extraction kickers and septum; vertical lines – quads; blue squares – dipoles.

Figure 3 shows the rms betatron sizes through the entire ring. As one can see the beam is round in the cooling solenoid and has constant radius through its entire length. The vertical beam size in the arcs is much smaller than the horizontal size. That greatly increases both the IBS and the mode 2 betatron tune shift due to beam space charge. Note that putting the smaller emittance corresponding to the mode 2 into the vertical plane in the arcs is the only way to obtain reasonably small IBS driven emittance growth for this mode. A study showed that optics with the circular modes in the arcs results in the emittance growth rate for the mode 2 orders of magnitude larger than the optics considered in this section. Similar planar-to-circular and circular-to-planar adapters are used in the second straight located on the other side of the ring. We will call it the technical straight. This introduction of circular modes in the technical straight increases the beam cross-section in the straight and, consequently, significantly reduces the IBS and the space charge betatron tune shift for the mode 2.

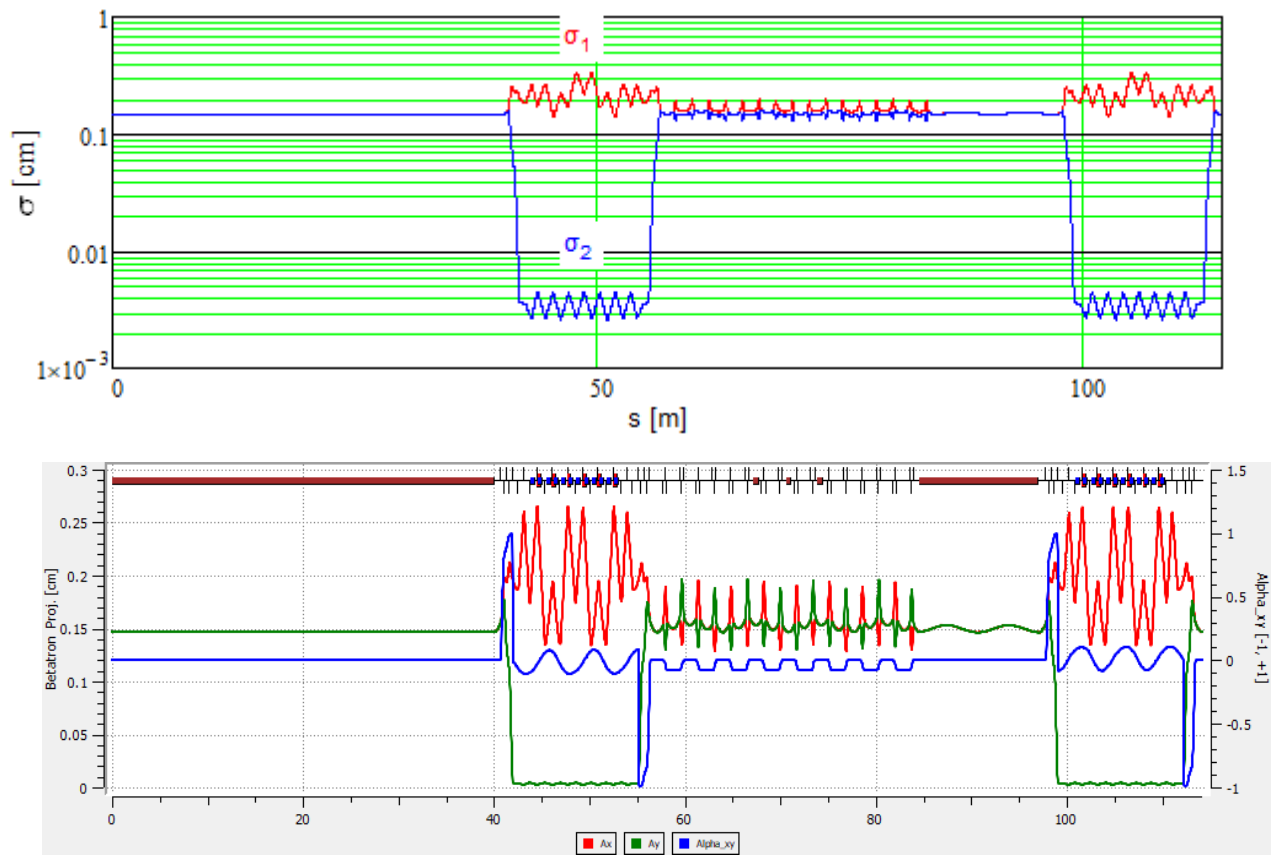

Figure 3: The rms betatron sizes (top) and their projections (bottom) along the entire ring starting from the cooling solenoid. The bottom plot also shows the parameter $\alpha = \langle xy \rangle / \sqrt{\langle x^2 \rangle \langle y^2 \rangle}$ characterizing the ellipse rotation relative to the $x$- and $y$-planes; the rms mode emittances are: $\varepsilon_1$=2.2 µm, $\varepsilon_2$=0.76 nm.

Figure 4 shows the beta-functions of the circular-to-planar adapter. It has five equidistant skew-quads. Such number of quads is chosen to minimize the beta-beating and to support the transformation in a wide range of magnetic field in the cooling solenoid. The left plot shows normal-uncoupled beta-functions for the case of not rolled quads. The initial uncoupled beta-functions ($\beta_x$, $\beta_y$) are twice larger than the 4-D-beta-functions in the solenoid. The focusing in the adapter is adjusted so that the uncoupled beta-functions and their derivatives for both planes at the adapter end would coincide and the difference of the betatron phase advances would be equal to 90º. After rotation of all quads by 45º one obtains a circular-to-flat adapter. The right pane presents corresponding 4D-beta-functions. At the adapter end two beta-functions ($\beta_{1x}$, $\beta_{2y}$) coincide with normal uncoupled beta-functions ($\beta_x$, $\beta_y$) and other two are equal to zero.

Both arcs have the same optics. Figure 5 shows the beta-functions and dispersions for one arc. The optics is symmetric relative to the arc center. There are 3 quads at each end which match the adapters to the arc proper. To minimize the space charge betatron tune shift introduced by the arcs to the mode 2 the arc optics was built to minimize the maximum of vertical beta-function. The betatron tune of the mode 2 (vertical mode in the arcs) can be adjusted by defocusing quads. The requirement of arc achromaticity determines the strength of horizontal quads leaving very little room for regulation of mode 1 betatron phase advance (horizontal mode in the arcs) in the arcs.

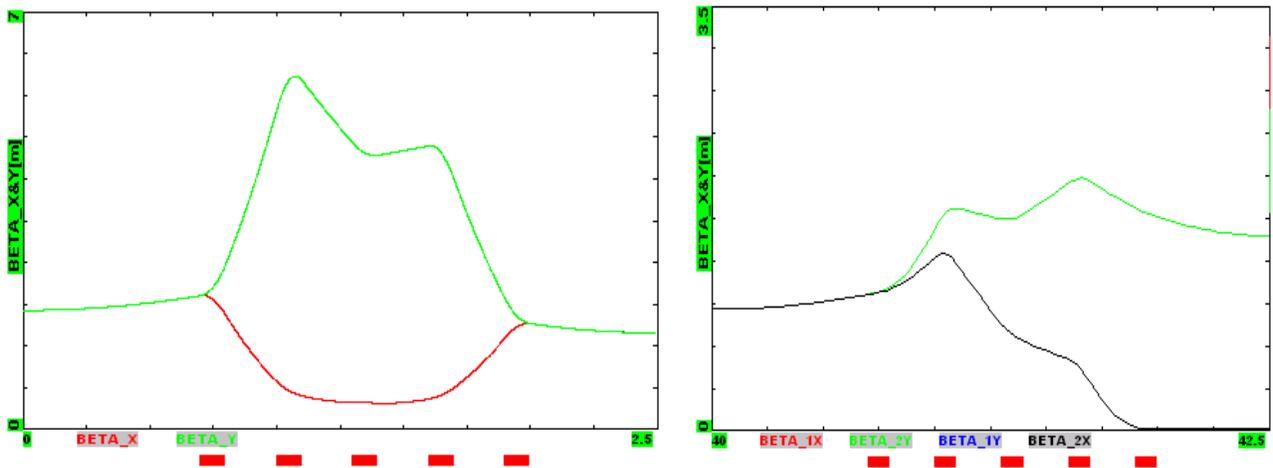

Figure 4: Beta-functions for the circular-to-planar adapter. The plots start from the cooling solenoid and proceed to the arc input; left - uncoupled beta-functions for unrolled quads; right – 4D-beta-functions with quads rolled by 45 deg. Locations and lengths of quads are shown at the bottom of the pictures.

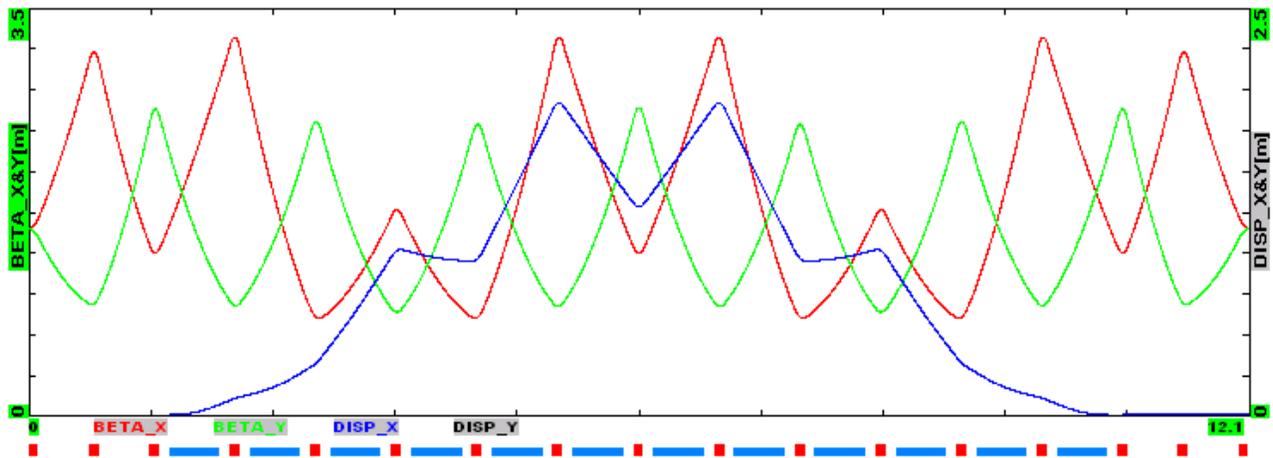

Figure 5: Beta-functions and dispersions for one arc.

To conserve the circular modes in the technical straight its optics should have equal betatron phase advances for both betatron planes. It is achieved by the quadrupole triplets with changing signs of quadrupoles. The coupled beta-functions in the straight may be seen in Figure 2. There are 8 periods of this triplet focusing followed by a 12.45 m solenoid aimed for the transverse beam stabilization as discussed in Section 9.2. Each period includes one positive and one negative triplet. The positive and negative triplets have the same strength of quadrupoles but opposite signs. The stabilization solenoid field is close to the field of cooling solenoid. Figure 6 presents the uncoupled beta-functions for one period of the technical straight. A phase advance per period of 0.25 (~45° per triplet) is set to be relatively small to minimize the beta-beating. The number of periods and phase advances per period were chosen to obtain the beta-functions at the ends to be the same as in the cooling straight.

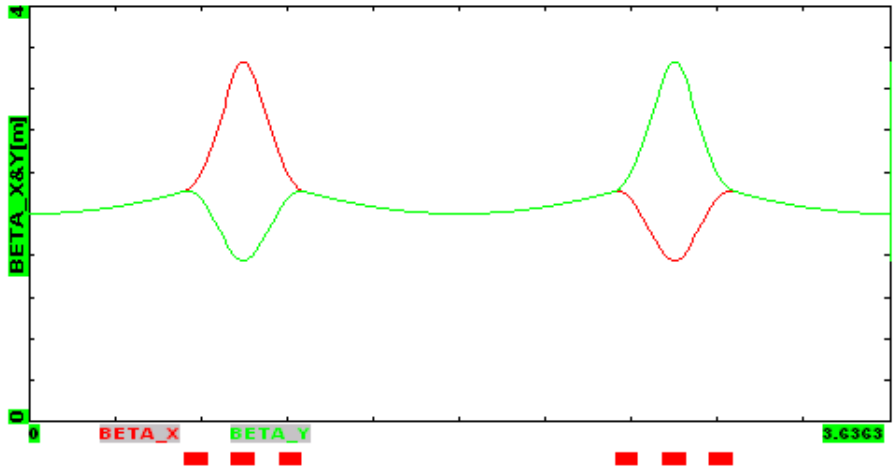

Figure 6: Uncoupled beta-functions for one period of technical straight.

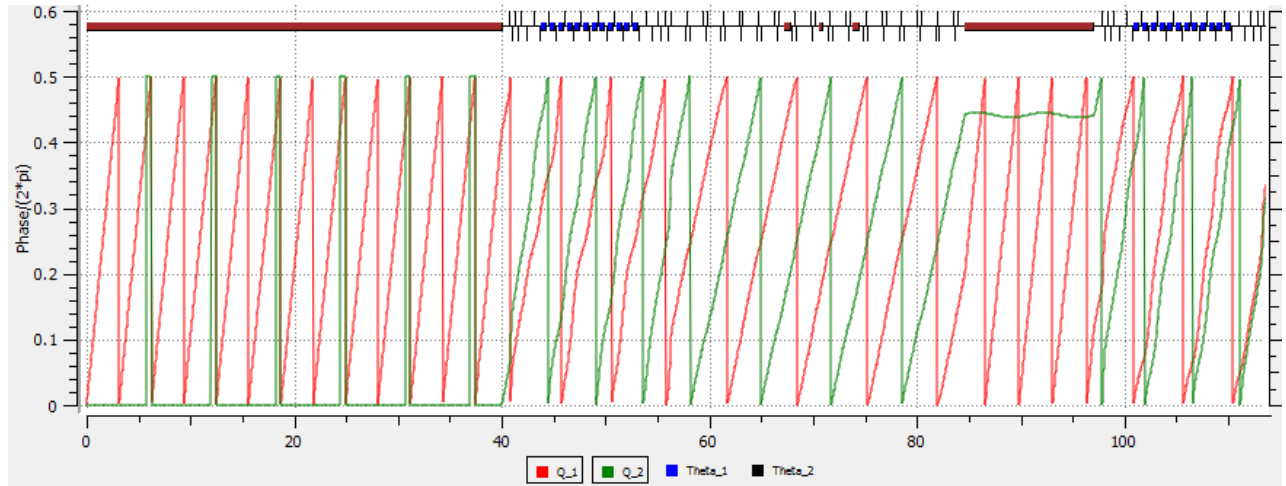

Figure 7: Betatron phase advances along the cooling ring plotted on modulo 0.5: red – mode 1, green – mode 2.

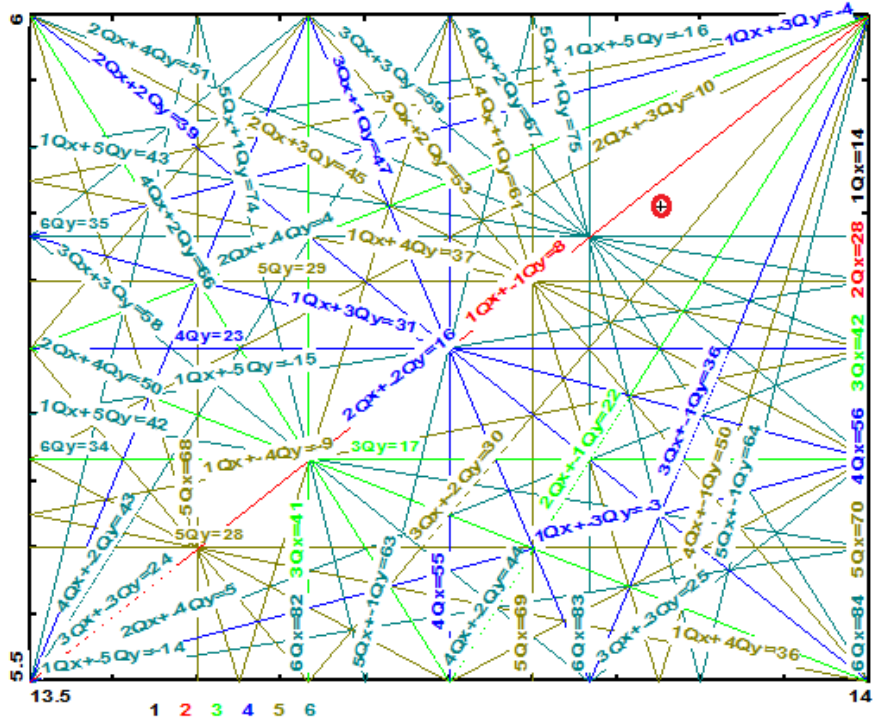

Figure 8: Tune diagram.

As will be seen below, the space charge effects are strong and therefore some flexibility in choosing betatron tunes is important. Figure 7 presents the betatron phase advances along the ring for both betatron modes. As one can see the phase advance of the mode 1 stays constant at the entire length of both cooling and stabilizing solenoids, while the mode 2 phase advance goes twice faster than in the technical straight section with triplets which has close values of beta-functions (see Figure 2) but does not have longitudinal magnetic field, and, consequently, *x-y* coupling. The polarities of solenoids and the adapters are chosen so that to put the mode 1 into the horizontal plane in the arcs and to obtain the desired tunes. The tunes are chosen to be: $Q_1$=13.8764, $Q_2$= 5.8548. This choice is justified by a requirement of avoiding the second and fourth order resonances excited by the beam space charge and avoiding the first order coupling resonance. For the chosen ratio of emittances, the space charge tune shift in $Q_2$ exceeds the change in $Q_1$ by an order of magnitude. Thus, the tune footprint represents a narrow vertically located region directed down from the basic tune marked by a circle in Figure 8. The chosen optics enables tune adjustments in a sufficiently large tune space. Tune adjustments can be achieved with minor change of solenoidal field (both for cooling and instability damping solenoids) which changes mostly $Q_1$ tune, and with changes of vertical focusing in arcs which mostly changes $Q_2$ tune.

Changes of triplet focusing in the technical straight changes both tunes by the same amount. Large tune changes (jumps) can be also achieved by changing polarity of rotators which changes are limited by the requirement of optics closure and setting the mode 1 into the horizontal plane in both arcs.

## 4. Optics Sensitivity to Changes in Focusing

One of the main requirement to optics stability is associated with minimization of variations of transverse temperature of electron beam in the cooling section. Acceleration of the beam in the induction linac results in a pancake velocity distribution in the cooling section, *i,e.* the longitudinal velocity spread is much smaller than the transverse ones. In this case the longitudinal cooling force for small longitudinal velocity of a proton, $v_{ps}$, depends on the rms velocity spreads in the electron beam, $\sigma_{vx}, \sigma_{vxy}, \sigma_{vs}$, as: $F_{\parallel}\big|_{v_{p\perp}=0} \propto v_{ps}/(\sigma_{vx}\sigma_{vy}\sigma_{vs})$ (see Section 6 for details). Therefore, we introduce the effective transverse angular spread as a square root of a product of rms transverse angular spreads in the coordinate frame aligned with the axes of transverse velocity ellipsoid. That yields:

$$\theta_{\perp eff}^2 = \theta_1 \theta_2 = 1/\sqrt{|\Xi_\theta|}, \tag{3}$$

where matrix $\Xi_\theta$ determines the transverse angular distribution in the electron beam: $f(\theta_x, \theta_y) \propto \exp(-\boldsymbol{\theta}^T \Xi_\theta \boldsymbol{\theta}/2)$ and $\boldsymbol{\theta} = [\theta_x, \theta_y]$. Using expressions for the matrix elements derived in Appendix A of Ref. [16]:

$$\Xi_{11} = \frac{\beta_{1x}}{\varepsilon_1} + \frac{\beta_{2x}}{\varepsilon_2}, \quad \Xi_{22} = \frac{\beta_{1y}}{\varepsilon_1} + \frac{\beta_{2y}}{\varepsilon_2}, \quad \Xi_{12} = \Xi_{21} = \frac{\sqrt{\beta_{1x}\beta_{1y}}\cos\nu_1}{\varepsilon_1} + \frac{\sqrt{\beta_{2x}\beta_{2y}}\cos\nu_2}{\varepsilon_2}, \tag{4}$$

we can rewrite Eq. (3) in the following form

$$\theta_{\perp eff}^2 = \frac{1}{\sqrt{\left(\dfrac{\beta_{1x}}{\varepsilon_1}+\dfrac{\beta_{2x}}{\varepsilon_2}\right)\left(\dfrac{\beta_{1y}}{\varepsilon_1}+\dfrac{\beta_{2y}}{\varepsilon_2}\right)-\left(\dfrac{\sqrt{\beta_{1x}\beta_{1y}}\cos\nu_1}{\varepsilon_1}+\dfrac{\sqrt{\beta_{2x}\beta_{2y}}\cos\nu_2}{\varepsilon_2}\right)^2}}. \tag{5}$$

Here the same as in Ref. [16] we use the eigen-vectors parameterization as:

$$\mathbf{v}_1 = \begin{bmatrix} \sqrt{\beta_{1x}} \\ -\dfrac{i(1-u)+\alpha_{1x}}{\sqrt{\beta_{1x}}} \\ \sqrt{\beta_{1y}}\,e^{i\nu_1} \\ -\dfrac{iu+\alpha_{1y}}{\sqrt{\beta_{1y}}}\,e^{i\nu_1} \end{bmatrix}, \quad \mathbf{v}_2 = \begin{bmatrix} \sqrt{\beta_{2x}}\,e^{i\nu_2} \\ -\dfrac{iu+\alpha_{2x}}{\sqrt{\beta_{2x}}}\,e^{i\nu_2} \\ \sqrt{\beta_{2y}} \\ -\dfrac{i(1-u)+\alpha_{2y}}{\sqrt{\beta_{2y}}} \end{bmatrix}. \tag{6}$$

For the circular betatron modes all beta-functions are equal, $u = 1/2$, and $|\nu_{1,2}| = \pm\pi/2$. Accounting that $\varepsilon_2 \ll \varepsilon_1$, assuming minor deviations of the beta-functions and the phases $\nu$ from their target values $\beta_{xx} = \beta_0 + \delta\beta_{xx}$, $\nu_{1,2} = \pi/2 + \delta\nu_{1,2}$ and leaving only leading terms in the obtained result we obtain:

$$\theta_{\perp eff}^2 \approx \frac{\varepsilon_2}{\beta_0 \left(1 + \frac{\delta\beta_{2x}}{2\beta_0} + \frac{\delta\beta_{2y}}{2\beta_0} - \frac{\delta\nu_2^2}{2}\right)} \quad . \tag{7}$$

Similarly, using results of Ref. [16] one obtains the beam cross-section (product of rms values of major axes of transverse size distribution):

$$\sigma_{\perp eff}^2 = (\varepsilon_1 \beta_{1x} + \varepsilon_2 \beta_{2x})(\varepsilon_1 \beta_{1y} + \varepsilon_2 \beta_{2y}) - \left(\varepsilon_1 \sqrt{\beta_{1x}\beta_{1y}} \cos\nu_1 + \varepsilon_2 \sqrt{\beta_{2x}\beta_{2y}} \cos\nu_2\right)^2 . \tag{8}$$

Expending and leaving leading terms one obtains:

$$\sigma_{\perp eff}^2 \approx \varepsilon_1 \beta_0 \left(1 + \frac{\delta\beta_{1x}}{2\beta_0} + \frac{\delta\beta_{1y}}{2\beta_0} - \frac{\delta\nu_1^2}{2}\right) . \tag{9}$$

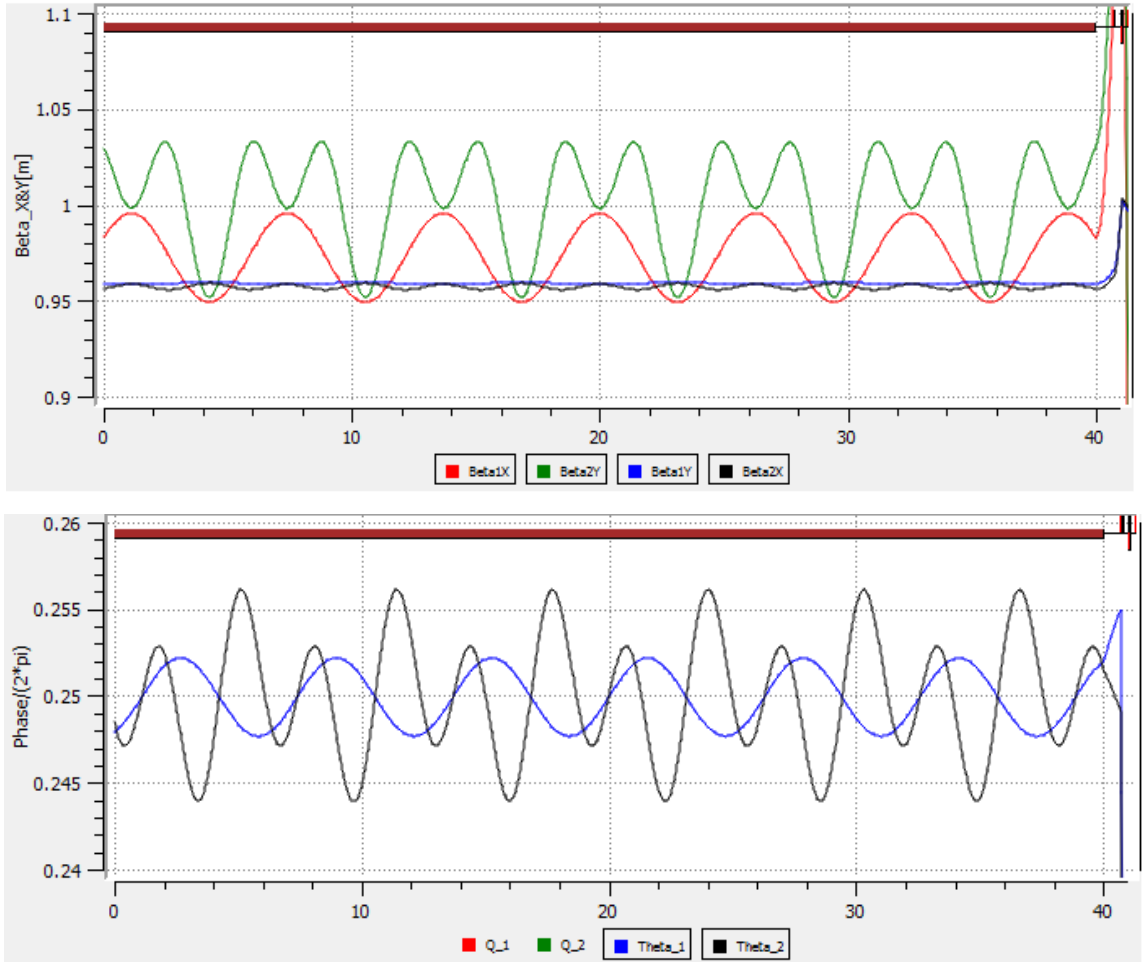

Figure 9: Beta-functions inside solenoid (top) and phase $\nu_1$ and $\nu_2$ in units $\nu/2\pi$ (bottom) for +1% momentum change in the absence of chromaticity correction.

As one can see from Eq. (7) the effective transverse temperature is determined by the emittance of mode 2 (smaller emittance) while the effective cross-section is determined by the emittance of mode 1 (larger emittance). Requiring the cross-section and the temperature variation be within ±10% one obtains that the beta-functions should be within ±10%, and $\delta\nu_1$ and $\delta\nu_2$ within 0.45 (or $\delta\nu/2\pi < 0.07$). That sets requirements to the optics correction accuracy and chromaticity of optics functions.

Figure 9 presents the beta-functions and the phases $\nu_{1,2}$ in the cooling solenoid for +1% momentum change which corresponds to ±5σ of rms momentum spread at the end of cooling cycle. As one can see the beta-functions oscillate near its value at nominal energy with about 6% amplitude. The variations of $\delta\nu_1$ and $\delta\nu_2$ are within ±0.07·2π ≈ 0.44. Thus, the chromatic variations of beta-functions are within requirements even in the absence of chromatic corrections. The natural tune chromaticities are: -14.09 for mode 1 and -5.53 for mode 2. It yields that the maximum tune shifts are 0.14 for the mode 1 and 0.055 for the mode 2. At the low boundary of momentum acceptance of -0.85% the tune for mode 1 approaches an integer. Also, the tunes cross at the momentum deviation of 0.253%. A chromaticity correction, performed with one family of sextupoles, addresses both problems. The sextupoles are built-in into focusing quads of the arcs at the locations with non-zero dispersion. Corresponding tune dependencies are presented in the right pane of Figure 10. After correction the chromaticities are about -7 for mode 1 and -8 for mode 2. The relative variations of beta-functions in the solenoid are within required ±10% in the entire momentum acceptance of the ring of ±0.85%. Focusing accuracy and stability of separate elements is within typical requirements to accelerator elements of about $2·10^{-4}$.

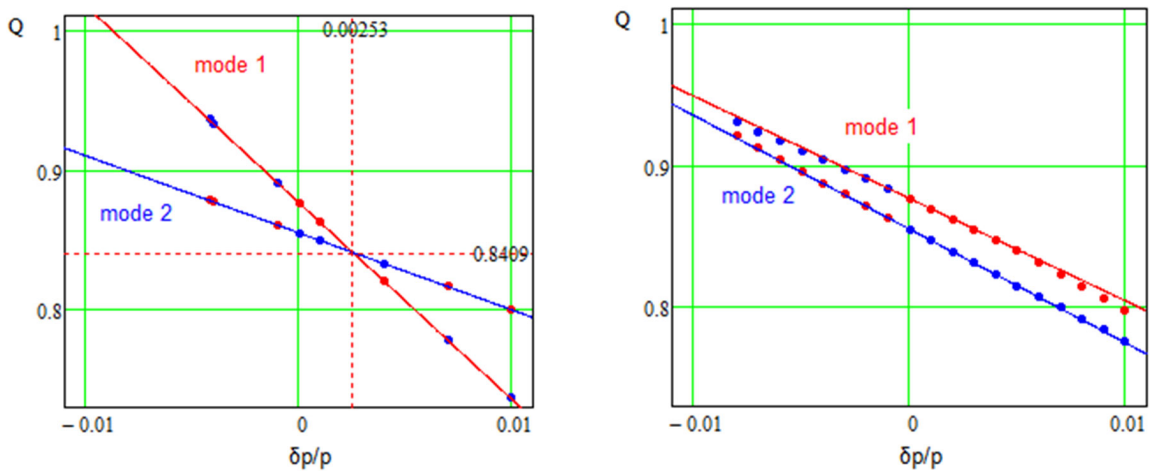

Figure 10: Dependencies of fractional parts of betatron tunes on particle momentum without (left) and with (right) chromaticity correction.

# 5. Intrabeam Scattering and Beam Heating

## 5.1. Multiple intrabeam scattering

Intrabeam scattering is analyzed by using the equations in Ref. [21]. Figure 3 shows the rms betatron beam sizes (minimum and maximum ellipse semi-axes) and their projections to the horizontal and vertical planes. Figure 3 also presents ellipse rotations which are characterized by parameter $\alpha$. Note that by design a significant difference between sizes and projections should appear only at locations of rotators (between the cooling solenoid and the arcs) where the beam ellipse is rotated in the *x-y* plane. Minor rotations visible in the arcs are related to imperfections in the optics of rotators. Some rotations are also present in the technical straight where the betatron modes are circular at the entrance and the exit but an absence of ideal "smooth" focusing results in minor misbalance of beam sizes and ellipse rotations. Note also that decoupling of betatron modes (putting them into horizontal and vertical planes) in the arcs reduces the beam cross-section by about 30 times resulting in that the arcs make the major contribution to the IBS.

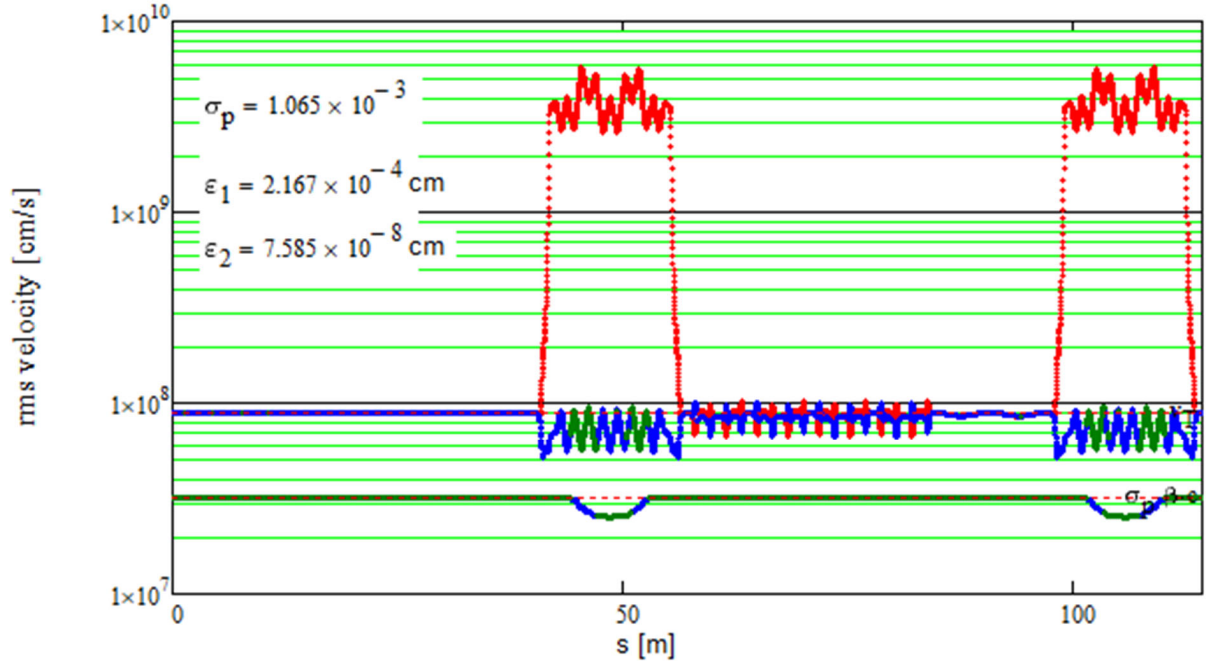

Figure 11: Local rms beam velocity spreads (major semi-axes of the beam ellipse in the space of velocities) in the beam frame for the entire ring for optics shown in Figure 2 and the beam parameters of Table 1.

Figure 11 shows corresponding local rms velocity spreads in the beam frame (major semi-axes of the beam ellipse in the 3D-space of velocities). The bottom (green) line corresponds to the longitudinal velocity spread and the two other lines to the transverse velocity spreads at places with zero dispersion where one of the ellipsoid axis is directed along the longitudinal axis of the coordinate frame. The top line (red points) corresponds to the mode 1 corresponding to the horizontal motion in the arc. As one can see the longitudinal velocity in the solenoid is significantly smaller than the transverse velocities which is required for effective electron cooling. In difference to the rotational modes in the straights, where the transverse velocity spread in the solenoid is determined by the emittance of mode 2 (see Eq. (7)), the decoupled betatron motion in arcs results that the horizontal velocity spread is determined by the larger emittance of mode 1. That yields about 50 times increase of horizontal velocity spread in the arcs. It partially compensates the IBS increase due to decrease of beam transverse cross-section.

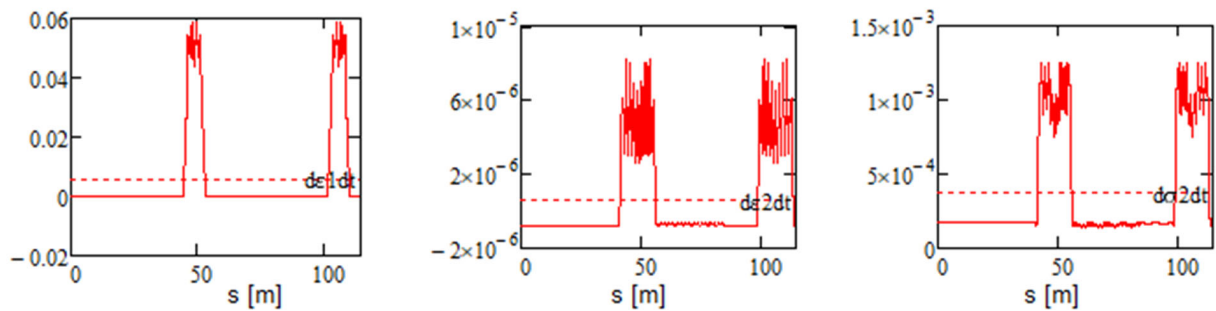

Figure 12: Local contributions to the emittance growth for entire ring: left – mode 1, center – mode 2, right – momentum spread; $\sigma_p=1.065 \cdot 10^{-3}$. Dashed lines show corresponding average values.

Figure 12 shows local contributions to the IBS growth rates determined as: $Cd^2\varepsilon/dtds$, where $C$ is the ring circumference. As one can see the major contribution to the mode 1 (corresponding to the horizontal motion in the arcs) comes from the arcs. It is associated with the non-zero horizontal dispersion and the small vertical beam size. The emittance growth for the mode 2 is suppressed by 4 orders of magnitude because of absence of the vertical dispersion. Similar to the mode 1 the arcs also make the major contribution to the momentum spread growth due to small vertical beam size and, consequently, small beam cross-section. The mentioned above increase of the horizontal velocity spread in the arcs reduces the difference between local contributions to longitudinal heating for arcs and straights to about 6 times. This 6-fold reduction of longitudinal IBS for the vortex modes is the major reason while the vortex modes are also used in the technical straight.

Figure 13 presents the dependences of beam emittances and the momentum spread on time for 50 A electron beam. The rms initial momentum spread of $1.06 \cdot 10^{-3}$ is set by a requirement to avoid a microwave instability (see details in Section 9). The transverse emittances are set by the thermal emittance of the gun and the magnetic field at the cathode. As one can see the beam emittances stay reasonably small in the course of 5 ms. That determines the required repetition rate of 200 Hz.

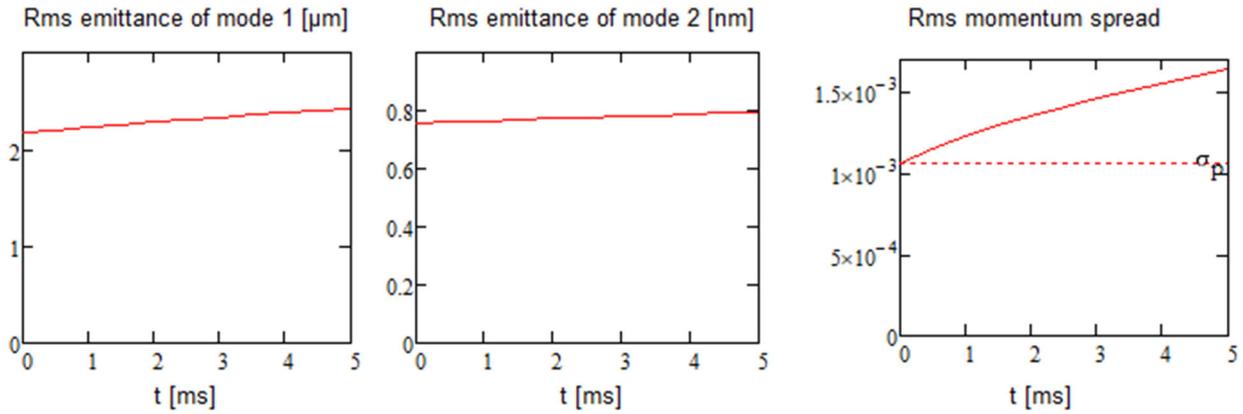

Figure 13: Dependencies of rms mode emittances and momentum spread on time for 50 A electron beam.

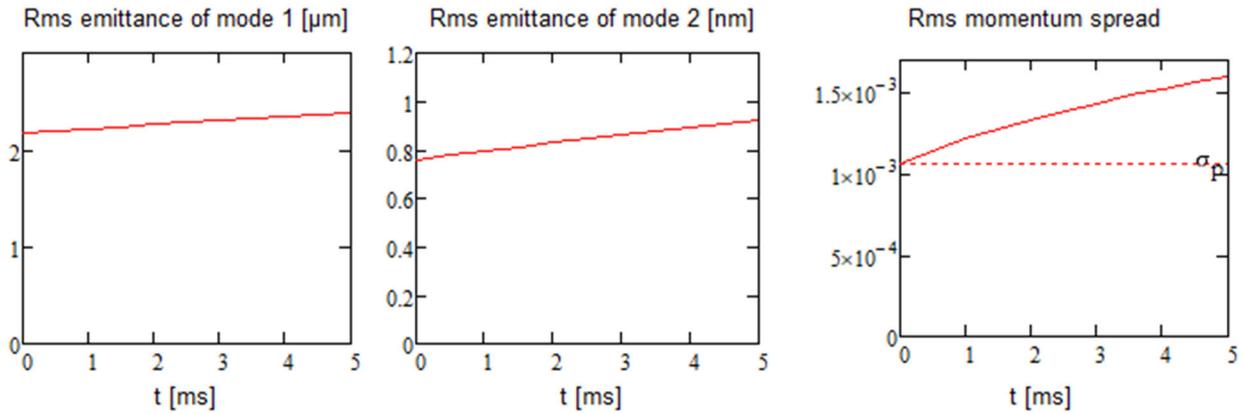

Figure 14: Dependencies of rms mode emittances and momentum spread on time for 50 A electron beam for perturbed optics.

### *4.2. Sensitivity of IBS growth rates to optics errors*

The emittances of modes 1 and 2 are different by more than 3 orders of magnitude. Therefore, an analysis of emittance growth rates on optics perturbations is required. Taking into account that the cooling force is determined by the mode 2 emittance one cannot accept its significant growth. We require the emittance growth of mode 2 due to optics errors to be less than 0.2 nm in 5 ms cycle what is about 1000 times smaller than for the mode 1. To estimate the required accuracy of beam optics we rolled one quadrupole at a high dispersion location by 0.1 deg. It resulted in that the auxiliary beta-functions, $\beta_{1y}$, $\beta_{2x}$, became non-zero in the arcs with peak values of about 2.5 mm, and an appearance of vertical dispersion propagating through the entire ring with amplitude of about 0.5 mm. Corresponding emittance growths are presented in Figure 14. As one can see the faster growth rate for the mode 2 slightly reduces the growth rates for the mode 1 and the momentum spread. There are two major mechanisms resulting in an acceleration of the vertical emittance growth. Both mechanisms are related to optics mismatches in the arcs. The first mechanism is associated with appearance of non-zero $\beta_{2x}$ in dispersive locations in the arcs. That couples this vertical-in-arc mode (mode 2) with the horizontal dispersion and pumps its emittance growth. It requires a suppression of relative value of the auxiliary beta-function, $\beta_{2x}/\beta_{2y}$, in the arcs better or about $10^{-3}$. Another mechanism is related to an appearance of vertical dispersion in the arcs. Accounting that the vertical emittance growth in arcs is proportional $D_y^2/\beta_y$ and that the emittance growth has to be suppressed by 1000 times in comparison with the emittance growth of mode 1 we obtain that the ratio of horizontal to vertical dispersions should be $D_x/D_y \leq \sqrt{1000} \approx 30$.

With usage of modern optics correction tools an achievement of required accuracy is certainly possible. If one uses the LOCO method [22] for the optics measurements the excitation of out-of-plane orbit distortion is about $\sqrt{\beta_{1y}/\beta_{1x}}$ or $\sqrt{\beta_{2x}/\beta_{2y}}$ times of the in-plane orbit distortion. Thus, to obtain the required suppression of auxiliary beta-functions of better than $10^{-3}$ one needs to measure the ratio of amplitudes for the off-plane to the in-plane orbit displacement of better than $\sqrt{\beta_{2x}/\beta_{2y}} \approx 1/30$. That is close to the mentioned above accuracy required for the dispersion suppression.

### 5.3. Touschek scattering

Usually single (Touschek) intrabeam scattering represents the main mechanism of beam loss in a high-intensity electron beam. Computations of Touschek scattering must account for strong *x-y* coupling presented in the ring. For the proposed electron cooling the momentum spread in the beam frame is non-relativistic. That allows one to simplify computations. Consequently, for a continuous beam they result in the following equation for the particle loss rate:

$$\frac{1}{N_e}\frac{dN_e}{dt} = \frac{N_e r_e^2 c}{2\beta^3\gamma^3 C}\left\langle \frac{1}{\sigma_1\sigma_2\sigma_{\theta1}\sigma_{\theta2}\theta_m^2}\int_{\theta_m/\gamma}^{\infty}\left(1-\frac{\theta_m^2}{\gamma^2\theta^2}\left(1+\ln\left(\frac{\gamma\theta}{\theta_m}\right)\right)\right)e^{-\frac{\theta^2}{2}\left(\frac{1}{\sigma_{\theta1}^2}+\frac{1}{\sigma_{\theta2}^2}\right)}I_0\left(\left|\frac{\theta^2}{2}\left(\frac{1}{\sigma_{\theta1}^2}-\frac{1}{\sigma_{\theta2}^2}\right)\right|\right)\right\rangle_s. \quad (10)$$

Here $N_e$ is the number of particles in the electron beam, $r_e$ is the classical radius of electron, $\beta$ and $\gamma$ are the beam Lorentz factors, $C$ is the ring circumference, $c$ is the light speed, $\theta_m$ is the momentum deviation above which the particle is lost, $\sigma_1$ and $\sigma_2$ are the major semi-axes of transverse particle distribution in the real space (rms values of beam sizes), $\sigma_{\theta1}$ and $\sigma_{\theta2}$ are the major semi-axes of particle distribution in the space of transverse angles (rms values of angular spreads), $I_0(\ldots)$ is the modified Bessel function, and $\langle\ldots\rangle_s$ denotes averaging over the ring circumference. $\theta_m \equiv (\Delta p/p)_{max}$ is determined by the momentum aperture in the ring which, in the general case, depends on the longitudinal location of scattering. In general, the momentum aperture depends both on the RF bucket size and aperture limitations. In the absence of RF, it is set by aperture limitations in the arcs. Touschek scattering instantly changes the longitudinal momentum of a particle. If the scattering happens in a location with non-zero dispersion this momentum change also excites the betatron motion with single particle emittance of

$$\varepsilon = \frac{D_x^2 + (\beta_x D' + \alpha_x D)^2}{\beta_x} \theta_\parallel^2 , \qquad (11)$$

where $D_x$ and $D'_x$ are the dispersion and its derivative at the scattering location, $\beta_x$ and $\alpha_x$ are the corresponding horizontal beta- and alpha-functions, $\theta_\parallel$ is the longitudinal momentum change, and we assume that the vertical dispersion and x-y coupling are absent at the scattering location. At places with large dispersion this excitation about doubles the particle maximum deviation from the reference orbit and reduces the momentum aperture in the same proportion. For a scattering in a location with lattice parameters $\beta_x$, $\alpha_x$, $D_x$ the maximum momentum deviation is:

$$\theta_m = \frac{a^*}{D^* + \sqrt{\frac{\beta^*}{\beta_x}\left(D^2 + (\beta_x D' + \alpha_x D)^2\right)}} , \qquad (12)$$

where $a^*$ is the aperture at the location which limits the maximum momentum deviation. For the chosen lattice the locations for $D^*$ and $\beta^*$ (dispersion and beta-function) coincide with the location of maximum dispersion. Note that it is usually the case for other lattices.

The aperture in the cooling ring is round with radius of 2.54 cm in its entire length except injection and extraction regions. In the computation of the Touschek loss rate we reduced the aperture radius by 5 mm. This reduction includes two items: (1) the energy spread of the beam ($D^*\sigma_p = 2$ mm), which is not accounted in Eq. (10), and (2) 3 mm of closed orbit distortion. Figure 15 shows a dependence of $\theta_m$ on a location in the ring for one of the two arcs. Performing computations in Eq. (10) we obtain the Touschek lifetime of about 17 s. Figure 16 presents the corresponding rms beam sizes, local momentum spreads and the local contributions to the Touschek loss rate. As one can see the arcs strongly dominate the particle loss. The Touschek scattering results in ~61 W power loss which is sufficiently large and has to be intercepted by cooled surfaces. About half of particles will be immediately lost in the arcs. Other potentially could be intercepted by collimators.

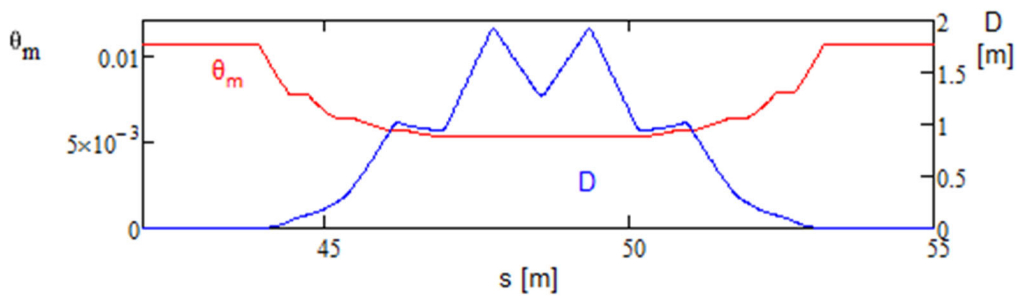

Figure 15: Dependence of $\theta_m$ (red line) and dispersion (blue line) on the loss location shown for one of two arcs.

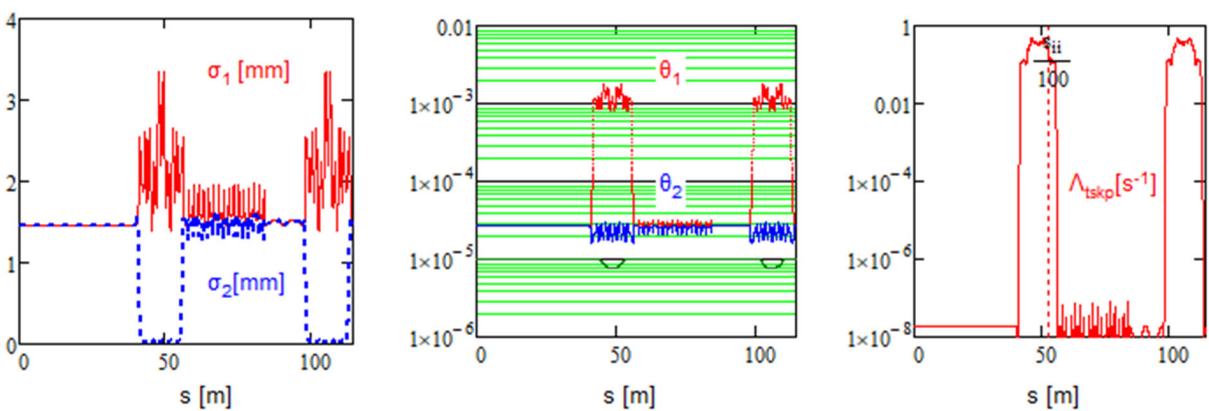

Figure 16: Dependencies of rms beam sizes (left), momentum spreads (center) and the local contributions to the Touschek loss rate (right) on the location along the ring. The bottom dark green line in the center figure presents effective longitudinal momentum spread in the beam frame ($\theta_s/\gamma$).

### 5.4. *Electron beam heating by the proton beam*

Electron beam heating by the proton beam has two sources. The first one is due to binary scattering of electrons at protons, and the second one is due to excitation of the coherent motion of electrons by the proton beam space charge. As will be shown below the first source is significantly smaller than the IBS scattering in the electron beam, while the second source may contribute significant heating if not properly addressed. Below we assume the proton bunch parameters of JLEIC [23]: the number of particles per bunch - $N_p = 1.1 \cdot 10^{10}$, the average bunch-to-bunch distance - $L_{bb}$=68 cm, and the rms bunch length - $\sigma_s$=2.5 cm.

To estimate effect of binary scattering of electrons at protons we compare it with the IBS scattering in the electron beam considered above in Section 5.1. Similar to the electron beam the longitudinal velocity spread in the proton beam is much smaller than the transverse one; and the dispersion in the cooling section of the electron ring is zero. These two conditions result in that the growth of longitudinal momentum spread due to binary collisions of electrons with protons significantly exceeds the growth rates for transverse degrees of freedom. Therefore, we only consider the longitudinal electron heating by binary collisions between protons and electrons. We also account that:

- The transverse velocity spread in the proton beam is twice smaller than in the electron one. That yields twice faster heating for collisions with protons.
- The rms size of the proton beam is ~3 times smaller than the electron beam, however the proton beam current (8.5 A) is ~10 times smaller than the electron one. That leaves the proton beam density being close to the electron beam density. Consequently, it leaves the corresponding heating terms being approximately equal for the IBS in the electron beam and the electron heating by protons.
- As one can see from Figure 12 the longitudinal heating rate in the solenoid is about half of the average heating rate; and the length of the section, where the electron and proton beams overlap, is ~3 times shorter than the electron ring circumference. Thus, these terms result in the proton heating to be about six times smaller than the IBS.

Combining results of the above statements, we conclude that the longitudinal heating rate due to binary collisions with protons should be about 30% of the IBS in the electron beam. Note also that the rotational modes in the solenoid and large difference between the mode emittances result in that the radius of a particle in the electron beam in the cooling solenoid stays approximately constant as over propagation through the cooling section as in the course of all following passages. Therefore, binary collisions with protons affect only particles located in the electron beam center which radial positions overlap with the proton bunches.

Now we find contributions from the proton beam space charge on the growth of emittances and momentum spread.

We start our consideration from the transverse planes. As it was already mentioned above due to large difference between mode emittances the Larmor radius of each electron in the cooling solenoid is much smaller than its average radial position. In this case we can use a perturbation theory and assume that the proton beam electric field is constant in the area of electron motion. Consequently, we can consider that an electron is moving in crossed electric and magnetic fields. Without limiting generality of consideration, we can assume that the initial vertical position is equal to zero. Consequently, we obtain the perturbations of the electron transverse angles and coordinates by the electric field of proton beam:

$$\delta\theta_x = -\frac{E}{B_0\beta}\sin(kL_s), \quad \delta x = \frac{E}{B_0\beta}\frac{\cos(kL_s)-1}{k},$$
$$\delta\theta_y = \frac{E}{B_0\beta}(\cos(kL_s)-1), \quad \delta y = \frac{E}{B_0\beta}\left(\frac{\sin(kL_s)}{k}-s\right), \tag{13}$$

where $E$ is the electric field of the proton bunch at location of the electron (positive for the force directed to the beam center), $B_0$ is the magnetic field in the cooling solenoid and $L_s$ is its length, $\beta = v/c$ is the dimensionless velocity of electron, and $k = eB_0/pc$ is the wave-number of electron Larmor motion. To find the corresponding change of the electron mode emittances we express the vector of particle 4D coordinates through the eigen-vectors and the single particle emittances (actions) [16]:

$$\mathbf{x} = \frac{1}{2}\left(\sqrt{2\varepsilon_1}\,\mathbf{v}_1 e^{i\mu_1} + \sqrt{2\varepsilon_2}\,\mathbf{v}_2 e^{i\mu_2} + CC\right). \tag{14}$$

Multiplying this equation by $\mathbf{v}_{1,2}{}^+\mathbf{U}$ at both sides and accounting the symplectic orthogonality condition,

$$\begin{aligned}\mathbf{v}_i{}^+\mathbf{U}\mathbf{v}_j &= -2i\delta_{ij}, \\ \mathbf{v}_i{}^+\mathbf{U}\mathbf{v}_j &= 0,\end{aligned} \quad i,j=1,2, \tag{15}$$

we obtain:

$$\varepsilon_{1,2} = \frac{1}{2}\left|\mathbf{v}_{1,2}{}^+\mathbf{U}\mathbf{x}\right|^2, \tag{16}$$

where $\mathbf{U}$ is the unit symplectic matrix. In the cooling solenoid the eigen-vectors and the vector of perturbation are:

$$\mathbf{v}_1 = \begin{bmatrix} \sqrt{\beta_0} \\ -\dfrac{i}{2\sqrt{\beta_0}} \\ -i\sqrt{\beta_0} \\ -\dfrac{1}{2\sqrt{\beta_0}} \end{bmatrix}, \quad \mathbf{v}_2 = \begin{bmatrix} -i\sqrt{\beta_0} \\ -\dfrac{1}{2\sqrt{\beta_0}} \\ \sqrt{\beta_0} \\ -\dfrac{i}{2\sqrt{\beta_0}} \end{bmatrix}, \quad \delta\mathbf{x} = \begin{bmatrix} \delta x \\ \delta\theta_x \\ \delta y \\ \delta\theta_y \end{bmatrix}. \tag{17}$$

Assuming that the perturbations of particle coordinates and angles are much smaller than the initial coordinates and angles, and performing averaging over initial phases of betatron motion we obtain average changes of mode emittances:

$$\delta\varepsilon_{1,2} = \frac{1}{2}\left|\mathbf{v}_{1,2}^{+}\mathbf{U}\delta\mathbf{x}\right|^2 = \frac{1}{2}\left(\left(\sqrt{\beta_0}\delta\theta_x \pm \frac{\delta y}{2\sqrt{\beta_0}}\right)^2 + \left(\sqrt{\beta_0}\delta\theta_y \mp \frac{\delta x}{2\sqrt{\beta_0}}\right)^2\right). \quad (18)$$

Here the top sign belongs to the mode 1 and the bottom sign to the mode 2. Substituting Eqs. (13) into Eq. (18) and dropping the terms related to the non-oscillating part in $\delta y$ we obtain:

$$\delta\varepsilon_{1,2} = \frac{9}{2}\beta_0\left(\frac{E}{B_0\beta}\right)^2 \sin^2\left(\frac{L_s}{2\beta_0}\right)\left(1 \mp \frac{1}{2}\right), \quad (19)$$

where we accounted that $k = 1/\beta_0$, and we need to note that the non-oscillating term in $\delta y$ ($\propto s$) appeared in Eq. (13) due to simplification where we considered the flat geometry instead of the cylindrical one. In accurate consideration it would be canceled in Eq. (18) by other terms appearing due to cylindrical geometry. As one can see from Eq. (19) the changes in the mode emittances are of the same order. Taking into account that the emittance for mode 2 is much smaller than for mode 1 we further consider the mode 2 only.

We assume round proton beam with Gaussian distribution. Then its transverse electric field is:

$$E \equiv E(r,s) = \frac{2eN_p}{\sqrt{2\pi}\sigma_s\gamma^2 r}e^{-s^2/2\sigma_s^2}\left(1-e^{-r^2/2\sigma_r^2}\right), \quad (20)$$

where $\sigma_s$ and $\sigma_r$ are the rms length and the rms transverse size of the proton beam, and $N_p$ is the number of particles in the beam. Substituting this equation into Eq. (19), performing averaging over longitudinal coordinate and multiplying it by number of turns we obtain:

$$\delta\varepsilon_2 = \frac{9r_e^2 N_p^2 \beta_0^2 n_\sigma^2 N_{turn}}{\sqrt{\pi}\beta^4\gamma^6\sigma_s L_{bb}\varepsilon_1}\left(\sigma_r\frac{1-e^{-r^2/2\sigma_r^2}}{r}\right)^2 \sin^2\left(\frac{L_s}{2\beta_f}\right). \quad (21)$$

Here $n_\sigma = \sqrt{\varepsilon_1\beta_0}/\sigma_r$ is the ratio of electron beam size to the proton beam size, $L_{bb}$ is the bunch to bunch distance, $\varepsilon_1$ is the emittance of mode 1, and $N_{turn}$ is the number of turns; and we assume that the revolution frequency in the electron beam and the bunch frequency are not harmonically related so that an appearance of proton bunches in the electron beam is uniformly distributed in time.

The maximum emittance increase, as a function of radial position, is achieved at $r/\sigma_r \approx 1.59$. Using this radius and $N_p = 1.1\cdot 10^{10}$, $L_{bb}$=68 cm, $\sigma_s$=2.5 cm, $\beta_0$=99.2 cm, $\varepsilon_1$=4.2 μm, $\varepsilon_2$=0.75 nm and

$N_{turn}$=13,000 we obtain:

$$\frac{\delta\varepsilon_2}{\varepsilon_2} \approx 1.4\sin^2\left(\frac{L_s}{2\beta_0}\right).$$

As one can see from the above equation, if uncompensated, the emittance growth is unacceptably large. If the beam interaction would happen in the solenoid only, then the complete compensation of the emittance increase would be achieved when the integer number of Larmor periods occurs in the cooling solenoid: $L_s = 2\pi n\beta_0$. However, in the real cooler there is also an interaction between beams near the solenoid entrance and exit. To minimize the beam interaction outside of the solenoid one needs to separate beams at distance much shorter than $\beta_0$. The required angular kick is ~5 mrad. It can be achieved by dipole corrector with $BdL$= 1 kG·cm or by vertical displacement of the solenoid axis from the beam center by 10 mm. In this case the required horizontal kick is produced by the edge field of the solenoid. To achieve good compensation for the entire interaction straight, which in addition to the solenoid includes these exit/entrance regions, one needs to use the betatron phase advance in the solenoid somewhat lower than an integer. Detailed optimization will be performed later when design of the solenoid and beam separation will be known. We also need to stress the importance of relative beam positioning to be accurate to better than 0.1 mm in the entire cooling section. Note that similar requirements are also necessary for effective cooling.

Next, we find a contribution of the proton beam space charge to the growth of longitudinal momentum spread. Due to large difference between radii of the vacuum chamber and the beams the longitudinal electric field does not change significantly at the electron bunch cross-section. Its averaged value can be approximated as:

$$E_\parallel \approx \sqrt{\frac{2}{\pi}} \frac{eN_p}{\sigma_s^2 \gamma^2} \ln\left(\frac{a}{1.5\sigma_r}\right)\left(\frac{s}{\sigma_s} e^{-s^2/2\sigma_s^2}\right), \quad (22)$$

where $a$ is the vacuum chamber radius. Then, for the relative longitudinal momentum change of an electron we obtain:

$$\delta\theta_\parallel \equiv \frac{\delta p_\parallel}{p} E_\parallel \approx \sqrt{\frac{2}{\pi}} \frac{r_e N_p L_s}{\sigma_s^2 \beta\gamma^3} \ln\left(\frac{a}{1.5\sigma_r}\right)\left(\frac{s}{\sigma_s} e^{-s^2/2\sigma_s^2}\right). \quad (23)$$

Assuming uniformly distributed appearance probability of proton bunches in the electron beam we find the average relative momentum spread increase to be:

$$\overline{\delta\theta_\|^2} \approx \frac{N_p^2 r_e^2 L_s^2 N_{turn}}{\sqrt{\pi}\beta^2\gamma^6\sigma_s^3 L_{bb}} \ln^2\left(\frac{a}{1.5\sigma_r}\right) \ . \tag{24}$$

Using introduced above proton beam parameters and the initial rms momentum spread of $\sigma_p$ = 1.065·10$^{-3}$ we obtain an increase of relative momentum spread $\sqrt{\overline{\delta\theta_\|^2}}/\sigma_p \approx 0.082$. Taking into account that the momentum growth rates are added as squares this value does not produce significant contribution into the growth of longitudinal momentum spread (~1%). However, it may start to dominate the electron beam heating with further increase of the proton beam intensity.

In conclusion we need to note:

- That the above estimates of electron beam heating do not account a collective response of the beam which becomes important if the beam is close to the instability threshold at the bunch frequencies. In this case the emittance growth rate is amplified as $\approx I_{th}^2/(I_{th}-I)^2$, where $I$ is the electron beam current, and $I_{th}$ is the instability threshold current.
- The electro-magnetic fields excited by the proton bunch due to its interaction with the vacuum chamber walls (mostly through the resistive wall impedance) are significantly smaller than the effects of the space charge for chosen beam energy; and therefore, they can be neglected for both the longitudinal and transverse excitations.

## 6. Cooling Rates

As can be seen in Table 3, the transverse and longitudinal temperatures of the electron beam are significantly larger than in the low energy magnetized cooling, based on the electrostatic acceleration of electron beam. Therefore, effects of magnetization in the electron cooling are small and we can neglect the magnetization. Consequently, the dependence of cooling force on the proton velocity in the beam frame, $\mathbf{v} = (v_x, v_y, v_z)$, is described by the following equation [24]:

$$\mathbf{F}(\mathbf{v}) = \frac{4\pi n'_e e^4 L_c}{m_e} \int f(\mathbf{v}') \frac{\mathbf{v} - \mathbf{v}'}{|\mathbf{v} - \mathbf{v}'|^3} d\mathbf{v}'^3 = \frac{4\pi n'_e e^4 L_c}{m_e} \nabla_\mathbf{v} \left( \int \frac{f(\mathbf{v}')}{|\mathbf{v} - \mathbf{v}'|} d\mathbf{v}'^3 \right). \tag{25}$$

Here $f(\mathbf{v})$ is the electron distribution function over velocity, $e$ is the electron charge, $n'_e$ is the density of electrons, $L_c$ is the Coulomb logarithm, and $\nabla_\mathbf{v}$ denotes the gradient in the space of velocities. For a Gaussian distribution function,

$$f(v_x, v_y, v_z) = \frac{1}{(2\pi)^{3/2} \sigma_{vx} \sigma_{vy} \sigma_{vz}} \exp\left( -\frac{v_x^2}{2\sigma_{vx}^2} - \frac{v_y^2}{2\sigma_{vy}^2} - \frac{v_z^2}{2\sigma_{vz}^2} \right), \tag{26}$$

Eq. (25) can be rewritten as [25, 26]:

$$\mathbf{F}(\mathbf{v}) = \frac{4\pi n'_e e^4 L_c}{m_e} \nabla_\mathbf{v} \left( \frac{2}{\sqrt{\pi}} \int_0^\infty \frac{\exp\left( -\frac{v_x^2 t^2}{1+2\sigma_{vx}^2 t^2} - \frac{v_y^2 t^2}{1+2\sigma_{vy}^2 t^2} - \frac{v_z^2 t^2}{1+2\sigma_{vz}^2 t^2} \right)}{\sqrt{(1+2\sigma_{vx}^2 t^2)(1+2\sigma_{vy}^2 t^2)(1+2\sigma_{vz}^2 t^2)}} dt \right). \tag{27}$$

Accounting that both transverse temperatures are equal, $\sigma_{vx} = \sigma_{vy} \equiv \sigma_{v\perp}$, we can write for the transverse and longitudinal forces:

$$F_\parallel(v_z) = \frac{4\pi n'_e e^4 L_c}{m_e} \left( \frac{4v_z}{\sqrt{\pi}} \int_0^\infty \exp\left( -\frac{v_z^2 t^2}{1+2\sigma_{vz}^2 t^2} \right) \frac{t^2 dt}{\left(1+2\sigma_{v\perp}^2 t^2\right)\left(1+2\sigma_{vz}^2 t^2\right)^{3/2}} \right),$$

$$\mathbf{F}_\perp(\mathbf{v}_\perp) = \frac{4\pi n'_e e^4 L_c}{m_e} \left( \frac{4\mathbf{v}_\perp}{\sqrt{\pi}} \int_0^\infty \exp\left( -\frac{v_\perp^2 t^2}{1+2\sigma_{v\perp}^2 t^2} \right) \frac{t^2 dt}{\left(1+2\sigma_{v\perp}^2 t^2\right)^2 \sqrt{1+2\sigma_{vz}^2 t^2}} \right),$$

(28)

where by definition the longitudinal/transverse force is defined for the pure longitudinal/transverse proton velocity.

**Table 3: Electron beam parameters in the beam frame**

| | |
|---|---|
| Electron density, $n'_e$ | $1.54 \cdot 10^9$ cm$^{-3}$ |
| Transverse electron temperature at the cooling start | 4.5 eV |
| Longitudinal electron temperature at the cooling start | 0.58 eV |
| Larmor radius | 27 μm |
| Debye radius | 420 μm |
| Average distance between electrons, $n'^{-1/3}_e$ | 8.9 μm |
| Rms transverse velocity at the cooling cycle start | $8.9 \cdot 10^7$ cm/s |
| Rms longitudinal velocity at the cooling cycle start | $3.2 \cdot 10^7$ cm/s |
| Larmor frequency, $\omega_L = eB_0/m_e c$ | $3.25 \cdot 10^{10}$ s$^{-1}$ |
| Plasma frequency, $\omega_p = \sqrt{4\pi n_{ebf} e^2 / m_e}$ | $2.14 \cdot 10^9$ s$^{-1}$ |
| Length of Larmor period in the lab frame, $2\pi pc/eB_0$ | 6.23 m |
| Length of plasma period in the lab frame, $2\pi\beta\gamma c/\omega_p$ | 94.8 m |

To find the cooling rates we need to average cooling over the proton distribution function. In the beam frame the average rate of particle kinetic energy change is:

$$\frac{d}{dt}\left(\frac{m_p \overline{v_i^2}}{2}\right) = \frac{1}{2}\int v_i F_i(\mathbf{v}) f_p(\mathbf{v}) d\mathbf{v}^3 , \quad i = x, y, z ,  \tag{29}$$

where $f_p(\mathbf{v})$ is the distribution function of the proton beam, the repeated index $i$ does not imply a summation, and the factor ½ in the right-hand side accounts the energy equipartitioning for linear oscillator as we assume that the proton beam is bunched in all three degrees of freedom. Substituting Eq. (27) into Eq. (29) and using the Gaussian distribution of the proton beam,

$$f_p(v_x, v_y, v_z) = \frac{1}{(2\pi)^{3/2} \sigma_{vpx}\sigma_{vpy}\sigma_{vpz}} \exp\left(-\frac{v_x^2}{2\sigma_{vpx}^2} - \frac{v_y^2}{2\sigma_{vpy}^2} - \frac{v_z^2}{2\sigma_{vpz}^2}\right) , \tag{30}$$

we obtain the emittance cooling rates for each degree of freedom:

$$\lambda_i \equiv \frac{1}{\sigma_{vpi}^2} \frac{\overline{dv_i^2}}{dt} = \frac{16\sqrt{\pi} n_e' e^4 L_c}{m_e m_p \sigma_{vpi}^2} \int \frac{d\mathbf{v}^3}{(2\pi)^{3/2} \sigma_{vpx}\sigma_{vpy}\sigma_{vpz}} \exp\left(-\frac{v_x^2}{2\sigma_{vpx}^2} - \frac{v_y^2}{2\sigma_{vpy}^2} - \frac{v_z^2}{2\sigma_{vpz}^2}\right)$$
$$\times v_i^2 \int_0^\infty \exp\left(-\frac{v_x^2}{2\sigma_{vx}^2} - \frac{v_y^2}{2\sigma_{vy}^2} - \frac{v_z^2}{2\sigma_{vz}^2}\right) \frac{t^2 dt}{\left(1+2\sigma_{vi}^2 t^2\right)^{3/2} \sqrt{1+2\sigma_{vj}^2 t^2} \sqrt{1+2\sigma_{vk}^2 t^2}} \quad , \tag{31}$$

where $i = x, y, z$, and indexes $i, j, k$ are different ($i \neq j \neq k, i \neq k$). Performing integration over proton velocity we obtain the emittance cooling rate in the beam frame:

$$\lambda_i = \frac{4\sqrt{2\pi} n_e' e^4 L_c}{m_e m_p} \int_0^\infty \frac{t^2 dt}{\left(1+\Sigma_{vi}^2 t^2\right)^{3/2} \sqrt{1+\Sigma_{vj}^2 t^2} \sqrt{1+\Sigma_{vk}^2 t^2}} \quad , \tag{32}$$

where

$$\Sigma_{vx} = \sqrt{\sigma_{vpx}^2 + \sigma_{vx}^2}, \quad \Sigma_{vy} = \sqrt{\sigma_{vpy}^2 + \sigma_{vy}^2}, \quad \Sigma_{vz} = \sqrt{\sigma_{vpz}^2 + \sigma_{vz}^2} \tag{33}$$

are the effective rms velocity spreads.

In the case of equal effective transverse temperatures, $\Sigma_{vx} = \Sigma_{vy} = \Sigma_{v\perp}$, the integral in Eq. (32) can be reduced to the elementary functions. First, we rewrite Eq. (32) in the following form:

$$\lambda_z = \frac{4\sqrt{2\pi} n_e' e^4 L_c}{m_e m_p \Sigma_{v\perp}^2 \Sigma_{vz}^2} \Psi_L\left(\frac{\Sigma_{vz}}{\Sigma_{v\perp}}\right), \quad \Psi_L(x) = \frac{x}{2}\int_0^\infty \frac{dt}{\left(t+x^2\right)^{3/2}(1+t)},$$
$$\lambda_\perp = \frac{2\sqrt{2\pi} n_e' e^4 L_c}{m_e m_p \Sigma_{v\perp}^3} \Psi_\perp\left(\frac{\Sigma_{vz}}{\Sigma_{v\perp}}\right), \quad \Psi_\perp(x) = \int_0^\infty \frac{dt}{\sqrt{x^2+t}(1+t)^2}. \tag{34}$$

Then, the computation of integrals yields:

$$\Psi_L(x) = \frac{1}{1-x^2}\left(1 + \frac{x}{2\sqrt{x^2-1}} \ln\left(\frac{x-\sqrt{x^2-1}}{x+\sqrt{x^2-1}}\right)\right),$$
$$\Psi_\perp(x) = \frac{1}{x^2-1}\left(x + \frac{1}{2\sqrt{x^2-1}} \ln\left(\frac{x-\sqrt{x^2-1}}{x+\sqrt{x^2-1}}\right)\right). \tag{35}$$

For practically useful case when $\Sigma_{vz} \leq 2\Sigma_{v\perp}$ the functions in Eq. (35) can be approximated within ±2% with the following simple equations:

$$\Psi_L(x) \approx \frac{1}{(1+1.083x)^{3/2}},$$
$$\Psi_\perp(x) \approx \frac{\pi}{2}\frac{1}{1+\sqrt{2}x}, \quad x \leq 2. \tag{36}$$

In the case of relativistic beams, we need to transform the above equations to the lab frame. In this transition we account that the both transverse temperatures in the electron beam are equal. Then from Eqs. (28) we obtain the longitudinal and transverse cooling forces:

$$\frac{1}{p}\frac{dp_{\parallel}}{ds} = \frac{16\sqrt{\pi}n_e r_e r_p L_c}{\beta^4 \gamma^2 \sigma_{ep}^3} \theta_{\parallel} \int_0^{\infty} \exp\left(-\frac{t^2}{1+2t^2}\frac{\theta_{\parallel}^2}{\sigma_{ep}^2}\right) \frac{t^2 dt}{\left(1+2t^2\right)^{3/2}\left(1+2\left(\frac{\gamma\theta_e}{\sigma_{ep}}\right)^2 t^2\right)},$$

$$\frac{1}{p}\frac{dp_{\perp}}{ds} = \frac{16\sqrt{\pi}n_e r_e r_p L_c}{\beta^4 \gamma^5 \theta_e^3} \theta_{\perp} \int_0^{\infty} \exp\left(-\frac{t^2}{1+2t^2}\frac{\theta_{\perp}^2}{\theta_e^2}\right) \frac{t^2 dt}{\left(1+2t^2\right)^{5/2}\sqrt{1+2\left(\frac{\sigma_{ep}}{\gamma\theta_e}\right)^2 t^2}}.$$

(37)

Here $\theta_{\parallel} = \Delta p_{\parallel}/p$ and $\theta_{\perp} = \Delta p_{\perp}/p$ are the relative momentum deviations of a proton, $\gamma$ and $\beta$ are the relativistic factors, $n_e = \gamma n'_e$ is the electron beam density in the lab frame, $r_e$ and $r_p$ are the classical electron and proton radii, and $\sigma_{ep} = \sigma_{vz}/(c\beta)$ and $\theta_e = \sigma_{vx}/(c\beta\gamma) = \sigma_{vy}/(c\beta\gamma)$ are the relative rms longitudinal and transverse momentum spreads in the electron beam, respectively.

Similarly, assuming that the both transverse velocity spreads in the proton beam are equal and using an approximation of Eqs. (35) in Eqs. (34) we obtain the emittance cooling rates:

$$\lambda_z \approx \frac{4\sqrt{2\pi}n_e r_e r_p L_c}{\gamma^4 \beta^4 \left(\Theta_{\perp} + 1.083\Theta_{\parallel}/\gamma\right)^{3/2} \sqrt{\Theta_{\perp}\Theta_{\parallel}}} L_{cs} f_0,$$

$$\lambda_{\perp} \approx \frac{\pi\sqrt{2\pi}n_e r_e r_p L_c}{\gamma^5 \beta^4 \Theta_{\perp}^2 \left(\Theta_{\perp} + \sqrt{2}\Theta_{\parallel}/\gamma\right)} L_{cs} f_0,$$

$$\frac{\Theta_{\parallel}}{\gamma\Theta_{\perp}} \leq 2.$$

(38)

Here $L_{cs}$ is the length of cooling section, $f_0$ is the revolution frequency of proton beam, $\Theta_{\parallel}$ and $\Theta_{\perp}$ are the effective relative rms momentum spreads so that

$$\Theta_{\parallel} = \sqrt{\sigma_{pe}^2 + \sigma_{pp}^2},$$

$$\Theta_{\perp} = \sqrt{\theta_e^2 + \theta_p^2},$$

(39)

$\sigma_{pp}$ is the rms momentum spread in the proton beam, and $\theta_p$ is the rms angular spread of protons in the cooling section.

Figure 17 presents plots of the cooling forces at the beginning and the end of cooling cycle computed with help of Eqs. (37). One can see that to the end of cooling cycle, changes in the longitudinal cooling force is larger than in the transverse one as it should be expected due to much larger change in the longitudinal temperature of electron beam (see Figure 13). The peaks of the transverse and longitudinal forces are reduced by 1.28 and 1.38 times respectively; and the slopes, which are responsible for the cooling rates, are changed by 1.32 and 2.01 times. Figure 18 presents the dependence of cooling rates on time within 5 ms cooling cycle for the case when the relative rms momentum spread and the angular spread in the proton beam are much smaller than in the electron beam (see Eq. (39)). It is good approximation for the case of the considered here cooling proposal. An increase of cooling time with time in the cooling cycle is related to the emittance growth due to IBS. It is implied in Figure 18 that the temperatures in the electron beam are determined by the thermal emittance at the cathode and the IBS. The cooling times averaged over cycle are about 30 and 14 minutes for the transverse and longitudinal planes, respectively.

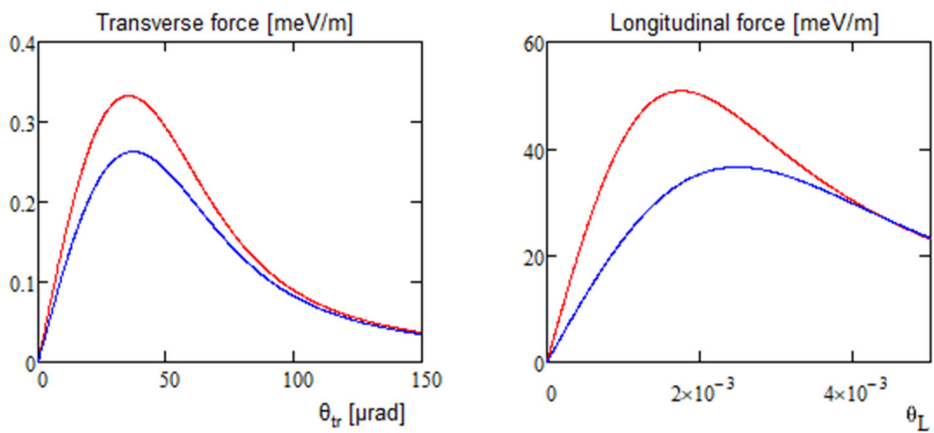

Figure 17: Dependencies of transverse (left) and longitudinal (right) cooling forces at the cooling cycle beginning (red curve, 0 ms) and at the cooling cycle end (blue curve, 5 ms). The initial electron beam emittances and momentum spread are determined in Table 1 and their values at the cycle end are determined by IBS (see Figure 13).

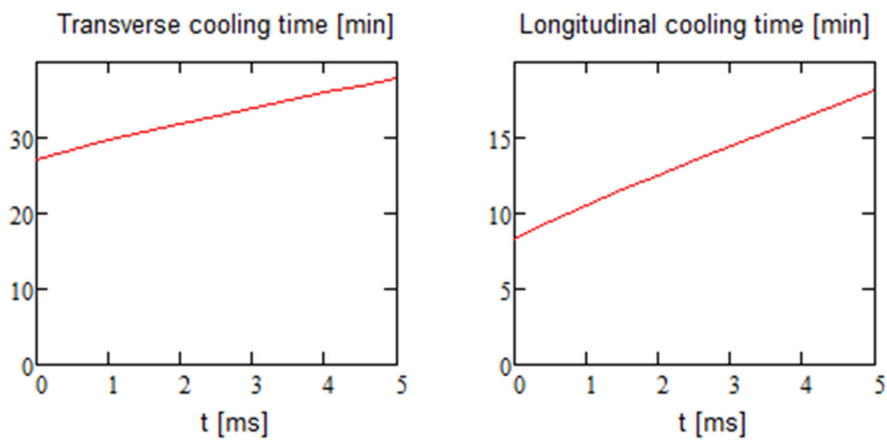

Figure 18: Dependencies of the transverse (left) and longitudinal (right) emittance cooling times on time within one cooling cycle for the case when the proton beam emittance and momentum spread in the proton beam are much smaller than for the electron beam.

Considering the above calculations being somewhat optimistic we further assume that in real operating conditions there are the following sources of the electron beam emittance growth: (1) 20% emittance growth of mode 2 in the electron gun and additional 20% in the induction linac, (2) 20% emittance growth of mode 1 in the gun and linac, (3) 20% emittance growth for each mode related to the optics mismatches in the ring and beam transfers, (4) 30% emittance growth of mode 2 due to beam interaction with protons, and (5) an angular misalignment of the electron beam of 10 μrad rms related to imperfect alignment of magnetic field lines of the solenoid. That yields the transverse temperature increase of 2.1 times and the electron beam radius increase of 1.2 times. The initial longitudinal momentum spread is artificially increased in the induction linac and therefore is not affected by imperfections. Consequently, we obtain an increase of the cooling times of about 2 times for both degrees of freedom, *i.e.* cooling times increase to 1 and 0.5 hour for the transverse and longitudinal planes, respectively. These values are presented in Table 1.

Finally, we consider how the transverse cooling force depends on the betatron amplitude of a proton. Assuming the electron beam is in a quasi-equilibrium Gaussian state and accounting that there is large difference between mode emittances the distribution function of electrons in the cooling solenoid can be presented as (see Eqs. (7) and (9)):

$$f(\boldsymbol{\theta}, \mathbf{r}_\perp) = \frac{1}{(2\pi)^{5/2} \theta_e^2 \sigma_\perp^2 \sigma_{pe}} \exp\left(-\frac{\theta_x^2 + \theta_y^2}{2\theta_e^2} - \frac{\theta_s^2}{2\sigma_{pe}^2} - \frac{x^2 + y^2}{2\sigma_\perp^2}\right) \quad (40)$$

where $\sigma_\perp = \sqrt{\varepsilon_1 \beta_0}$ and $\theta_e = \sqrt{\varepsilon_2 / \beta_0}$ are the rms transverse size and angular spread of electron beam, and $\sigma_{pe}$ is its rms momentum spread. We will reference the proton beam betatron motion to the center of cooling section where we also assume that the rms transverse beam sizes are equal and achieve their minimum. Then the proton positions are:

$$x_p = \sqrt{2I_{px}\beta_p} \cos\mu_x, \quad y_p = \sqrt{2I_{py}\beta_p} \cos\mu_y,$$
$$\theta_{xp} = \sqrt{\frac{2I_{px}}{\beta_p}} \sin\mu_x, \quad \theta_{yp} = \sqrt{\frac{2I_{py}}{\beta_p}} \sin\mu_y. \quad (41)$$

To find the cooling rate for given particle actions ($I_{px}$, $I_{py}$) we will use the cooling force of Eq. (37) where we additionally need to account for changing (at each turn) electron density averaged over particle path and to perform averaging over betatron motion. Using Eqs. (40) and performing integration over cooling section we can write the following equation for the average electron density:

$$\overline{n}_e = \frac{n_e}{L_s} \int_{-L_s/2}^{L_s/2} \exp\left(-\frac{(x_p + s\theta_{xp})^2 + (y_p + s\theta_{yp})^2}{2\sigma_\perp^2}\right) ds . \tag{42}$$

Integrating we obtain:

$$\overline{n}_e = \frac{\sqrt{\pi} n_e \sigma_\perp}{\sqrt{2} L_s \sqrt{\theta_{yp}^2 + \theta_{xp}^2}} \exp\left(-\frac{(x_p \theta_{yp} - y_p \theta_{xp})^2}{2\sigma_\perp^2 (\theta_{yp}^2 + \theta_{xp}^2)}\right)$$
$$\times \left( \mathrm{erf}\left( \frac{\sqrt{\theta_{yp}^2 + \theta_{xp}^2}}{\sqrt{2}\sigma_\perp} \left( \frac{L_s}{2} + \frac{x_p \theta_{xp} + y_p \theta_{yp}}{\theta_{yp}^2 + \theta_{xp}^2} \right) \right) - \mathrm{erf}\left( \frac{\sqrt{\theta_{yp}^2 + \theta_{xp}^2}}{\sqrt{2}\sigma_\perp} \left( \frac{x_p \theta_{xp} + y_p \theta_{yp}}{\theta_{yp}^2 + \theta_{xp}^2} - \frac{L_s}{2} \right) \right) \right), \tag{43}$$

where $\mathrm{erf}(x) = (2/\sqrt{\pi}) \int_0^x e^{-t^2} dt$ is the error function.

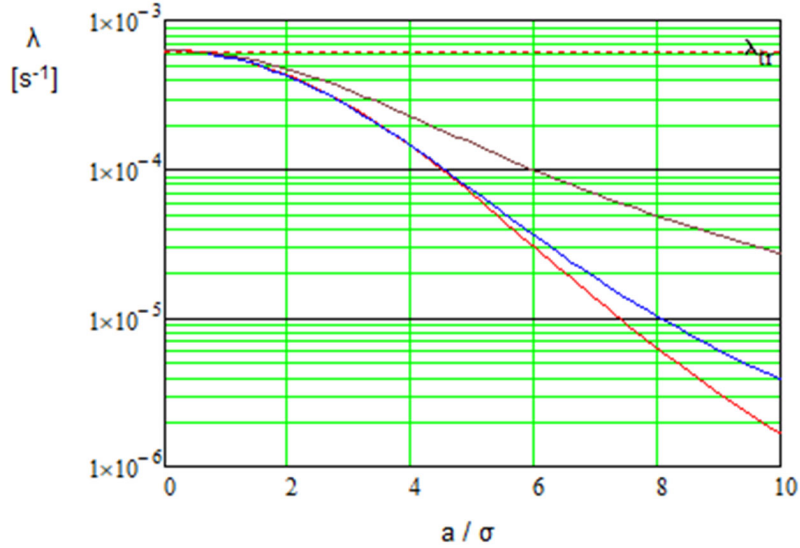

Figure 19: Dependence of cooling rate on the dimensionless particle betatron amplitude; brown (top) line – the electron beam size is much larger than the amplitude of betatron motion of a proton, blue (center) line – the rms electron beam size of 1.47 mm and betatron oscillations are in one plane, red (bottom) line – the rms electron beam size of 1.47 mm and the betatron oscillations have equal amplitudes for the horizontal and vertical planes $a_x = a_y = a/\sqrt{2}$. The reference rms emittance of proton beam is 9.6 µm, corresponding to the rms size in the center of cooling section (1σ) of 0.61 mm.

The emittance damping rate per turn is obtained by the averaging over betatron motion. For the horizontal damping we have:

$$\lambda_x = \frac{\beta_p}{4\pi^2 I_p} \int_0^{2\pi} \int_0^{2\pi} \theta_{xp} \delta\theta_{xp} d\mu_x d\mu_y . \tag{44}$$

Here

$$\delta\theta_{xp} = \left(\frac{1}{p}\frac{dp_\perp}{ds}\right)\overline{\frac{n_e}{n_e}}L_s \ , \tag{45}$$

is the change of transverse angle after passing the cooling section, and the term $\left((dp_\perp/ds)/p\right)$ is determined by Eq. (37). An equation for the vertical cooling rate is obtained by exchanging *x* and *y* indices.

Figure 19 shows dependences of the cooling rate on the particle betatron amplitude obtained by numeric integration of Eq. (44). The top line presents the cooling rate for the case of electron beam transverse beam size being much larger than the betatron motion amplitude and with the uniform electron beam density equal to the design density in the electron beam center. In this case a reduction of cooling rate with amplitude happens due to reduction of cooling force with transverse velocity (see Figure 17). The center line presents the cooling rate for pure horizontal (or vertical) motion. One can see that the cooling rate decreases additionally starting from about 1σ amplitude. This reduction happens due to the finite value of electron beam transverse size. The bottom line presents the cooling rate for the case when the horizontal and vertical betatron amplitudes are equal. Here we also assume that $a_x = a_y = a/\sqrt{2}$. In this case an additional reduction of cooling rate happening above 4.5σ is associated with reduction of probability of a particle to enter the electron beam in the case of two dimensional betatron motion. As one can see at 6σ the reduction of cooling rate is about 20 times. That will lead to creation of non-Gaussian tails in the proton distribution.

## 7. Beam Space Charge Tune Shifts

To compute the betatron tune shifts due to beam space charge in the case of strongly coupled optics we use equations developed in Ref. [27]. For a continuous beam the tune shifts of the betatron modes can be written in the following form:

$$\delta v_{SCk} = -\frac{r_e N_e}{2\pi\gamma^3 \beta^2 C} \oint ds \mathbf{v}_k^+ \mathbf{U}\hat{\mathbf{M}}\mathbf{v}_k^+ , \quad k = 1, 2 , \tag{46}$$

where $C$ is the ring circumference, $\mathbf{v}_k$ is the eigen-vectors determined by Eq. (6), $N_e$ is the number of particles in the beam,

$$\mathbf{U} = \begin{bmatrix} 0 & 1 & 0 & 0 \\ -1 & 0 & 0 & 0 \\ 0 & 0 & 0 & 1 \\ 0 & 0 & -1 & 0 \end{bmatrix}, \quad \hat{\mathbf{M}} = \begin{bmatrix} 0 & 0 & 0 & 0 \\ \Theta_{11} & 0 & \Theta_{12} & 0 \\ 0 & 0 & 0 & 0 \\ \Theta_{12} & 0 & \Theta_{22} & 0 \end{bmatrix}, \quad \hat{\Theta} = \begin{bmatrix} \Theta_{11} & \Theta_{12} \\ \Theta_{12} & \Theta_{22} \end{bmatrix}. \tag{47}$$

Matrix $\hat{\Theta}$ is related to the $\Sigma$-matrix presenting second order moments of the rms beam sizes which matrix elements are:

$$\begin{aligned}
\hat{\Sigma}_{11} &= \overline{x^2} = \varepsilon_1 \beta_{1x} + \varepsilon_2 \beta_{2x} + D_x^2 \sigma_{pe}^2 , \\
\hat{\Sigma}_{22} &= \overline{y^2} = \varepsilon_1 \beta_{1y} + \varepsilon_2 \beta_{2y} + D_y^2 \sigma_{pe}^2 , \\
\hat{\Sigma}_{12} &= \hat{\Sigma}_{21} = \overline{xy} = \varepsilon_1 \sqrt{\beta_{1x}\beta_{1y}} \cos v_1 + \varepsilon_2 \sqrt{\beta_{2x}\beta_{2y}} \cos v_2 + D_x D_y \sigma_{pe}^2 .
\end{aligned} \tag{48}$$

Then, we diagonalize the $\Sigma$-matrix by matrix $\mathbf{T}$ so that

$$\mathbf{T}^T \Sigma \mathbf{T} = \begin{bmatrix} \sigma_1^2 & 0 \\ 0 & \sigma_2^2 \end{bmatrix}. \tag{49}$$

That enables to express the matrix $\hat{\Theta}$ in the following form

$$\hat{\Theta} = \frac{1}{\sigma_1 + \sigma_2} \mathbf{T} \begin{bmatrix} 1/\sigma_1 & 0 \\ 0 & 1/\sigma_2 \end{bmatrix} \mathbf{T}^T . \tag{50}$$

In the solenoid all 4D- beta-functions are equal and both dispersions are equal to zero. Accounting that $\varepsilon_2 \ll \varepsilon_1$ we obtain from Eq. (46) that the solenoid contributions into the space charge tune shifts are equal for both modes:

$$\delta v_s = -\frac{r_e N_e}{2\pi\gamma^3 \beta^2 \varepsilon_1} \frac{L_s}{C}, \quad \beta_{1x} = \beta_{2x} = \beta_{1y} = \beta_{2y} , \tag{51}$$

where $L_s$ is the solenoid length. In the arcs where coupling is absent, Eq. (46) is reduced to the

well-known equations:

$$\delta v_{x,y} = -\frac{r_e N_e}{2\pi\gamma^3\beta^2 C}\int\frac{\beta_{x,y}ds}{\sigma_{x,y}(\sigma_x+\sigma_y)} \quad . \tag{52}$$

Figure 20 presents local contributions to the betatron tune shifts for modes 1 and 2. The total space charge tune shifts are: $\Delta v_{SC1} = -0.019$ and $\Delta v_{SC2} = -0.236$. The major contribution to the mode 2 which corresponds to the vertical motion in the arcs comes from the arcs where the vertical beam size is very small. The tune shift for the mode 2 is quite large. However, it is close to what is presently achieved in proton synchrotrons with similar number of turns. Further increase of beam current above 50 A is questionable due to an increase of the space charge tune shift for the mode 2.

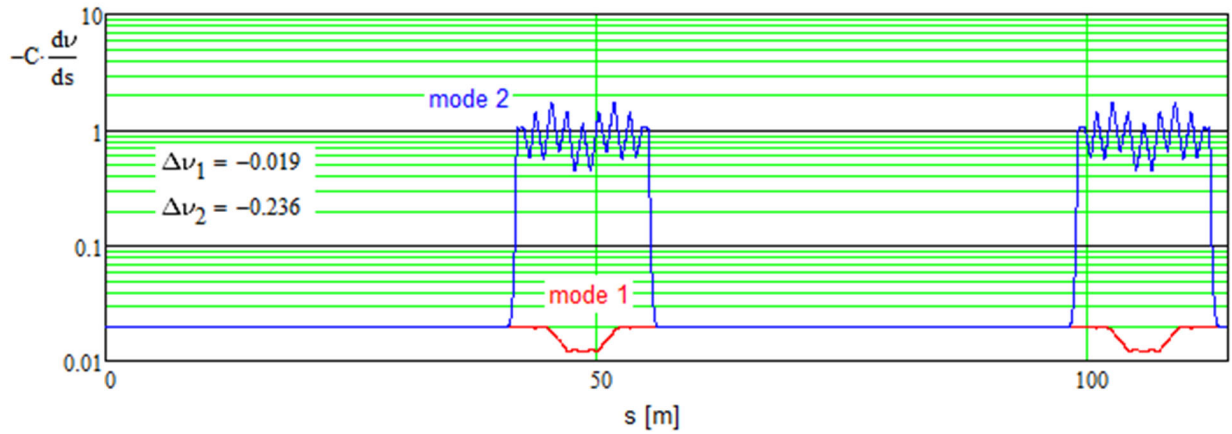

Figure 20: Local contributions to the betatron tune shifts due to beam space charge at the cooling cycle beginning for beam current of 50 A; red line – mode 1, blue line – mode 2.

# 8. Technical Systems of the Cooling Ring

## 8.1. Injection and extraction

Initially we considered a usage of the barrier bucket RF to create a short abort gap which would enable a lossless beam extraction. However, an analysis showed that an introduction of such system would require not only extremely powerful RF system required to counteract the beam loading but also would greatly increase the ring longitudinal impedance. Therefore, we refused to use the barrier bucket RF.

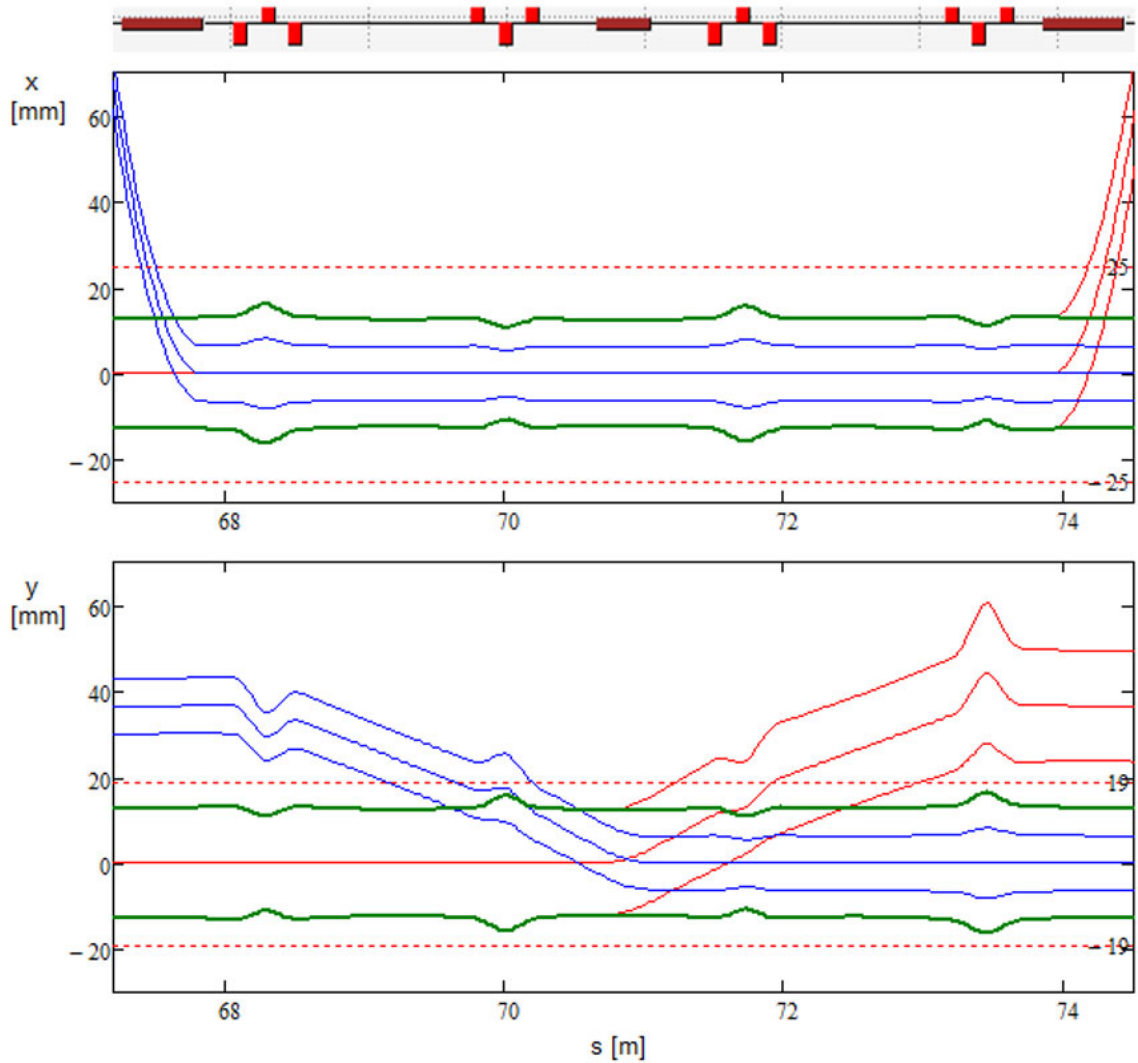

Figure 21: Horizontal (top) and vertical (bottom) beam envelopes for the injected (blue) and extracted (red) beams in the injection-extraction aria. Green lines show the envelope for the betatron acceptances. The beam direction is from left to right. The beam envelopes for the injected beam are shown for 4σ-size. Touschek scattering fills the entire horizontal acceptance. Therefore, the beam envelope of the extracted beam coincides with acceptance. Dashed horizontal lines mark the beampipe boundaries.

A usage of sufficiently short rise time of the extraction kicker pulse allowed us to achieve acceptable beam loss at the beam extraction of continuous beam. The injected beam duration is shorter than the revolution time and therefore the finite value of the fall time does not produce the beam loss.

To minimize the ring impedances both the beam injection and extraction are performed simultaneously by the same kicker and the same kicker pulse as shown in Figure 21. The injected beam is bent horizontally by the injection Lambertson septum and then set to the closed orbit by the vertical kicker. At the same time the extracted beam is kicked vertically by the same kicker and is extracted horizontally by the extraction Lambertson septum.

**Table 4: Parameters of the injection/extraction system**

| Kicker bending angle | 17.7 mrad |
|---|---|
| Kicker length | 60 cm |
| Kicker gap (distance between plates) | 38 mm |
| Kicker voltage | ±15.4 kV |
| Beam displacement at the extraction septum entrance (exit for injection) | 36.5 mm |
| Kicker pulse duration, FWHM | 385 ns |
| Injected beam duration | 376 ns |
| Kicker characteristic impedance (per plate) | 25 Ω |
| Kicker current (per plate) | 615 A |
| Septum bending angle | 200 mrad |
| Septum length | 60 cm |
| Septum magnetic field | 612 G |

Table 4 presents major parameters of the extraction/injection system. The duration of the kicker pulse flat top is equal to the revolution period of 381 ns. With 4 ns rise and fall times it implies that the kicker pulse duration (FWHM) is about 5 ns longer than the revolution period; and the duration of injected beam is 5 ns shorter. To reduce the kicker voltage its vertical aperture and the aperture in the septum magnets are reduced to 38 mm. To minimize the number of transitions in the aperture size we keep constant the vertical aperture in the extraction area with exception of the injection and extraction channels. To reduce the ring impedance the transition to the normal round vacuum chamber of the ring is tapered.

With 60 cm kicker length its filling time will be 2 ns. We expect that the rise and fall times of the kicker pulser will be about 2 ns. That yields the effective rise and fall times of the kicker to be about 4 ns. Then the fraction of the beam which will not be clean-extracted is about 1%, and, consequently, the power of ~2 kW will be lost in the extraction area.

The 4 ns injection gap will be filled by particles after ~100 μs (~250 turns).

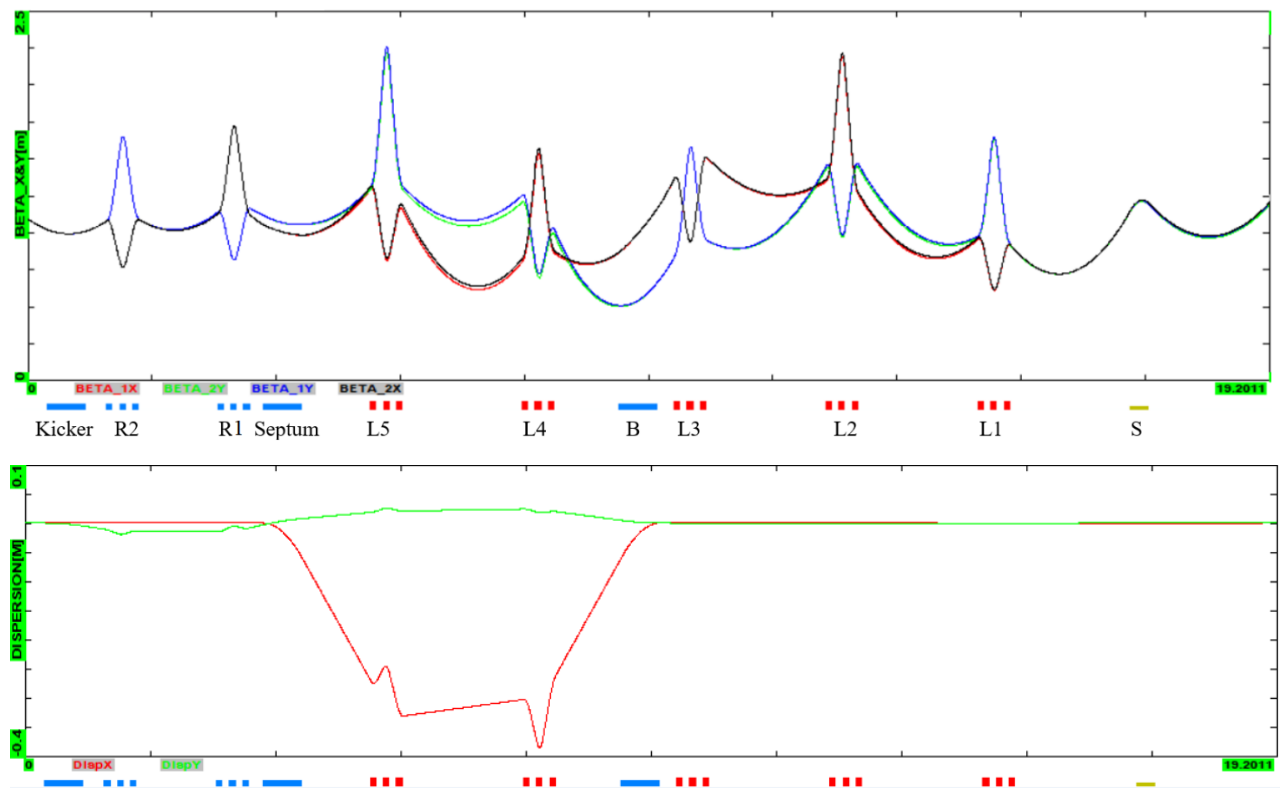

Figure 22: Dependence of 4D-beta-functions (top) and dispersions (bottom) along the injection line plotted in the inverse direction (from the injection kicker, to the last solenoid of the induction linac); L1- L5 mark locations of the transfer line triplets, and R1- R2 mark locations of the ring triplets.

## 8.2. Linac-to-ring beam transport

The beam injection is happening at the location where the beam phase space is presented by the circular modes which are also present at the exit of the induction linac. Therefore, to prevent an emittance growth the beam transport from the linac to the ring has to be rotationally invariant and achromatic. Figure 22 presents dependences of the 4D beta-functions and dispersions along the injection line. The bending angles of the septum magnet and the dipole B are chosen to be 11. 47 deg. That simplifies a compensation of vertical dispersion and yields enough space to separate quadrupoles of the ring and the transfer line. The triplet L5 is also shifted in the upstream direction to separate it longitudinally from the nearby triplet of the ring. Matching of the horizontal and vertical dispersions is achieved by triplets L4, L5 and minor rolls of the septum and the dipole B of 4.276 and -4.282 deg. respectively. The difference in the angle compensates the residual vertical kick of the injection kicker. Triplets L1 – L3 matches the beta-functions and the phase advances. Figure 23 shows dependence of the rms beam sizes along the beam line. As one can see the beam transport leaves the beam sizes being approximately equal which together with relatively small phase advance per sell (~65 deg.) is greatly helpful in reduction of space charge effects in the course of beam transport.

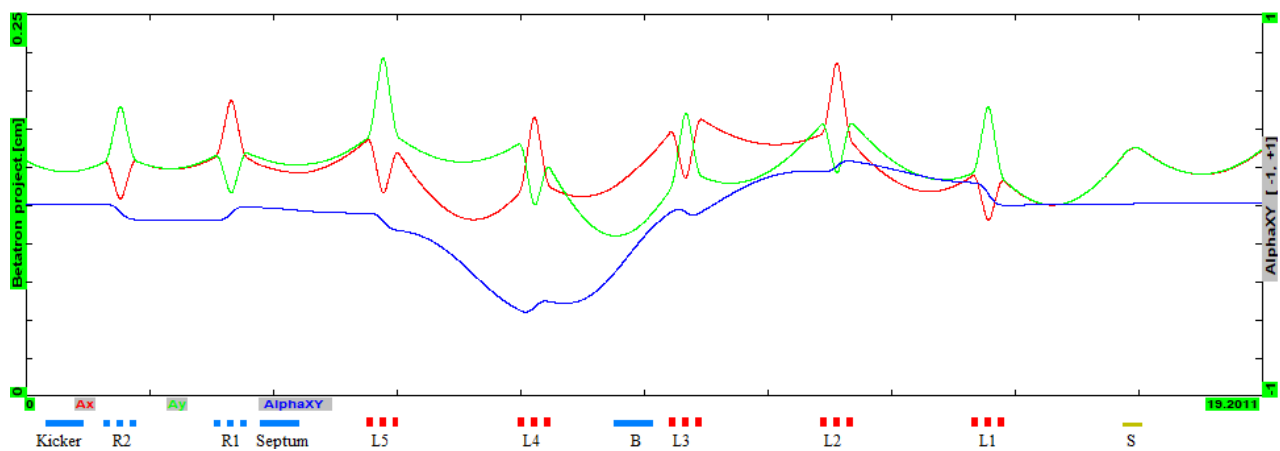

Figure 23: Dependence of the beam projections to the horizontal and vertical planes and parameter $\alpha = \overline{xy}/\sqrt{\overline{x^2}\,\overline{y^2}}$ along the injection line plotted in the inverse direction (from the injection kicker, to the last solenoid of the induction linac).

A mismatch in the transfer line optics may result in an emittance growth for the mode 2. First, we find the emittance growth induced by the dispersion mismatch while the assuming perfect betatron match. In this case the vector of particle transverse 4D-coordinates for a particle at the injection point can be expressed through the particle initial ($\varepsilon_1$, $\varepsilon_2$) and final ($\varepsilon_1+\delta\varepsilon_1$, $\varepsilon_2+\delta\varepsilon_2$) mode emittances and the relative momentum deviation ($\theta_s$):

$$\frac{1}{2}\left(\sqrt{2(\varepsilon_1+\delta\varepsilon_1)}\mathbf{v}_1 e^{i(\psi_1+\delta\psi_1)} + CC\right) + \frac{1}{2}\left(\sqrt{2(\varepsilon_2+\delta\varepsilon_2)}\mathbf{v}_2 e^{i(\psi_2+\delta\psi_2)} + CC\right) + \theta_s \mathbf{D} = \\ = \frac{1}{2}\left(\sqrt{2\varepsilon_1}\mathbf{v}_1 e^{i\psi_1} + CC\right) + \frac{1}{2}\left(\sqrt{2\varepsilon_2}\mathbf{v}_2 e^{i\psi_2} + CC\right) + \theta_s(\mathbf{D}+\boldsymbol{\delta D}) \quad . \tag{53}$$

Here $\psi_k$ and $\psi_k+\delta\psi_k$ are the initial and final particle phases ($k$ = 1,2), $\mathbf{D}=(D_x, D'_x, D_y, D'_y)^T$ is the vector of dispersions of the ring at the injection point, and $\boldsymbol{\delta D}$ is the vector characterizing the difference of dispersions of the ring and the transfer line. Multiplying both sides of this equation by $\mathbf{v}_k^+\mathbf{U}$, and using the symplectic orthogonality condition of Eq. (15) we obtain:

$$\frac{1}{2}\left(-2i\sqrt{2(\varepsilon_k+\delta\varepsilon_k)}e^{i(\psi_1+\delta\psi_1)}\right) = \frac{1}{2}\left(-2i\sqrt{2\varepsilon_k}e^{i\psi_1}+CC\right)+\theta_s\left(\mathbf{v}_k^+\mathbf{U}\boldsymbol{\delta}\mathbf{D}\right), \quad k=1,2. \tag{54}$$

Multiplying this equation by the complex conjugate and performing averaging over betatron phases we finally obtain the emittance increase due to betatron mismatch:

$$\delta\varepsilon_k = \frac{1}{2}\sigma_{pe}^2\left|\mathbf{v}_k^+\mathbf{U}\boldsymbol{\delta}\mathbf{D}\right|^2. \tag{55}$$

The eigen-vector at the injection point can be presented by Eq. (17). Performing multiplications in the above equation we obtain

$$\delta\varepsilon_1 = \frac{\sigma_{pe}^2}{8\beta_0}\left(\left(\delta D_x + 2\beta_0\delta D'_y\right)^2 + \left(\delta D_y + 2\beta_0\delta D'_x\right)^2\right),$$

$$\delta\varepsilon_2 = \frac{\sigma_{pe}^2}{8\beta_0}\left(\left(\delta D_x - 2\beta_0\delta D'_y\right)^2 + \left(\delta D_y - 2\beta_0\delta D'_x\right)^2\right). \tag{56}$$

Requiring $\delta\varepsilon_2/\varepsilon_2 \leq 0.01$ for each type of errors we obtain requirements for the dispersion errors: $\delta D_{x,y} \leq 0.72\,\text{cm}$, $\delta D'_{x,y} \leq 0.0037$. That is within accuracy which can be achieved with modern tools for optics measurements and correction.

Now we consider the emittance growth related to the betatron mismatch. Similar to above we express the vector of particle transverse 4D-coordinates for a particle at the injection point through initial ($\varepsilon_1$, $\varepsilon_2$) and final ($\varepsilon_1+\delta\varepsilon_1$, $\varepsilon_2+\delta\varepsilon_2$) mode emittances and the eigen-vectors of ring ($\mathbf{v}_1$, $\mathbf{v}_2$) and transfer line ($\mathbf{v}_1+\boldsymbol{\delta}\mathbf{v}_1$, $\mathbf{v}_2+\boldsymbol{\delta}\mathbf{v}_2$):

$$\frac{1}{2}\left(\sqrt{2(\varepsilon_1+\delta\varepsilon_1)}\mathbf{v}_1 e^{i(\psi_1+\delta\psi_1)}+CC\right)+\frac{1}{2}\left(\sqrt{2(\varepsilon_2+\delta\varepsilon_2)}\mathbf{v}_2 e^{i(\psi_2+\delta\psi_2)}+CC\right)=$$
$$=\frac{1}{2}\left(\sqrt{2\varepsilon_1}\left(\mathbf{v}_1+\boldsymbol{\delta}\mathbf{v}_1\right)e^{i\psi_1}+CC\right)+\frac{1}{2}\left(\sqrt{2\varepsilon_2}\left(\mathbf{v}_2+\boldsymbol{\delta}\mathbf{v}_2\right)e^{i\psi_2}+CC\right) \tag{57}$$

A condition of symplectic normalization of the perturbed vector,

$$\left(\mathbf{v}_k+\boldsymbol{\delta}\mathbf{v}_k\right)^+\mathbf{U}\left(\mathbf{v}_k+\boldsymbol{\delta}\mathbf{v}_k\right) = -2i\delta_{ij}, k=1,2,$$

results in that in the first approximation

$$\mathbf{v}_k^+\mathbf{U}\boldsymbol{\delta}\mathbf{v}_k = 0. \tag{58}$$

Similar to the above, multiplying Eq. (57) by $\mathbf{v}_2^+\mathbf{U}$, using the symplectic orthogonality condition, and then multiplying each part of the obtained equation by its complex conjugate and performing averaging over betatron phases $\psi_1$ and $\psi_2$ we obtain:

$$\delta\varepsilon_2 = \frac{\varepsilon_1}{4}\left(\left|\mathbf{v}_2^+\mathbf{U}\overline{\boldsymbol{\delta}\mathbf{v}_1}\right|^2 + \left|\mathbf{v}_2^+\mathbf{U}\boldsymbol{\delta}\mathbf{v}_1\right|^2\right) + \frac{\varepsilon_2}{4}\left|\mathbf{v}_2^+\mathbf{U}\overline{\boldsymbol{\delta}\mathbf{v}_2}\right|^2 , \qquad (59)$$

where we also used Eq. (58). Similar equation can be obtained for $\delta\varepsilon_1$ from Eq. (59) by exchanging indices 1 and 2. Taking into account that $\varepsilon_1 \gg \varepsilon_2$ we can neglect the second addend in the below estimate and conclude that optics errors for the mode 1 (horizontal in arcs) make major contribution to the emittance growth. Next, we decompose $\boldsymbol{\delta}\mathbf{v}_1$ in the base of the eigen-vectors and their complex conjugates which are linearly independent:

$$\boldsymbol{\delta}\mathbf{v}_1 = a_1\overline{\mathbf{v}_1} + a_2\mathbf{v}_2 + a_3\overline{\mathbf{v}_2} , \quad a_i \ll 1, \quad i = 1, 2, 3 , \qquad (60)$$

where we also used Eq. (58). Substituting the above equation into Eq. (59) we obtain:

$$\delta\varepsilon_2 = \varepsilon_1\left(|a_2|^2 + |a_3|^2\right) . \qquad (61)$$

Requiring $\delta\varepsilon_2/\varepsilon_2 \leq 0.01$ for each type of errors we obtain: $|a_2|, |a_3| \leq 0.002$. Tuning optics to this accuracy is expected to be extremely challenging task. For uncoupled optics it corresponds to the accuracy of beta-function measurements at ~0.4% level.

### 8.3. Cooling ring instrumentation

To achieve the discussed above cooling rates the ring optics has to be measured and tuned to the extremely high accuracy. That will require the beam instrumentation with exceptional accuracy. The instrumentation should include:

- Two coordinate Beam Position Monitors (BPMs) installed at minimum near each quad or triplet in the straight line and at each 60° of betatron phase inside solenoids.
- Beam profile monitors
- High accuracy beam current monitor (DCCT)
- Schottky monitors for measurements of momentum spread and transverse emittance measurements.

A small amplitude RF cavity will be also required to support BPM measurements. The cavity has to have very small impedance (both wide and narrow band).

## 8.4. Requirements to vacuum

The force acting on a particle from the beam direct space charge is summed from contributions of beam electric and magnetic fields. For the relativistic beam the electric field repulsion is almost compensated by magnetic attraction so that the force is $eE/\gamma^2$. This compensation is broken if ions, produced by the beam from the residual gas, are stored in the beam. In the absence of ion removal their density will be close to the density of electron beam, but in difference to the beam the ion velocity is small and ions do not create a magnetic field. Consequently, the betatron tune shifts produced by ions will exceed the space charge betatron tune shifts (see Section 7) by $\gamma^2$ times. Therefore, a storage of the ions in the electron beam is unacceptable. It can be prevented by separation in time between the injection of new beam and the extraction of the old one, so that ions stored in one 5 ms cycle could leave the beam between extraction and injection.

For relativistic beam the ionization cross-section of molecular hydrogen is ~$5 \cdot 10^{-20}$ cm$^{-2}$. The ionization cross-section of helium is close to this value and it grows $\propto \sqrt{Z}$ for heavier atoms. Assuming that the molecular hydrogen makes the major contribution to the residual gas density and requiring the ion compensation level to be smaller than $5 \cdot 10^{-5}$ we obtain that the pressure of the residual gas should be better or about $2 \cdot 10^{-10}$ Torr. That requires entire vacuum chamber to be treated as an ultra-high vacuum chamber.

The ions are produced in the field of electron beam and therefore their typical kinetic energy is about quarter of potential energy difference between the beam center and its boundary (1.5 kV). When the electric beam disappears, the ions conserve their velocities (~$2 \cdot 10^7$ cm/c for atomic oxygen), and they need ~100 ns to reach the vacuum chamber wall. Consequently, a delay between beam injection and extraction has to be ~100 ns (about quarter of revolution period).

## 9. Beam Stability

### 9.1. Longitudinal instabilities

For the longitudinal motion the stability of continuous beam for the harmonic $n$ is determined by the following criterion [28]: the complex value

$$A_n = \frac{1}{2\pi\beta\eta\sigma_p^2} \frac{eI_e}{pc} \frac{Z_n}{n} \tag{62}$$

has to be within the area bounded in the complex plane by the curve described by parametric equation

$$A(y) = \left( -i\sigma_p^2 \int_{\delta \to +0} \frac{df/dx}{y+x-i\delta} dx \right)^{-1}, \quad y \in [0,\infty]. \tag{63}$$

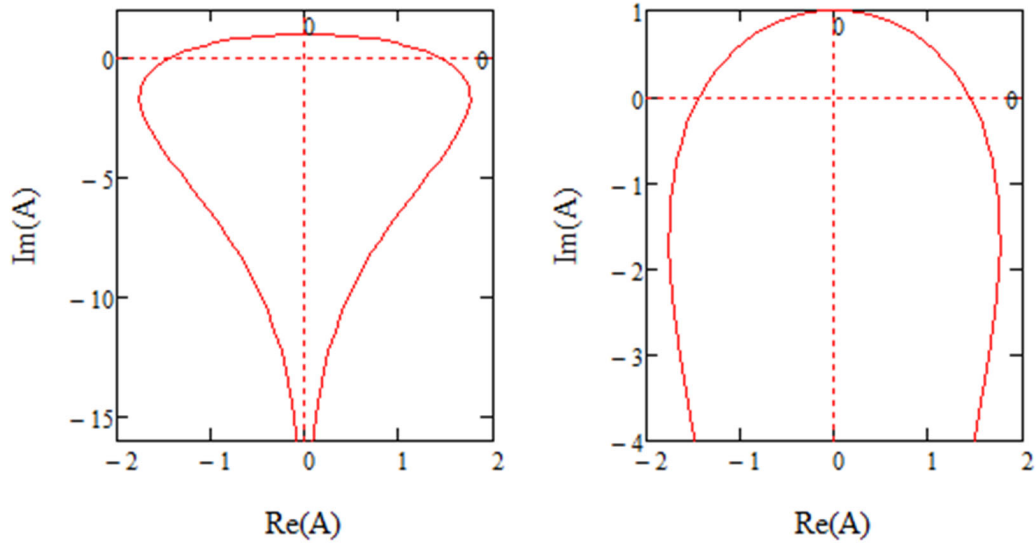

Figure 24: Stability area in the complex plane for Gaussian distribution. Right plot presents the same curve with better resolution along the imaginary axis.

Here $\beta = v/c$ is the dimensionless beam velocity, $e$ is the particle charge, $p$ is the momentum of reference particle, $x=\Delta p/p$ is the relative momentum deviation, $\sigma_p$ is the relative rms momentum spread, $\eta = \alpha - 1/\gamma^2$ is the slip-factor, $f(x)$ is the longitudinal distribution normalized as $\int f(x)dx = 1$, $I_e$ is the beam current, and $Z_n$ is the ring longitudinal impedance at the $n$-th revolution harmonic. We also assume that the impedance does not change in a frequency band covered by the spread of particle revolution frequencies at the given harmonic. In further estimates we assume the Gaussian distribution. Figure 24 presents corresponding stability area. The cooling ring operates above transition. In this case the most dangerous impedance is the capacitive one for which $A_n \propto i$. In this case the stability condition is $\text{Im}(A_n) = |A_n| \leq 1$.

Our approach to the beam stability assumes that the stability at low frequencies is achieved by the longitudinal damper while at high frequencies the stabilization is achieved by Landau damping which stability boundary is determined by Eqs. (62) and (63).

*Beam stability at high frequencies*

Our analysis shows that at high frequencies (more than few GHz) the most dangerous impedances are the space charge impedance and the impedance due to coherent synchrotron radiation (CSR).

The space charge impedance is:

$$\frac{Z_n}{n} \approx i \frac{Z_0}{\beta \gamma^2} \ln\left(\frac{a}{1.5 \sigma_\perp}\right), \quad \frac{a}{\sigma_\perp} \geq 3, \tag{64}$$

where $Z_0 = 4\pi/c \approx 377\ \Omega$ is the impedance of free space, $\gamma$ is the relativistic factor of the beam, $a$ is the vacuum chamber radius, and $\sigma_\perp$ is the beam rms size. We assume that the beam and vacuum chamber are round. The factor 1.5 in Eq. (64) appeared due to averaging the longitudinal electric field across the beam cross-section. For $\sigma_\perp = 1.5$ mm and $a = 25.4$ mm we obtain $Z_n/n \approx 79i$ m$\Omega$. The impedance stays constant up to very high frequency of about $2\pi c\gamma / a \approx 10$ THz above which it starts to decrease.

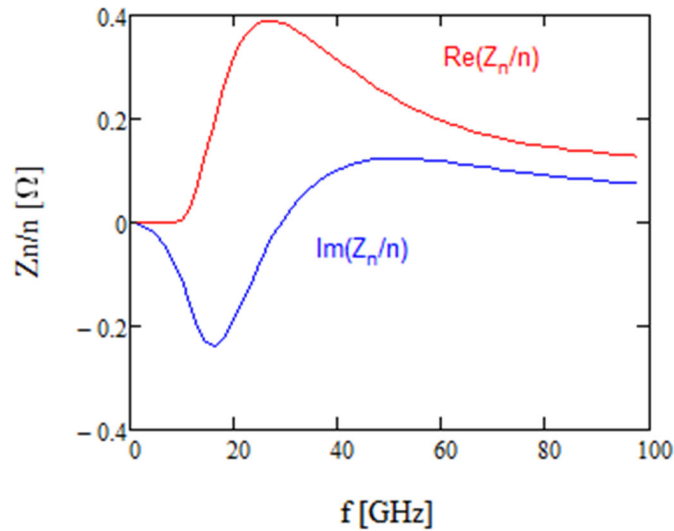

Figure 25: Real and imaginary parts of longitudinal impedance due to CSR.

The impedance of CSR was computed using results of Ref. [29] with help of Mathematica spreadsheet written by G. Stupakov [30]. This model accounts the SCR screening by vacuum chamber in the model of the beam between two parallel plates. The results of computations are presented in Figure 25. As one can see the impedance achieves its maximum at frequency of ~25 GHz.

We need to stress that such model, although captures major features of the impedance, does not represent an accurate solution of the task. However, we believe that it better represents the problem than a model of a waveguide developed in Refs. [31] and [32], which accounts that the wave is reflected from the waveguide walls resulting in an accumulation of e.-m. energy with bunch propagation along the bend and a resonant particle interaction with the wave. The dipoles are short relative to the bending radius (50 cm versus 191 cm) so that their length is close to the formation length of the wave at the resonance frequency ($f \approx 36$ GHz):

$$L_{form} = 2\sqrt{a\rho} \ . \tag{65}$$

For the bending radius $\rho = 191$ cm and aperture $a = 2.54$ cm we obtain $L_{form} = 44$ cm. Thus, the short length of the dipole does not allow a resonant amplification of the radiated wave. On the way to the next dipole the wave and the particle loose synchronization. Difference in the particle velocity and the phase velocity of wave in a round straight waveguide results in a phase shift between the wave and the particle. For the wave frequency of $f$ it is:

$$\delta\phi \approx \frac{L_{dd}c}{4\pi f a^2}\mu_{nm}^2 \ , \quad f \gg \frac{c}{2a} \ , \tag{66}$$

where $f$ is the frequency of the wave, $\mu_{nm}$ is the $m$-th root of $J_n(x)$ Bessel function, and $m$ and $n$ characterize the mode of electro-magnetic wave in a circular waveguide. For the distance between dipoles of $L_{dd} \approx 30$ cm we obtain for the lowest harmonic ($\mu_{01} \approx 2.405$) the phase shift of 1.8 rad at the resonance frequency. It will be larger for higher harmonics. The structure of the wave excited by a particle in a bend is represented by a mixture of modes of a straight waveguide. That will increase desynchronization of particle and wave after passing a straight between dipoles. Thus, passing through the long straight section completely destroys any synchronization accumulated in the arc. Overall accumulation of electromagnetic energy in the wave is prevented by the wave damping due finite resistivity of vacuum chamber walls. For the wave frequency well above the cut-off the damping length is:

$$L_d = 2a\sqrt{\frac{\sigma_R}{f}}, \qquad (67)$$

where $\sigma_R$ is the wall conductivity in Gauss units. For copper vacuum chamber at 20 C⁰ the resistivity of 16.8 nΩ·m corresponds to $\sigma_R = 5.35 \cdot 10^{17}$ s⁻¹. We also account that the wall roughness increases wall resistivity at ~30 GHz by about 5%. That yields that at the resonance frequency the damping length is $L_d \approx 190$ m which prevents multiturn e.-m. energy accumulation. The described above consideration is well supported by the results of numerical simulations presented in Ref. [33] which were carried out for the parameters close to considered inhere.

Figure 26 shows the stability diagram for 50 A beam in the range of high frequencies (4 - 100 GHz) where the CSR and space charge impedances are accounted. The beam loses stability at 56.5 A at the frequency of 36 GHz which we called above the resonance frequency. The beam current and the momentum spread at the cycle beginning were chosen to leave 10% margin.

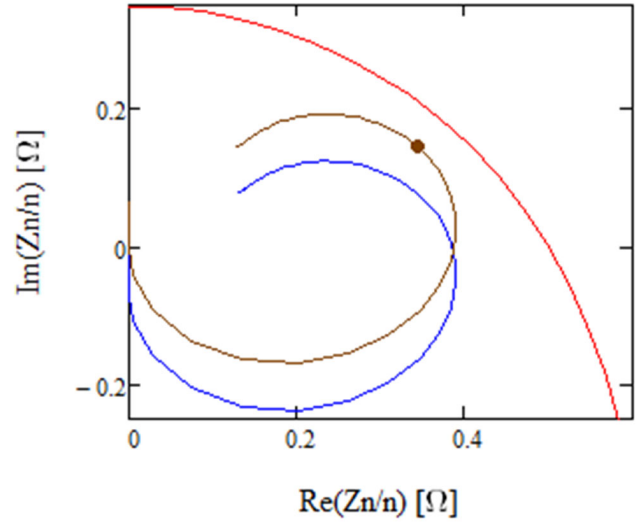

Figure 26: Stability diagram in the complex impedance space: red line – the stability boundary, blue line – trajectory of $Z_n/n$ with frequency change for the CSR impedance only; brown line – the same with additional accounting of the space charge impedance, brown dot shows $Z_n$ at the resonant frequency of 36 GHz. The beam current is 50 A.

*Beam stability at low frequencies*

The major contributors to the ring impedance at low frequencies ($f < 4$ GHz) are: the space charge (already discussed above), the wall resistivity, the beam position monitors, the RF cavity, the injection/extraction kicker, the kickers of the transverse dampers and pumping ports.

The resistive wall impedance is:

$$\frac{Z_n}{n} \approx (1-i\,\text{sgn}(n))\frac{Z_0 C}{4\pi a}\sqrt{\frac{f_0}{n\sigma_R}}\,, \tag{68}$$

where $f_0$ the revolution frequency. For the copper vacuum chamber and the first harmonic ($n=1$), it is equal to $Z_n/n \approx (1-i)\cdot 0.287\,\Omega$ and decreases as square root of frequency.

For the impedance of button BPMs we carried out two independent estimates which yield close results for frequencies below 1 GHz. In the first estimate we assumed the BPMs are short strip lines loaded by 50 $\Omega$ and based our estimate on the expression presented in Ref. [34] (see Eq. (70)). In the second case, which is used below, we assumed 4-button BPMs with the button radius of $r_b = 5$ mm in the round vacuum chamber with radius $a = 25.4$ mm. Each button is loaded by a resistor with resistance of $R_b = 30$ k$\Omega$ and has capacitance to the ground $C_b = 5$ pF. The corresponding time constant $\tau_b = R_b C_b = 150$ ns allows to observe all frequencies in the beam above the revolution frequency. The total impedance of one 4-button BPM was estimated using the following equation:

$$Z_{BPM}(\omega) = \frac{16 r_b^4 \omega}{\pi^4 c^2 a^2 C_b}\left(\frac{\omega \tau_b}{1+\omega^2 \tau_b^2} - 32i\right). \tag{69}$$

This estimate is justified for frequencies below 1 GHz. For frequencies above revolution frequency the impedance is mostly inductive, and $Z_n/n$ stays constant with frequency. The total impedance of 100 BPMs is: $Z_n/n \approx -i17$ m$\Omega$.

A low voltage RF cavity is used to induce density modulation in the continuous beam to make it visible for BPMs. We choose the harmonic number to be 31 ($f = 81.4$ MHz). Then the required cavity voltage amplitude is below 20 V. That allows to make cavity with very low impedance. Choosing R/Q=5 $\Omega$ and loaded Q=5 we obtain the shunt impedance of 25 $\Omega$.

For an estimate of impedance for the injection/extraction kicker (see Section 8.1) we use an expression from Ref. [34] for the strip-line BPM:

$$Z_{kick}(\omega) = 2Z_c\left(\frac{\phi_0}{2\pi}\right)^2 \left(2\sin^2(\omega L/c) - i\sin(2\omega L/c)\right), \tag{70}$$

where $Z_c = 25$ $\Omega$ is the characteristic impedance of a plate, and $\phi_0$ is the angle subtended by a plate from the kicker center.

There are two transverse damper kickers. One for each plane. Each kicker has the same length as the injection kicker. Such length supports the damper bandwidth of 250 MHz. To minimize the kicker impedance each plate has characteristic impedance of 12.5 Ω (half of injection kicker). Therefore, the total impedance of these two kickers is equal to the impedance of the injection kicker.

To estimate the impedance of pumping ports we assume that each port is shielded, so that for the beam each port is visible as $N_s$=15 slits with $w_s$=2 mm width and $L_s$=5 cm length. Using Ref. [34] we write:

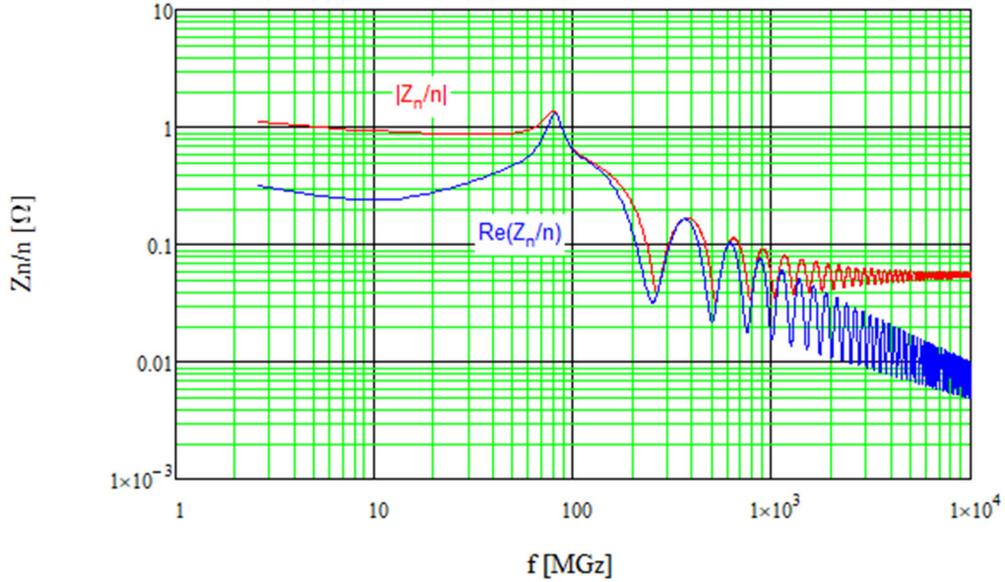

Figure 27: The dependence of the modulo and the real part of longitudinal ring impedance on frequency.

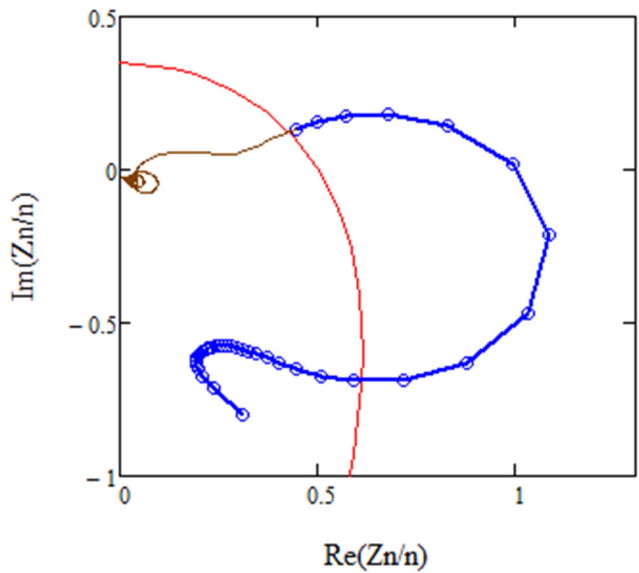

Figure 28: The stability diagram in the complex impedance space. Red line bounds the stable area. Blue line shows the trajectory of $Z_n/n$ with frequency change for frequencies below 97 MHz (harmonic 37), and brown curve the same for frequencies above 97 MHz. Blue dots mark first 37 revolution harmonics.

$$Z(\omega) = -\frac{i\omega Z_0}{4\pi^2 a^2 c} w_s^3 \left( 0.814 - 0.0344 \frac{w_s}{L_s} \right) N_s N_p \ , \tag{71}$$

where $N_p$=30 is the number of pumping ports. The total impedance of all pumping ports is: $Z_n/n \approx -i2.4$ m$\Omega$.

Summing all contributions, we obtain the total longitudinal impedance of the ring. Its dependence on frequency is presented in Figure 27. The corresponding stability diagram is shown in Figure 28. As one can see the Landau damping stabilizes the beam above 97 MHz. The stabilization at lower frequencies will be achieved with the longitudinal damper. Taking into account that the impedance of actual machine may be larger we conservatively estimate that the longitudinal damper has to have ~250 MHz bandwidth, *i.e.* capable to damp first 100 harmonics of the revolution frequency.

## *9.2. Transverse instabilities*

*Incoherent tune shifts*

First, we consider the incoherent betatron tune shifts excited by the beam due to its current reflections in the poles of dipoles. In the vertical plane the beam current reflections in the poles create horizontal magnetic field described by following equation:

$$B_x = \frac{2\pi^2 I_e}{3cg^2} \left( 1 + \frac{\pi^2}{15} \frac{y^2}{g^2} + \frac{2\pi^4}{315} \frac{y^4}{g^4} + \ldots \right) y \ , \tag{72}$$

where $g$ is the gap between poles. The corresponding first three non-zero multipoles are[2]:

$$G \equiv B_1 = \frac{2\pi^2 I_e}{3cg^2}, \quad O \equiv B_3 = -\frac{4\pi^4 I_e}{15cg^4}, \quad B_5 = \frac{32\pi^6 I_e}{63cg^6} \ . \tag{73}$$

The first addend results in the linear betatron tune shifts which values are determined by the following equation:

$$\begin{bmatrix} \delta v_1 \\ \delta v_2 \end{bmatrix} = \frac{\pi r_e N_e}{6c\gamma} \int\limits_{\substack{\text{over} \\ \text{dipoles}}} \begin{bmatrix} \beta_x \\ -\beta_y \end{bmatrix} \frac{ds}{g^2} \ . \tag{74}$$

Here $\beta_x$ and $\beta_y$ are the beta-functions, $N_e$ is the total number of particles in the beam, and we accounted that the beam is uncoupled in the arcs. For $g$ = 5.38 cm we obtain for the multipoles: $G$ = 1.14 G/cm, $O$ = -155 mG/cm$^3$, and $B_5$ = 10 mG/cm$^5$. The linear tune shifts are: $\delta v_1$ = 0.104, $\delta v_2$ = –0.094. While these values do not look large a non-uniform nature of the perturbation along the ring strongly distorts its optics making the beam unusable for cooling. A complete optics correction has to include not only the correction of betatron tunes but also the correction of beta-functions and dispersions. In this paper, we use only two families of quadrupoles: one focusing and one defocusing corresponding to the focusing and defocusing quads in the arcs. Such choice is sufficient for correction of beta-functions and dispersions. After correction we obtain that the tune for the high emittance mode (horizontal in arcs) stays at approximately the same value. However, the correction does not return the tune of small emittance mode to its initial position, although its change is reduced from 0.094 down to ≈0.04. This value is sufficiently small and should not represent a problem. Actually, it results in a better separation of tunes and should be helpful for reduction of emittance exchange due to mode coupling by the beam space charge. To test how the beam induced non-linearities of Eq. (73) affect a single particle motion stability we performed tracking with these non-linearities included. The tracking showed that the dynamic aperture exceeds the physical aperture by about 2 times.

---

[2] Here we use the standard definition for multipole coefficients: $B_y(x)\big|_{y=0} = \sum_{n=0}^{\infty} \frac{B_n x^n}{n!}$

The time for establishing the beam induced magnetic field is determined by the time required for magnetic field to penetrate through the vacuum chamber and then to penetrate inside steel of the magnetic poles to sufficient depth. The time for penetration of the magnetic field through the vacuum chamber is determined by its thickness, $d$, and conductivity:

$$\tau \approx \frac{2\pi \sigma_R d^2}{c^2} . \qquad (75)$$

For copper vacuum chamber with $d$ = 1 mm we obtain $\tau \approx$ 40 μs. The time of magnetic field penetration into poles is determined by the skin depth equal to $g/2\mu$ when the magnetic field flux generated by the beam current can be passing inside steel. For $\mu \approx$ 500 we obtain $g/2\mu \approx$ 50 μm, and the penetration time ≈6 μs. Thus, if the beam interruptions do not exceed ~1 μs and pulse to pulse current fluctuations do not exceed ~1% we can consider the beam induced fields being induced by DC.

Another mechanism of incoherent tune shifts is related to the beam center offsets from the center of vacuum chamber. For round vacuum chamber the beam offset generates the electric field induced by the chamber. For the horizontal beam offset $\delta x_c$ the electric field in the horizontal plane is:

$$E_x(x) = \frac{2eI_e}{c\beta} \frac{1}{a^2/x_c - x} = \frac{2eI_e x_c}{c\beta a^2}\left(1 + \frac{xx_c}{a^2} + \left(\frac{xx_c}{a^2}\right)^2 + ...\right). \tag{76}$$

The first term in the above equation represents the uniform electric field which value is equivalent to the magnetic field of 0.31 G for 2 mm beam offset. The second term represents the defocusing field (but focusing in the vertical plane) which results in the tune shifts. With *x-y* coupling accounted the tune shifts are:

$$\begin{bmatrix}\delta\nu_1 \\ \delta\nu_2\end{bmatrix} = -\frac{r_e N_e}{2\pi\beta^2\gamma a^4 C}\int_C\begin{bmatrix}(\beta_{1x}-\beta_{1y})(x_c^2-y_c^2) + 4x_c y_c\sqrt{\beta_{1x}\beta_{1y}}\cos\nu_1 \\ (\beta_{2x}-\beta_{2y})(x_c^2-y_c^2) + 4x_c y_c\sqrt{\beta_{2x}\beta_{2y}}\cos\nu_2\end{bmatrix}ds. \tag{77}$$

Here we also assume that the tunes are well separated and skew-quadrupole field excited by the beam does not affect the tunes. For the fixed horizontal offset of 2 mm along the entire ring we obtain the tune shifts of -0.06 and -0.04 for modes 1 and 2, respectively. In addition to the tune shifts these beam offsets affect the beam optics. Thus, to avoid uncontrolled beam optics variation one has to keep the beam at the vacuum chamber center with accuracy better than 1-2 mm. Note also that the time response of this mechanism is close to the time response for the considered above case with beam current reflections in the poles of dipoles. Immediately after beam injection the beam magnetic field is kept inside vacuum chamber which suppresses the beam induced electric force by $\gamma^2$ times. With time the magnetic field of vacuum chamber decays leaving only electric field acting on the beam.

Another serious limitation comes from the beam interaction with vacuum chamber in the cooling section. It is related with beam deflection by the beam induced electric field in the round vacuum chamber of the cooling section. The beam offset from the vacuum chamber center results in a dipole electric field described by the first addend of Eq. (76). A drift motion of the beam in crossed electric and magnetic fields yields additional transverse angles of $E/(\beta B_0)$. Here $B_0$ is the magnetic field in the cooling section. Requiring these angles do not exceed half of the rms transverse angle in the cooling section, or 13 μrad, yields that the beam must be centered on the vacuum chamber within ~0.2 mm. Note that this requirement can be significantly mitigated if the electron beam trajectory can be measured and corrected within ~10 μrad.

*Transverse ring impedances and transverse coherent tune shifts*

The major contributors to the ring transverse impedance are: the beam interaction with round vacuum chamber, and the injection/extraction and transverse damper kickers. Other sources like the beam position monitors (button type), vacuum ports (shielded), *etc.* contribute much less.

The beam current interaction with vacuum chamber walls has two contributions. The first one is related to the direct beam interaction with round vacuum chamber:

$$Z_\perp(\omega) = Z_x(\omega) = Z_y(\omega) = -i\frac{Z_0 C}{2\pi a^2 \beta \gamma^2} \ . \tag{78}$$

This impedance is pure imaginary. Therefore, it changes the betatron tunes but does not contribute to the instability growth rates. The second contribution is related to the wall conductivity:

$$Z_\perp(\omega) = Z_x(\omega) = Z_y(\omega) = \frac{Z_0 C}{2\pi a^3}\left(\mathrm{sgn}(\omega)-i\right)\delta_R(|\omega|) \ , \tag{79}$$

where $\delta_R(\omega) = c/\sqrt{2\pi\sigma_R \omega}$ is the skin depth in the chamber material.

The transverse impedance of injection/extraction kicker is mostly vertical. For its estimate we use an expression from Ref. [34] for the strip-line BPM:

$$Z_y(\omega) = \frac{4c Z_{kick}(\omega)}{a^2 \omega}\left(\frac{2}{\phi_0}\sin\frac{\phi_0}{2}\right)^2 , \quad Z_x(\omega) = 0 , \tag{80}$$

where $Z_{kick}(\omega)$ is determined by Eq. (70).

There are two kickers of transverse dampers: one horizontal and one vertical. They have the same length as the injection kicker, but their characteristic impedances were chosen to be half of the injection kicker. Therefore, the impedance of each kicker is half of the injection kicker impedance, but their impedances are directed in different planes.

To estimate the impedance of 4-button BPM we use a relationship which bounds the longitudinal and vertical impedances for axially symmetric vacuum chamber with constant radius *a*:

$$Z_\perp(\omega) = Z_x(\omega) = Z_y(\omega) = \frac{2c}{a^2 \omega}Z_\parallel(\omega) \ , \tag{81}$$

and, then, use Eq. (69) which determines the longitudinal impedance of the BPMs.

Figure 29 presents different contributions to the transverse impedance. One can see that the kickers and the wall resistivity strongly dominate.

For the case of strong *x-y* coupling the coherent tune shits for the modes 1 and 2 are determined by the following equation [35]:

$$\delta\nu_{\lambda n} = -i\frac{eI_e}{4\pi mc^2 \gamma \beta^2}\int_C\left(\beta_{\lambda x}Z_x(\omega_{\lambda n}) + \beta_{\lambda y}Z_y(\omega_{\lambda n})\right)\frac{ds}{C}, \quad \lambda = 1,2 . \tag{82}$$

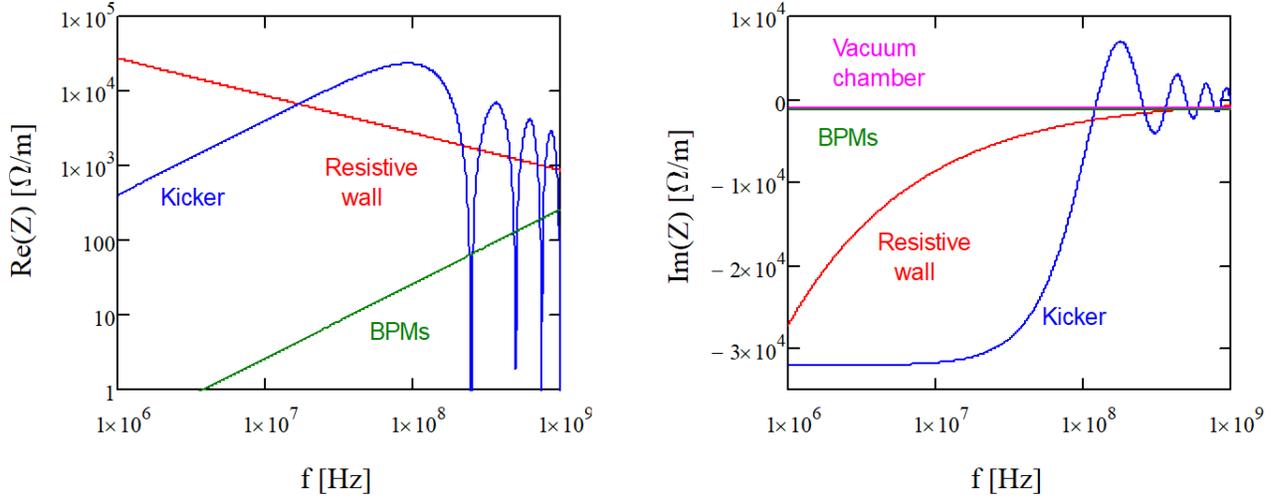

Figure 29: Major contributions to the transverse impedance of the cooling ring.

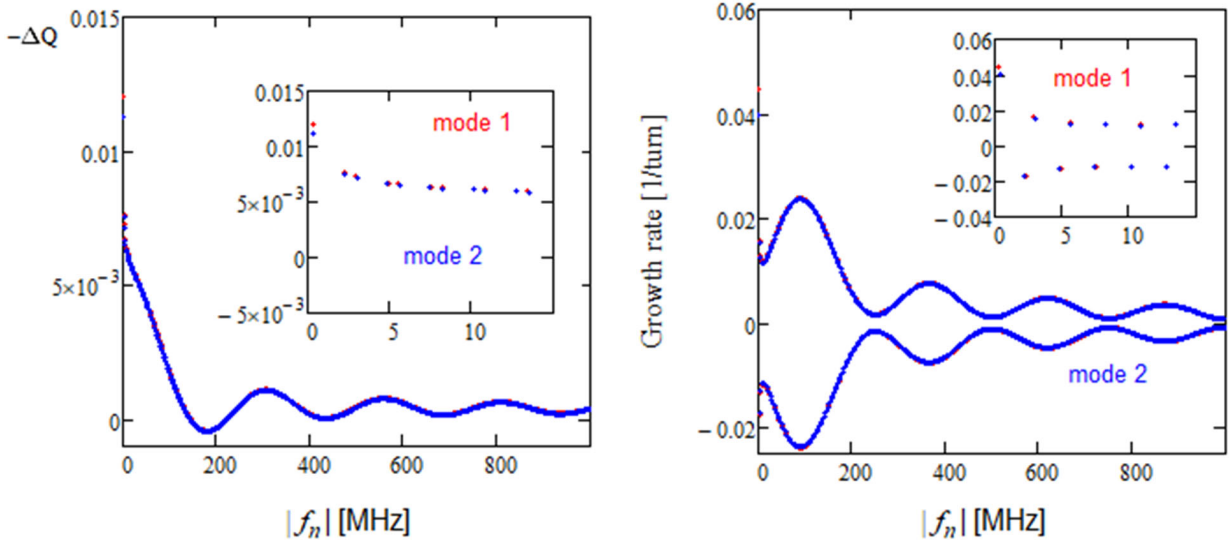

Figure 30: Dependence of real part of coherent tune shifts (left) and the instability growth rates (right) on mode frequency. Insets show low frequency parts of the plots.

Here $\omega_{\lambda n} = \omega_0(\nu_\lambda - n)$ are the $n$-th mode frequencies for betatron modes 1 and 2, and $\nu_\lambda$ are the tunes of the betatron modes, and $\omega_0$ is the circular frequency of particle motion in the ring. The real parts of the coherent shifts represent the betatron tune shifts for the corresponding modes. While $2\pi \operatorname{Im}(\delta Q_{\lambda n})$ represent the instability growth rates of the corresponding betatron modes in the absence of feedbacks and Landau damping. The growth rates are negative (damping) for $\omega_{\lambda n} > 0$ and positive for $\omega_{\lambda n} < 0$. Figure 30 presents the tune shifts for different modes. Note that because all 4D-beta-functions are equal in the kickers they make equal contributions to both modes.

*Suppression of transverse instabilities*

Same as for the longitudinal degree of freedom we assume that the transverse damper has ~250 MHz bandwidth. To damp the low frequency modes the damper must have the damping gain of ≥0.05 turn$^{-1}$. The frequencies above 250 MHz have to be Landau damped with damping rate of ~0.008 turn$^{-1}$.

Landau damping can be provided by two mechanisms: (1) by the chromatic dependence of the particle tunes and (2) by the transverse nonlinearity of the machine optics. Note that for coasting beams the nonlinearity due to space charge does not contribute to Landau damping; moreover, the space charge suppresses it, separating single-particle and coherent tunes [36].

According to Ref. [37] for continuous beam with Gaussian distribution in all degrees of freedom the instability is stabilized if:

$$|v'_\lambda + n\eta| > \frac{0.6\delta v_{SC\lambda}}{\sigma_p \ln(\delta v_{SC\lambda} / \operatorname{Im} \delta v_{\lambda n})} \ . \tag{83}$$

Here $\eta = \alpha - 1/\gamma^2$ is the slippage factor, $v'_\lambda$ is the tune chromaticity of the corresponding mode, and $\delta v_{SC\lambda}$ is the SC tune shift of the mode $\lambda$ determined by Eq. (52). The right-hand side in Eq. (83) has a weak dependence on the frequency (mode number) and for the more unstable mode 2 is about 20. That exceeds the left-hand side in the frequency range from 0 to ~1.5 GHz making the beam unstable in this frequency range with exception of 0 to 250 MHz band where the beam is stabilized by damper.

Thus, the only way to achieve the beam stability is an introduction of non-linearities in the transverse particle motion so that to introduce a dependence of tunes on the betatron amplitudes. Usually it is achieved by introduction of octupoles. However, an introduction of non-linearity by the Landau electron lens [37] would be much more effective means for the stabilization. Let us assume that the lens of length $L_l$ is installed in the technical straight of the ring. The lens transverse density distribution is Gaussian and is identical to the beam's one with rms size of $\sigma_l$. For the lens current $I_l$ and the electron velocity $\beta_l c$ opposite to the beam's one, the maximal incoherent tune shift the lens provides for the mode $\lambda$ is:

$$\Delta \hat{v}_\lambda = -\frac{r_e I_l L_l (\beta_{\lambda x} + \beta_{\lambda y})}{4\pi ce\beta^2 \gamma \sigma_l^2} \left(\frac{1}{\beta_l} + \beta\right), \quad \lambda = 1,2 \ . \tag{84}$$

Because all 4D-beta-functions are equal at the lens location the lens provides the same tune shift for both modes.

Landau damping of collective motion is provided by energy transfer from the coherent mode to incoherent motion of those particles which individual frequencies are in resonance with the collective tune. That is why the direct space charge, which increases separation of coherent and incoherent tunes, suppresses Landau damping: indeed, while it shifts the incoherent tunes down, its influence on the coherent spectrum is much weaker. Thus, the mode 2 is more vulnerable for the instability, since its space charge tune shift is much larger than one of the mode 1. According to Ref. [36], the lens provides Landau damping rate for the coherent mode 2

$$\Lambda_2 = -\pi \langle \Delta v_2 \rangle \Delta \hat{v}_c \int dJ_1 dJ_2 \frac{\partial f}{\partial J_2} J_2 \delta(\Delta v_2(J_1, J_2) - \Delta \hat{v}_c). \tag{85}$$

Here $\Delta v_2(J_1, J_2)$ is the dependence on actions of tune shift for the mode 2 due to direct beam space charge and by definition $\Delta v_2(0,0) = \Delta v_{SC2}$, $\langle \Delta v_2 \rangle \approx \Delta v_{SC2}/2$ is the average incoherent tune shift for the mode 2 due to direct beam space charge, the integrals over actions are taken over the entire phase space with the distribution function normalized so that $\iint f dJ_1 dJ_2 = 1$, and $\Delta \hat{v}_c \approx \Delta \hat{v}_2/2$ is the coherent tune shift due to the electron lens for mode 2. According to this formula, Landau damping is possible due to the negative tune shift provided to the collective mode by the electron lens, so that beam electrons at far enough tails may be in the resonance with the collective mode, as it reflected by the δ-function argument. To determine the dependence of the space charge tune shifts on actions we account that the major contribution to $\Delta v_2(J_1, J_2)$ comes from arcs where the beam is decoupled. In this case we can use the equation of Ref. [37] derived for Gaussian distribution with very different transverse emittances. The space charge tune shift for the mode with much smaller emittance is:

$$\Delta v(J_1, J_2) = \frac{\Delta v_{SC2}}{2} \exp\left(-\frac{J_1}{2}\right) I_0\left(\frac{J_1}{2}\right) \int_0^1 \exp\left(-\frac{J_2 t}{2}\right) \left(I_0\left(\frac{J_2 t}{2}\right) - I_1\left(\frac{J_2 t}{2}\right)\right) \frac{dt}{\sqrt{t}}. \tag{86}$$

Here $I_0(x)$ and $I_1(x)$ are the modified Bessel functions, the actions $J_{1,2}$ are dimensionless; each one is taken in units of its own emittance, so that average values are 1 for each of them, $\iint J_{1,2} f dJ_1 dJ_2 = 1$. To compute Landau damping, one may note that for the halo particles the space charge tune shift can be approximated as $\Delta v_2(J_1, J_2) \simeq 0.6 \Delta v_{SC2}/\sqrt{J_1 J_2}$. With this approximation, the Landau damping integral can be taken with the saddle-point method, yielding:

$$\Lambda_2 \simeq \pi^{3/2} \Delta v_{SC2} \zeta^{5/2} e^{-2\zeta}, \quad \zeta = 0.6\Delta \frac{\Delta v_{SC2}}{\Delta \hat{v}_c}. \tag{87}$$

Exponential dependence of the Landau damping rate is shown in Figure 31. According to the above estimations, the instability growth rate does not exceed $0.01\Delta v_{2m}$; thus, according to this plot, to suppress the instability one needs $\zeta \simeq 5$, which corresponds to $\Delta \hat{v}_\lambda \approx \Delta v_{SC2}/4$.

For $L_l = 10$ m, $\beta_{l\lambda} = 1$ m, $\sigma_l = 1.47$ mm and the beam energy of 5 kV we obtain the lens stabilizing current of 18 mA. Large length of the lens, $L_l \gg \beta_{\lambda x}, \beta_{\lambda y}$, was chosen to minimize the resonance driving terms for the given values of lens's tune shifts.

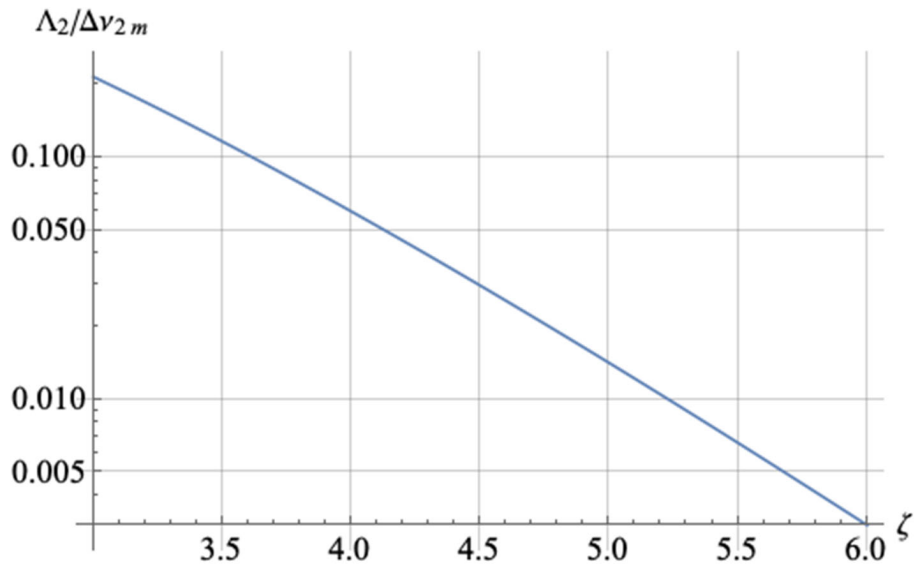

Figure 31: Dimensionless Landau damping rate versus the tune shift parameter $\zeta$.

## 10. Induction Linac

Beam acceleration up to injection energy is provided by an induction linac. The Linac parameters are listed below:

**Table 5. Main requirements for the Induction Linac systems**

| Energy | 54.5 MeV |
|---|---|
| Beam current | 50 A |
| Pulse width | 380 ns |
| Repetition rate | 200 s$^{-1}$ |

The main challenge is to generate the beam with emittance close to thermal one and preserve it to the end of the Linac. The system consists of the Injection Unit providing low – emittance beam, and Linear Induction Accelerator (LIA).

The <u>Injection Unit</u> has the following components:
 - Pulsed power system;
 - Electron gun;
 - Matching section (solenoid, iron yoke and focusing pole piece).

The LIA unit consists of the following sub-systems:
 - Accelerating cells;
 - Pulsed power system;
 - Focusing System (FS);
 - FS power source;
 - Cooling system.

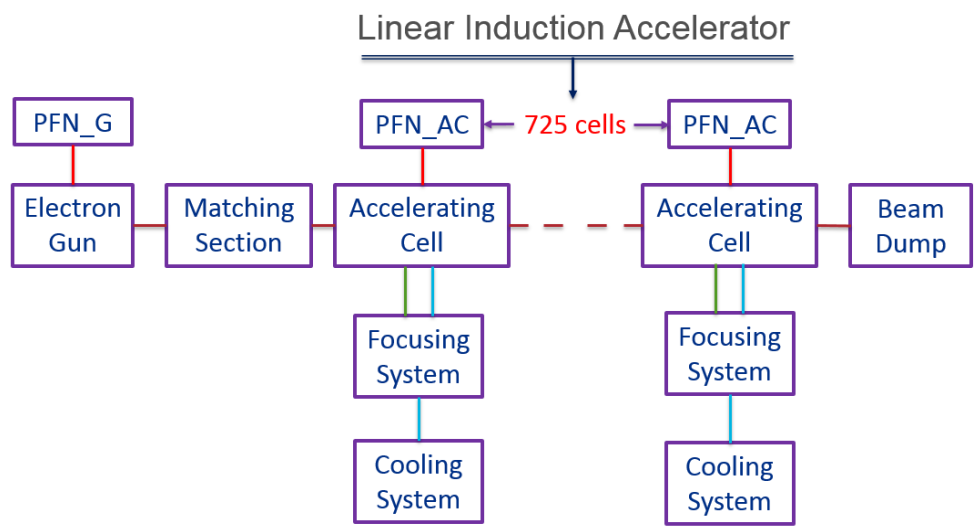

Figure 32: Block system of the accelerating system.

Block scheme of the system is shown in Figure 32. Main parameters of the sub-systems are estimated below for the beam current of 50 A.

## 10.1. Electron gun.

The low-emittance high current diode electron guns have been developed, built, tested and successfully exploited in the past for 7-GHz, 11.4-GHz, and 34.3-GHz magnicons [38-40]. All of these electron guns demonstrated reliable operation and the repetition rate of up to 10 Hz and the pulse width of up to 1 µs. The gun sketch is shown in Figure 33. However, in order to operate at a higher repetition rate, the anode and cathode electric fields should be reduced significantly. In addition, the gun voltage should be smaller than ~300 kV to provide reliable operation of the HV source.

Our concept proposes to:

a. Generate the beam with a uniform transverse current distribution
b. Operate at much lower beam area compression than the guns of Ref. [38-40], while keeping the same concept (Figure 33);
c. Use micro-perveance ~1.-1.5 micro-perv.

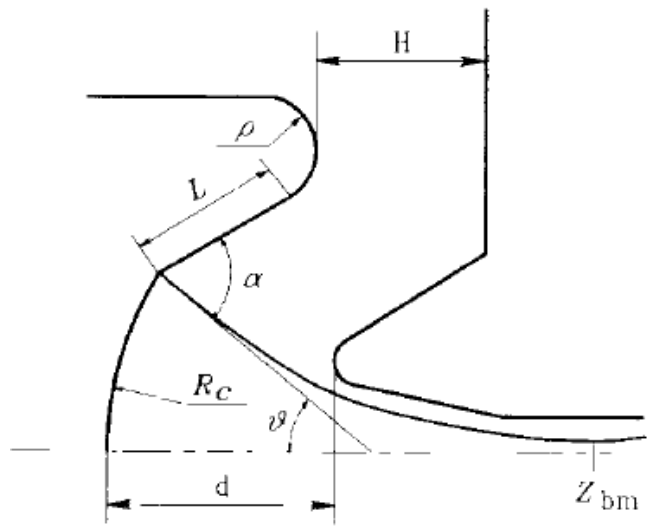

Figure 33: Scheme of diode gun with the high beam area compression and enlarged electric strength.

Table 6 shows the main gun parameters (the gun voltage - $U$, the beam pulsed power - $P_{pulsed}$ and the beam average power - $P_{average}$) for different values of micro-perveance $P_\mu$ at the beam current of 50 A.

**Table 6: Main Parameters of Gun**

| Pμ  | U, kV     | P$_{pulsed}$, MW | P$_{average}$, kW |
|-----|-----------|------------------|-------------------|
| 0.1 | 629.9605  | 31.49803         | 2.39385           |
| 0.3 | 302.8534  | 15.14267         | 1.150843          |
| 0.5 | 215.4435  | 10.77217         | 0.818685          |
| 0.7 | 172.153   | 8.607651         | 0.654181          |
| 0.9 | 145.5967  | 7.279837         | 0.553268          |

Micro-perveance of 0.3 μperv looks reasonable, considering that even higher micro-perveance (~1.2 μperv) was used successfully for the gun presented in Ref. [38]. The gun voltage in this case is 300 kV. The gun has a modest compression ratio with reasonable surface electric fields, which is required at high repetition rate. In order to obtain reasonable cathode lifetime (20,000-30,000 hours), the dispenser cathode diameter is chosen to be 25 mm, limiting loading to 12 A/cm$^2$, which is the same as in the gun for 34 GHz magnicon [40]. Taking into account the cathode current density inhomogeneity this is the maximal current density necessary for anode aberration compensation [38]. This gun is optimized for low emittance and contains a special near-cathode molybdenum electrode (ring following the cathode edge), having a Pierce angle with respect to the cathode surface. It prevents emittance dilution caused by the gap between the electrode and the cathode (required for thermal insulation). The dispenser cathode has the operating temperature of 1050 C, which provides acceptable normalized thermal emittance of < 4 μm. The cathode has the surface roughness, which contribution to the effective beam temperature $\Delta T$ may be estimated by the following formula [40]: $\Delta T[°C] \approx kj^{2/3}[A/cm^2]d^{4/3}[\mu m]$, where $j$ is the cathode current density, and $k$ is a factor of order unity. For dispenser cathodes, surface roughness scale $d$ is a few microns, so for a current density of 12 A/cm$^2$ the effective temperature increase caused by the cathode surface roughness is less than 50°C, and therefore, thermal emittance increase caused by the cathode roughness is negligible. The beam in the accelerating channel has diameter of 10 mm, which gives very modest beam area compression in the gun, 6.5:1, and consequently, reasonable surface fields. The gun is combined with the trimming coil [40], which provides 12 Gs magnetic field on the cathode to create magnetized beam required for electron cooling. The magnetic field also matches the beam into the focusing coil. The yoke pole piece aperture and its location relative to the cathode should provide magnetic such field distribution that the magnetic force lines coincide the beam trajectories. It provides minimal beam scalloping and minimal emittance increase. The concept of the beam matching to the coil is shown in Figure 34.

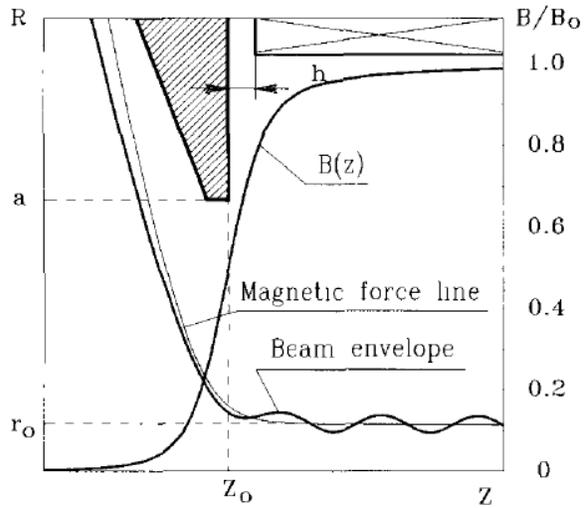

Figure 34: The beam matching to the first focusing solenoid.

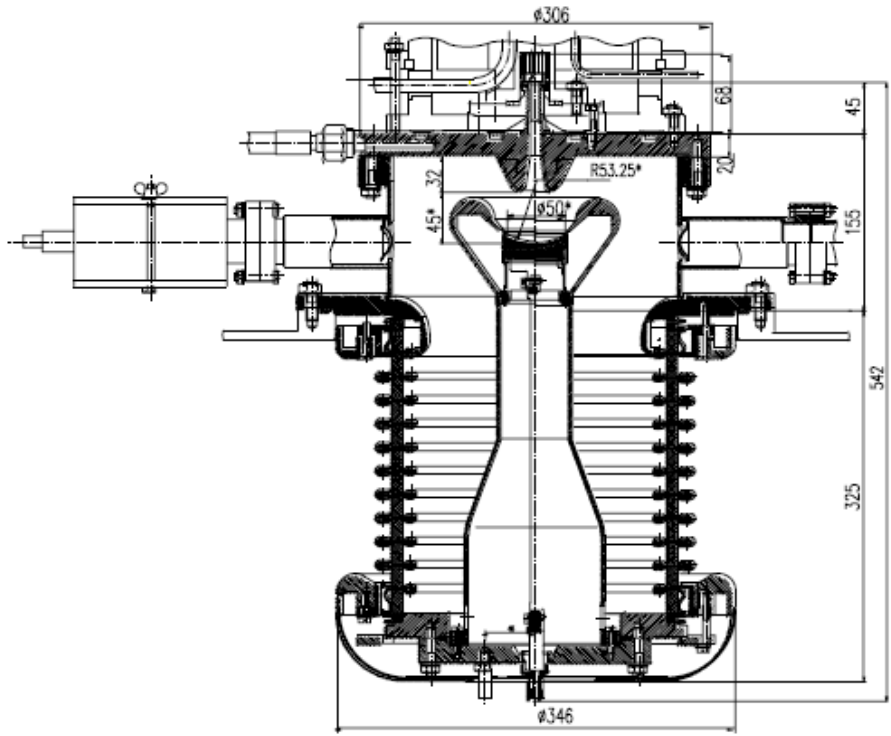

Figure 35: The gun design concept.

The design concept of the gun (according to Ref. [40]) is shown in Figure 35. All critical elements should be fabricated and assembled with accuracy < 0.1 mm.

As the first step, a preliminary optimization of the gun geometry was made. It required matching the gun optics into the linac magnetic system. Corresponding simulation of the beam dynamics during acceleration was performed in order to understand whether it is feasible to achieve the beam normalized emittance close to thermal one at the output of the linac. At this step, we did not pay too much attention to the surface fields in the gun. The goal was to prove that it is feasible to achieve optics in the entire linac with acceptable emittance dilution.

Note, that it is a considerable challenge to simulate the optics with high precision so that to resolve small beam emittance comparable to the thermal one because the simulation codes have noise caused by the space charge approximation. Different codes have different artificial effects, which lead to emittance overestimation. We used 3D MICHELLE code [41] and 2D WinSAM code [42]. The MICHELLE code has a modern interface but has significant artificial aberrations and noise. Therefore, we used the MICHELLE code for initial optimizations, and WinSAM code for the final precise simulations to minimize the emittance. WinSAM code is based on a more accurate numerical model and provides much smaller artificial aberrations and noise. 3D MICHELLE code also has been used for tolerance analysis (optics sensitivity versus the anode electrode offset and tilt) to be sure that the assembly tolerances required for emittance preservation are reasonable. Simulation to optimize the gun characteristics has been done, using the dimensions shown in Figure 36. The design has been optimized for 50 A current and 300 kV voltage. The optimization criteria were aimed at minimization of normalized emittance and obtaining the uniform distribution of beam current density. It means that the optics aberration should be accurately compensated. The current density distribution over the cathode surface has minimum on the axis, which allows one to compensate anode aberrations and to obtain the "geometrical" emittance (*i.e.* rms emittance without considering of the thermal effects) lower than the thermal one.

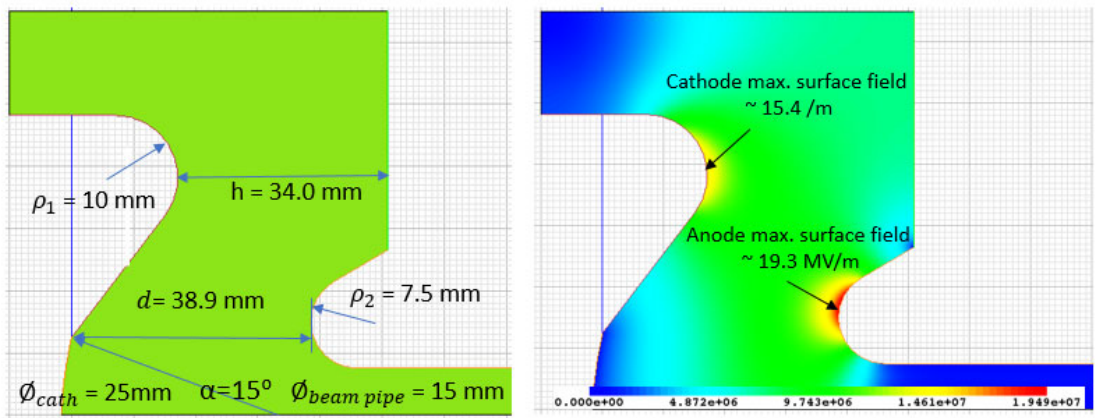

Figure 36: Main dimensions, electric field distribution and field surface values.

Figure 36 shows the main dimensions and electric field distribution in the optimized version.

Note that the shape of the focusing electrode is not optimized yet to provide fast gun conditioning and reliable operation at high repetition rate at small breakdown rate. However, the maximal electric field there does not exceed 15 MV/m, which is close to acceptable [38-40]. The field at the anode nose of ~20 MV/m is also acceptable.

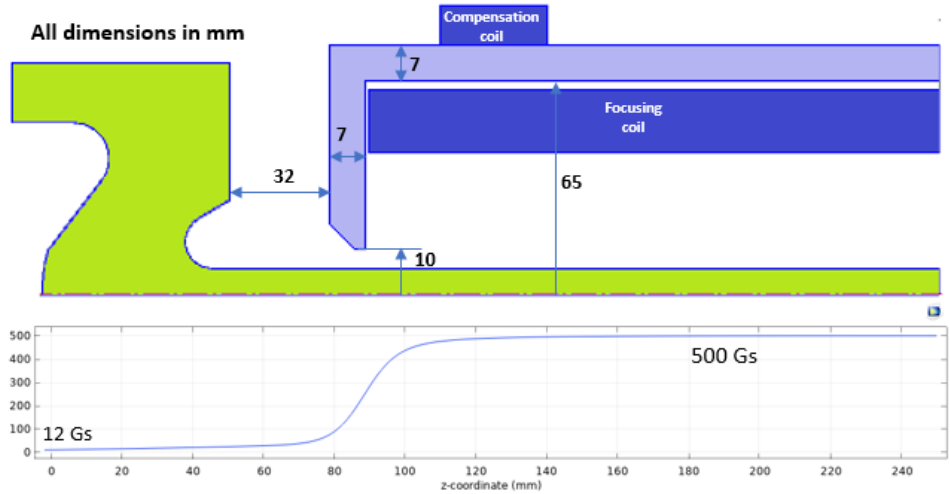

Figure 37: Layout of the focusing system and distribution of magnetic field along axis starting from the cathode.

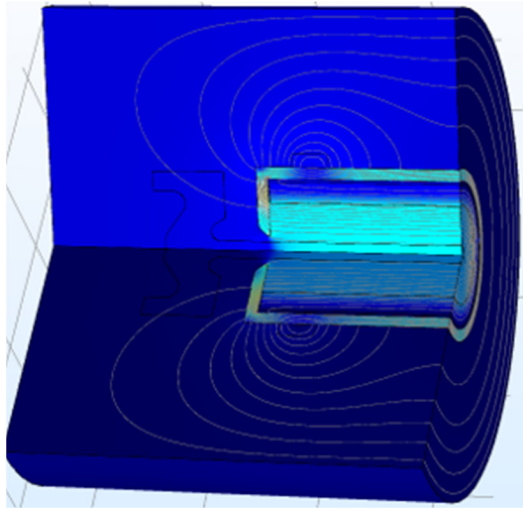

Figure 38: Contour of magnetic flux density

At the first stage the simulations have been performed without external focusing magnetic field. That resulted in the minimal normalized geometrical emittance $\varepsilon_G$ of ~ 2 μm with a uniform current density distribution and the beam radius of 5 mm. At the next step the simulations were performed with external magnetic field matched to the beam. The focusing coil is combined with the compensation coil providing 12 G magnetic field on the cathode. Figure 37 and Figure 38 show the layout of the focusing and compensation coils and magnetic flux distribution.

Initially, the beam dynamics was optimized without taking into account the thermal spread of the electron velocities on the cathode. In Figure 39 the beam trajectories are shown for this case as well as the mesh used for space charge approximation. In the right upper corner, the current distribution over the cathode is shown. Details of current density distribution is shown in Figure 40.

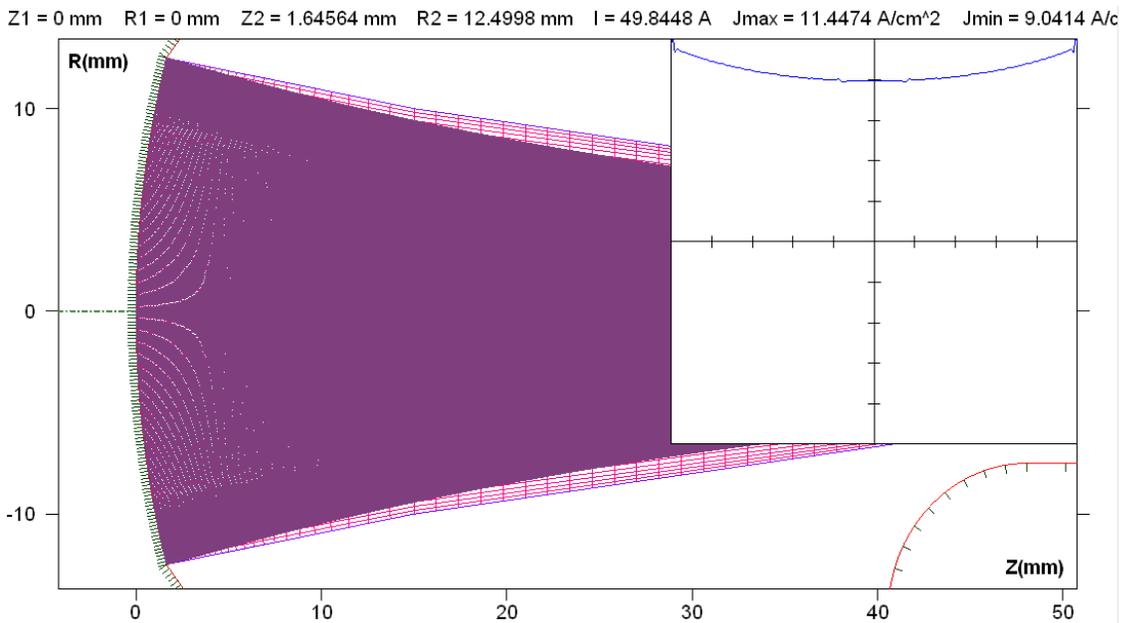

Figure 39: The beam trajectories (in purple) in the gun gap and the current density distribution on the cathode. The mesh for space charge approximation is shown also (in red).

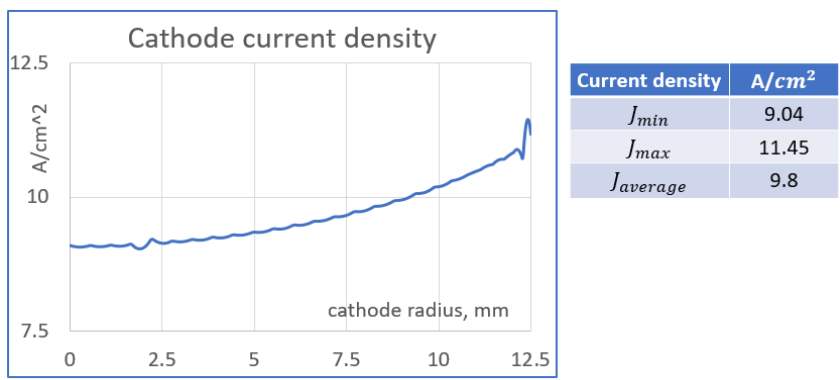

Figure 40: The current density distribution over the cathode. Noise is caused by space charge approximation. Maximal current density on the cathode edge does not exceed 12 A/cm².

The beam trajectories in the presence of magnetic field and thermal spread of velocities are shown in Figure 41 for optimized gun optics. One can see that in this case the beam envelop scalloping amplitude does not exceed 3.5%. The zoomed mesh for space charge approximation used in WinSAM is shown in Figure 42. The mesh is denser in critical areas to maximize accuracy.

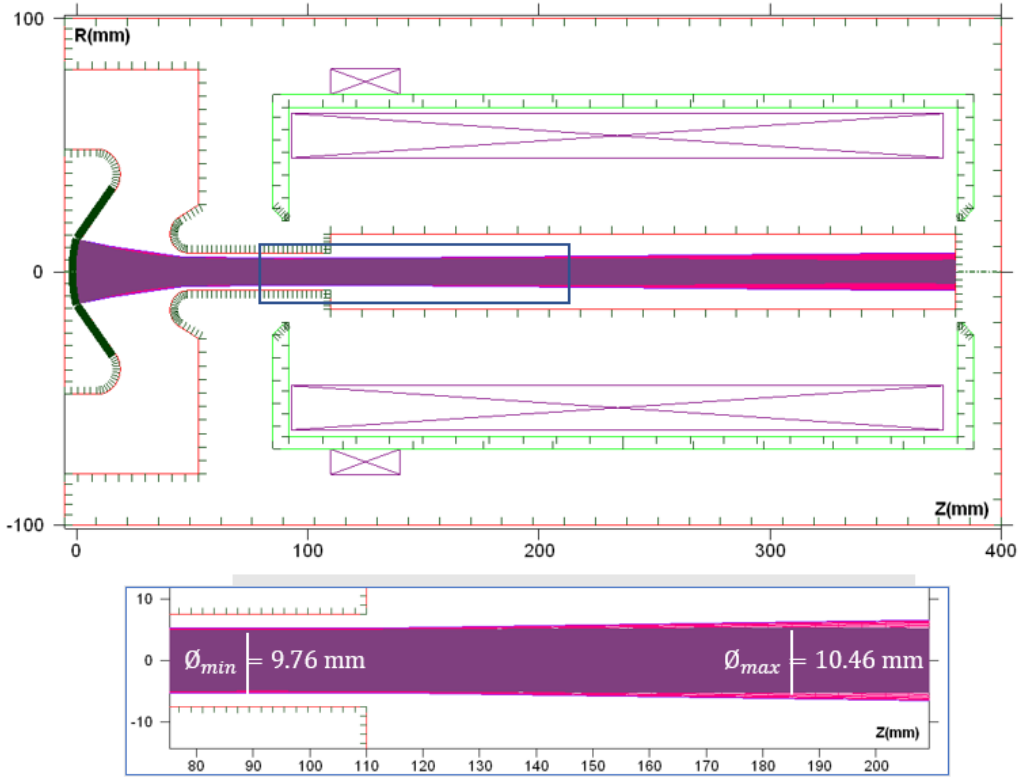

Figure 41: The gun and matching system layout and the beam trajectories (in purple) in presence of magnetic field and thermal spread. The mesh is shown in red. Below a zoomed part is shown to demonstrate the beam scalloping.

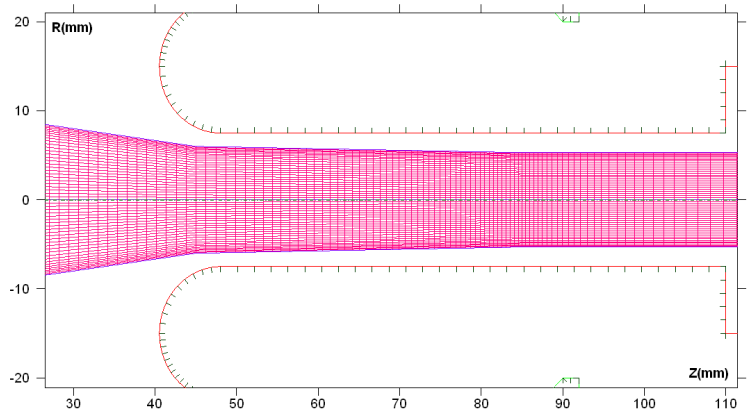

Figure 42: Zoom of mesh used in WinSAM simulations.

Note that the beam optics depends on the gun voltage, and one should make sure that there will not be current interception by the anode electrode during the gun voltage rise and fall times. In Figure 43 the beam trajectories (in purple) are shown for the voltage of 200 kV and at the voltage of 50 kV.

One can see that in both cases there is no current interception by the anode electrode. The beam at 50 kV voltage is not matched, there is a strong scalloping. For lower voltage, where the beam is not relativistic, the beam dynamics does not depend on the beam voltage. In the left lower figure, the normalized rms beam emittance (actually, it is square root of four-dimensional emittance) is shown at different distances for different beam voltage. Right lower figure shows the beam current as a function of the beam voltage.

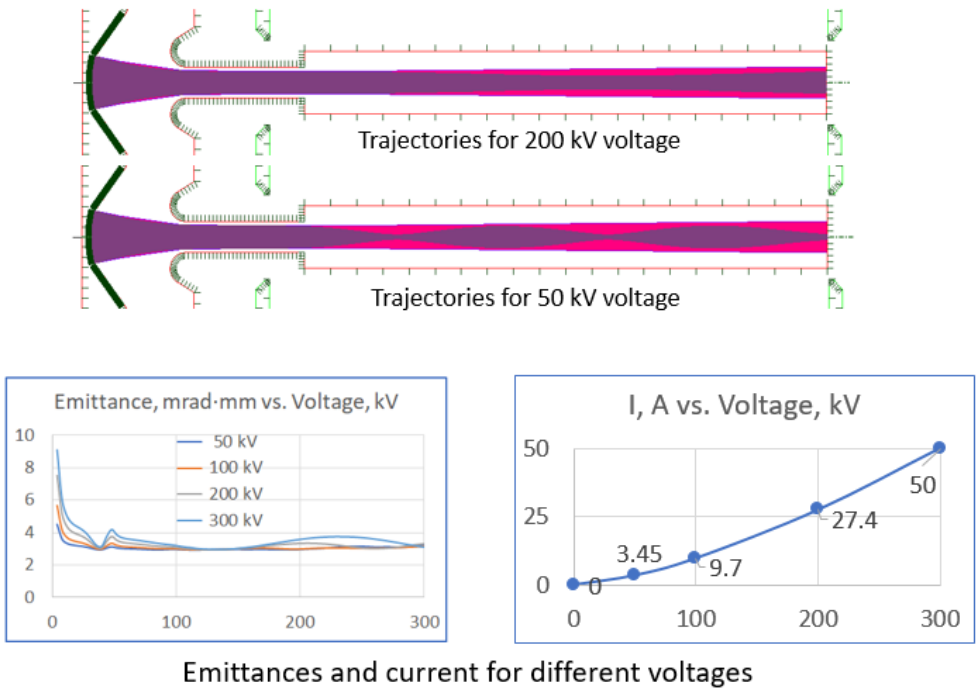

Figure 43: The beam trajectories for 200 kV and 50 kV (top); the beam normalized rms emittance versus longitudinal coordinate for different beam voltages (bottom, left) and beam current versus the beam voltage (bottom, right).

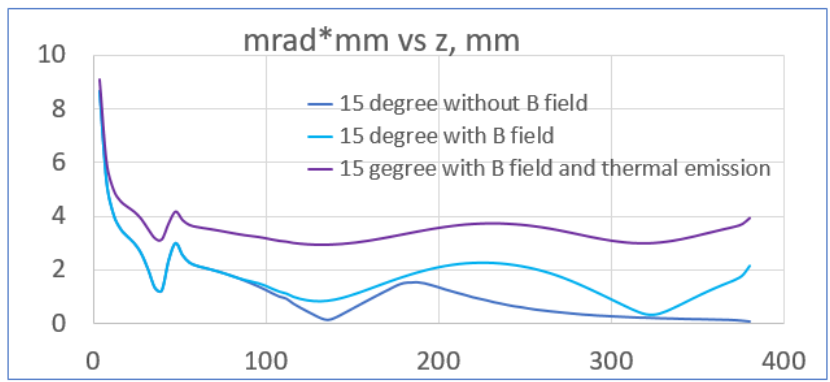

Figure 44: The beam normalized rms emittance along the beam axis.

The evolution of normalized rms emittance (square root of four – dimensional emittance) along the gun axis is shown in Figure 44 for optimal cathode curvature angle (15°) for three cases: (1) the beam without thermal velocity spread and without external magnetic field; (2) the beam without thermal velocity spread in presence of magnetic field, and (3) the beam in presence of thermal spread and external magnetic field.

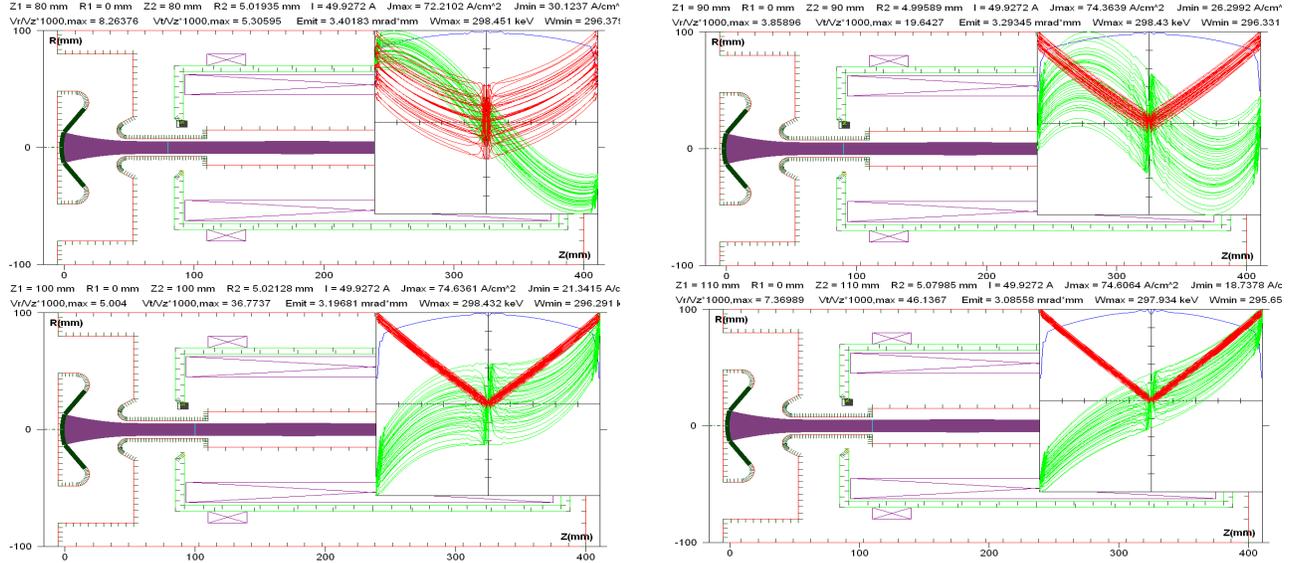

Figure 45: Beam current density distribution (blue) and dependence of the radial (red) and azimuthal (green) velocities versus radius, upper right corner.

Figure 45 presents the beam evolution in different axial cross sections: current density, radial and azimuthal velocities versus radius for the beam in presence of initial thermal velocity spread. The gun parameters are summarized in the Table 7.

## *10.2. Focusing system*

In the linear induction accelerator (LIA), electric (accelerating) field is concentrated in the gaps between the accelerating cells. Focusing magnetic field for the electron beam transport is generated by a system of coils located in each cell (Figure 46). Due to the gaps between the coils, the magnetic field is not uniform and there is some ripple of the beam envelope. To reduce the amplitude of the ripple, the gap between the coils must be small relative to the length of the coil and its diameter. To get the radius of the beam close to the Brillouin limit the strength of the focusing magnetic field at the beginning of the accelerator must be ~450 G. It will be gradually reduced to ~50 G towards the end of the accelerator so that the beam diameter stays the same over the entire length of the beam line. During initial stage of the accelerating cell design, with some safety margin, a 700 G starting magnetic field was chosen. To obtain this field in the front sections of the LIA, each coil will be wound using AWG14 copper wire with the total amount of turns of ~600. The radial space that this coil will occupy is ~20 mm. Each coil will dissipate ~200 W of heat and cooling will be needed to remove it. However, the required cooling rate is small compared to the expected losses in the cell cores as will be shown below.

**Table 7: Main parameters of the gun**

| | |
|---|---|
| U, kV | 300 |
| I, A | 50 |
| Aperture radius, mm | 7.5 |
| Perveance | 0.3 |
| R cathode, mm | 12.5 |
| T cathode, ºC | 1050 |
| Cathode $J_{min}$, A/ $cm^2$ | 9.04 |
| Cathode $J_{max}$, A/ $cm^2$ | 11.45 |
| Cathode $J_{average}$, A/ $cm^2$ | 9.8 |
| $\varepsilon_T$, mm·mrad | 3.75 |
| $\varepsilon_G$, mm·mrad | 2.25 |
| B solenoid, Gs | 500 |
| B cathode, Gs | 12 |
| Beam radius, mm | 4.9 |
| Scalloping amplitude | 3.3% |
| Maximal cathode E field | 15.4 MV/m |
| Maximal anode E field | 19.3 MV/m |
| Solenoid yoke, apt, mm | 20 |
| Yoke thickness, mm | 7 |

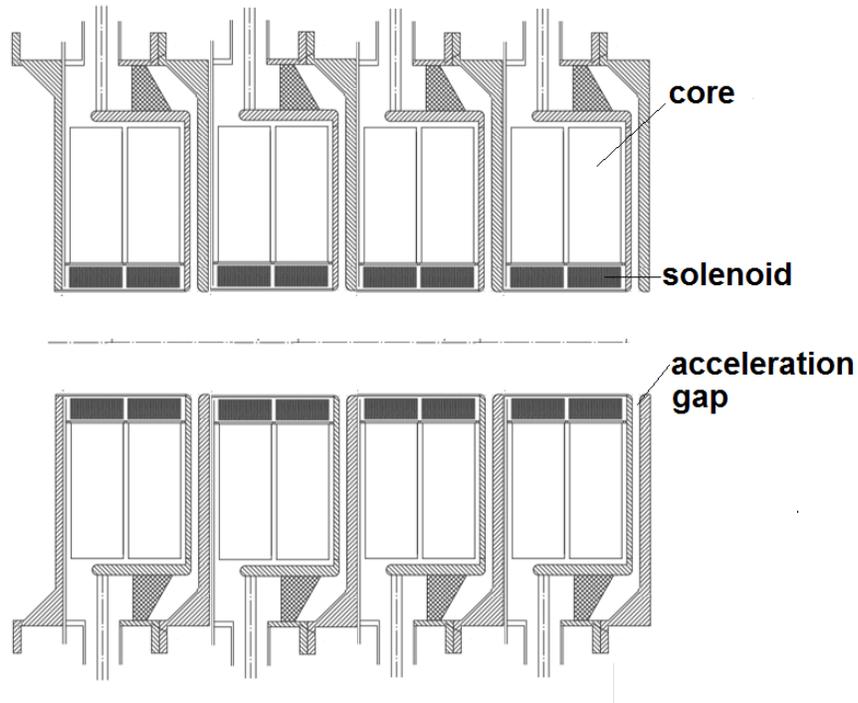

Figure 46: Schematic of the induction accelerator showing accelerating cells. Each cell contains two cores, two focusing solenoids and an acceleration gap.

## 10.3. *Accelerating cells*

To configure accelerating cell of the LIA, several factors must be taken into account and proper choice of major cell parameters must be made (at least acceptable as a first iteration in the design). One of major decisions to make at the initial stage of the accelerating cell design is the choose of material for the cell's core. The choice is driven by power loss and the fabrication cost. Results of the study of the materials for large scale linear induction accelerators summarized in [43] point to Metglas 2605SC as the most preferable candidate for the core production. This material can provide the maximum flux swing of ~3.1 T. To reduce the power loss, the cores must be annealed in magnetic field after winding. Example of the difference in the power loss between annealed and "as cast" material 2605SA1 (which is the 2605SC optimized for the 60 Hz frequency) can be seen in Figure 47. The hysteresis loop for the 2605SC material is shown in Figure 48. Empirical expression for the core loss, which summarizes the findings in [43] is:

$$U[\text{J/m}^3] = 850 \left(\frac{\Delta B}{2.5}\right)^2 \left(\frac{1}{T_p}\right)\left(\frac{t}{25.4}\right) , \qquad (88)$$

where $\Delta B$ is in Tesla, pulse duration $T_p$ is in µs, and the thickness of the Metglas ribbon $t$ is in µm.

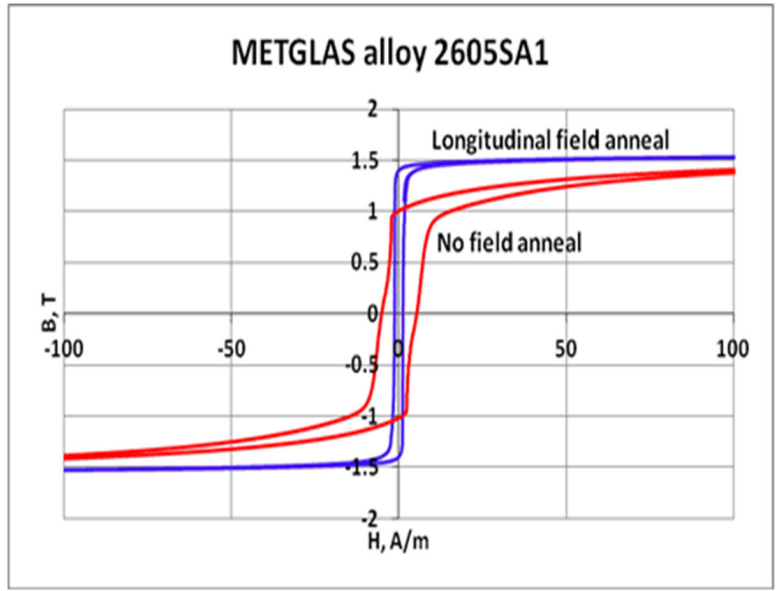

Figure 47: Impact of annealing on the core loss.

The amount of the magnetic material needed for each cell is defined by its magnetic properties, accelerating voltage, and the duration of the pulse. Knowing the required diameter of the beam channel (~100 mm) and the projected size of the focusing coils (see Section 10.2) the inner diameter of the Metglas cores in the accelerating cells of the LIA was chosen to be ~155 mm.

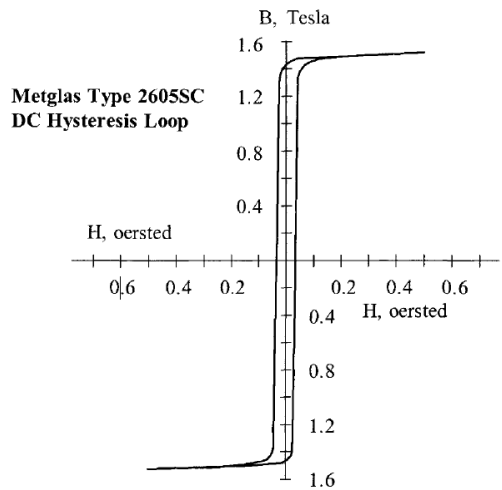

Figure 48: DC hysteresis loop for the 2605SC material (81% Fe, 13.5% B, 3.5% Si).

Evaluation of the power loss in the core of a cell can be made as following:

- Let's assume that the width of the ribbon is 50 mm and that two cores are used in each cell.
- Let's accept accelerating voltage of the cell $V_{acc}$ = 75 kV and the active cross-section of the Metglas 2605SC in the cell $S$ = 0.01 m².

- With the active length of the pulse of 380 ns, the required swing of the magnetic flux density in the core is $\Delta B = U \cdot T_p / S$, or 2.85 T. That agrees well with an acceptable swing for the material.
- Let's also accept radial filling factor in each core of 0.77 (this filling factor was demonstrated). That yields the radial size of the cell to be ~130 mm.
- Accepting the 25 µm thickness of the ribbon and 2.85 T flux density swing we obtain, according to Eq. (88), the energy loss in one cycle $U \approx 2300$ J/m$^3$. This results in ~30 J/cycle power loss in the volume of the core.
- With the required 200 Hz repetition rate, the average power dissipated in one cell is ~6 kW.

Forced cooling is required to remove this power. If transformer oil (assumed specific heat $c$ = 1860 J/(kg·K), density $\rho$ = 0.88 kg/liter) is used with the allowed temperature rise $\Delta T$ = 20 K the required oil flow is: $dV/dt = (1/\rho)(dm/dt) = P/(c\rho\Delta T) = 0.183$ l/s or ~11 l/min (~3 GPM).

Sketch of the accelerating cell is shown in Figure 49. To prevent sparking the gap between the core and the outer portion of the input loop in oil (breakdown electric field $E_m$ = 20 kV/mm) must greater than ~4 mm. 7.5 mm accelerating gap shown in the figure is sufficient to withstand 75 kV voltage in vacuum.

The cell must be structurally sound as the cores and the winding will weight together ~100 kg. Oil is pumped into the coil region and goes radially in the space between the cores. Flow of the oil must be arranged carefully to ensure consistent cooling.

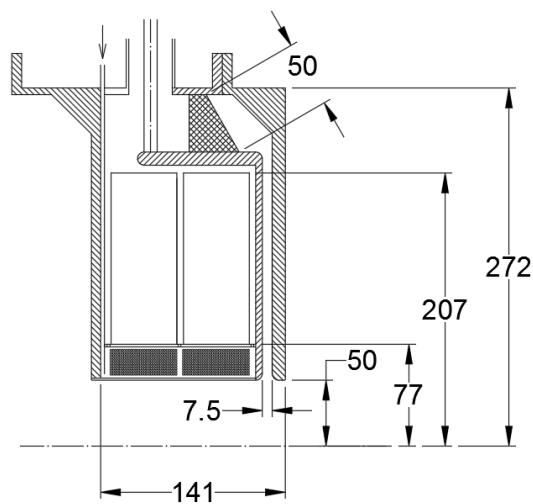

Figure 49: Acceleration cell concept with the solenoids.

The length of the vacuum surface of the insulator as shown in Figure 49 is ~50 mm; further optimization is possible.

Part of the cell impedance associated with the power loss in the core is $R_{core} = U^2/P_{max} \approx 94$ Ohm; the impedance to the beam is ~1500 Ohm. Together these two loads bring the effective impedance of the cell to $R_{eff} \approx 88.5$ Ohm.

At this design stage, a 50-Ohm effective impedance of the power delivery system is considered for each cell. This choice provides some flexibility for future design work. To get this impedance, a 125 Ohm resistance needs to be installed at the input of each cell with sufficient power handling capability (~1.8 kW average power). This brings the total instant power generated by the PFN at the input of the cell to ~112.5 MW. With the pulse duration 380 ns and 200 Hz repetition rate, the average power per one PFN is ~8.6 kW.

To ensure some symmetry of the power delivery, at least two inputs must be used.

To get the required final energy of the electrons, 727 accelerating cells must be used. With the average 8.6 kW power consumption per one PFN, the power consumption in the LIA without the injection section is ~6.2 MW.

### *10.4. Pulse forming network (PFN) for the accelerating cell*

A pulser that feeds the cell must provide the needed voltage within the required pulse length at the stated repetition rate. The best way to approach a solution for the pulser would be to design the power system after obtaining the pulse transfer function of the cell known. Preliminary evaluation is presented below.

The function of the pulse-forming network (PFN) is to get the needed energy from the power source into the primary storage capacitance, to get the required voltage on the intermediate storage element, to generate a rectangular 75 kV voltage pulse, and to provide needed rise and fall times of the output pulse. A concept of a pulse forming network (PFN) considered initially had 500 Hz repetition rate and higher electron beam current (Figure 50).

A need for some the re-assessment (and simplification) of the conceptual schematic came from the fact that the repetition frequency was reduced to 200 Hz and the beam current was reduced from 200 to 50 A. Besides there was a need to evaluate power loss in elements of the PFN and to ensure the charge time of the PFL is long enough to provide uniform distribution of the charge along the line to avoid unwanted pulsation at the top of the voltage pulse at the input of the cell (e.g. see [44]).

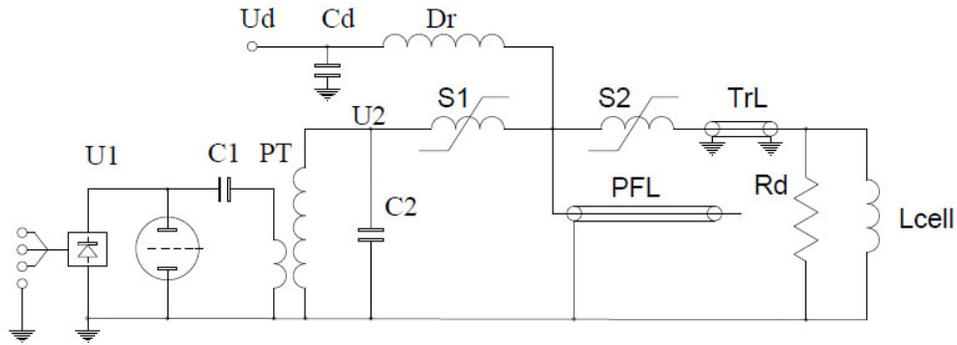

Figure 50: Conceptual schematic of the PFN proposed at the fist iteration of the study.

**Pulse-Forming Line (PFL)**

The length of the forming line (PFL) within the PFN is defined by the required 380 ns pulse length and the properties of insulating material:

$$l_{PFL} = \frac{Tc}{2\sqrt{\varepsilon}}.$$

where $l_{PFL}$ = 38.4 m, and $\varepsilon$ = 2.2 if transformer oil is used.

Cross-section of the PFL is defined by the required impedance (50 Ohm), the maximum PFL voltage, and the insulating properties of the transformer oil (~20 kV/mm). If a coaxial configuration is chosen for the PFN, its impedance is:

$$Z = \frac{60}{\sqrt{\varepsilon}} \ln(b/a).$$

With some reserve, the inner radius $a$ = 10 mm and the outer radius $b$ = 34.4 mm have been chosen in this study. The capacitance of this line (38. 4 m long) is ~3.8 nF.

It is commonly accepted that the charging time of a PFL must be at least three times longer than the pulse length [44]. Choosing the charging time $\tau_{PFL}$ = 2 μs (corresponding frequency of the charging pulse is $f_{charge} \approx$ 250 kHz), gives plenty of time for the charge to redistribute evenly along the PFL. This long charging time is only possible if mineral oil is used as dielectric in all elements of the PFL; water-filled lines, commonly used in the PFL-s of LIA, require much shorter charging times.

On the other hand, the long charging time provides an opportunity to avoid using the saturation switch S1 between the step-up transformant and the PFL (see Figure 50). The PFL can immediately follow the transformer, which is separated from the PFL by a choke that regulates the frequency of the charging wave. Corresponding equivalent circuit is shown in Figure 51.

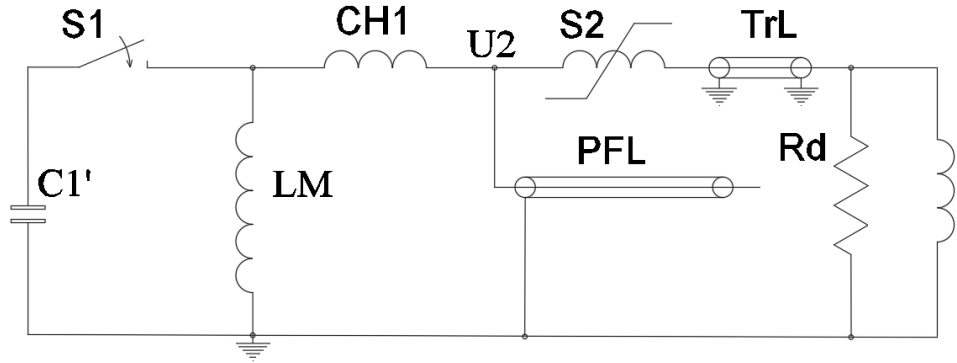

Figure 51: Equivalent circuit during the PFL charging stage.

The fringe inductance of the step-up transformer will serve as one part of the needed circuit inductance. The additional choke CH1 is used to bring the charging time to the desired value. The combined inductance $L$ of the choke and the transformer in combination with the storage capacitances of the primary storage C1 and the PFL must be chosen based on the expected shape of the voltage across the PFL:

$$U_{PFL} = \frac{U_2}{2}(1-\cos\omega t),$$

where $U_2$ is the charging voltage (150 kV), and $\omega$ is the current oscillation frequency:

$$\omega = \sqrt{\frac{2}{LC_{PFL}}}.$$

The expressions above are for the case when the maximum voltage on the PFL is equal to that on the primary storage capacitor with accounting the step-up ratio $k$: $U_2 = k \cdot U_1$. In this case, for the scheme in Figure 51, C1' = $C_{PFL}$.

As the capacitance of the PFL is chosen to store the required energy ($C_{PFL} \approx 4$ nF), we can evaluate the needed inductance of the charging contour: $L \approx 0.2$ mH.

Main components of the PFN in Figure 51 are the primary energy storage capacitor C1', the switch S1, the step-up transformer (marked here by its magnetization inductance and the fringe inductance), the choke CH1, the pulse forming line (PFL), and the saturation switch S2. Transmission line TrL is used to deliver the pulse to the cell equipped with the additional resistor Rd at the input.

To get the 75 kV voltage at the input of the accelerating cell, the PFL must be charged to 150 kV. This high voltage makes it difficult to find a reliable switching element. A step-up transformer is introduced to bring the switching voltage to the level where traditionally used switching elements can be employed. Two major parameters of interest are the maximum voltage and current; the maximum frequency and the number of cycles are also very important parameters to consider. The maximum current through the switching device in the primary circuit is defined by the step-up ratio $k$ of the transformer, the maximum voltage on the line, $U_0$, and the charging time:

$$I_{1max} = \frac{kU_0}{\sqrt{2L/C_{PFL}}} .$$

For known values of $L$, $C_{PFL}$, and $U_0$ we obtain that the maximum charging current in the secondary circuit does not exceed 500 A. Setting the step-up ratio to 5 (to get the ~30 kV charging voltage in the primary circuit), we get the maximum current there ~2500 A. This makes it possible to use a thyratron as the main switch. For example, for the PerkinElmer HY-5 thyratron, the maximum voltage is 40 kV and the peak current 5000 A. 200 Hz repetition rate is well within the capability of this device. Main parameters of this thyratron are shown in Table 8.

Table 8: HY-5 thyratron specifications

| Peak forward anode voltage | 40 kV |
|---|---|
| Peak Forward anode current | 5 kA |
| Rms current | 125 A |
| Cathode heater power | 6.3V @ 30 A |
| Reservoir heater power | 4.5 V @ 11 A |
| Control grid open circuit peak trigger voltage (min.) | 1.3 kV |
| Control grid trigger source impedance (max.) | 100 Ohms |
| Max. average anode current | 8 A |

The capacitance of a primary storage capacitor (as accepted earlier) must be ~$k^2 \cdot C_{PFL}$ = 0.1 µF. Good fit for the primary storage capacitor is the series DE High Voltage Pulse Discharge Capacitors of General Atomics, part # 37218 (see http://www.ga.com/series-de-high-voltage-pulse-discharge-capacitors). With the capacitance C = 0.01 µF it is rated to 45 kV with peak current up to 25 kA, the average current is 25 A, internal inductance is ~60 nH, and the design lifetime of ~$3*10^9$ cycles. Dimensions of the capacitor are 102x153x330 mm (see Figure 52). The weight of the capacitor is 4 kg. It is configured as a double-ended plastic case, which works well for the chosen schematic presented in Figure 51.

Ten capacitors are required to store all required energy for one cycle for one cell. Connected in parallel, they will have the positive end connected to the thyratron, and the negative end – to the primary coil of the step-up transformer.

The capacitors must be charged between the LIA pulses. The required average charging current is ~0.6 A. This current can be used to reset the initial (negative) saturation field in the transformer core. According to Figure 48, the required magnetic field, $H$, that brings the core to the negative saturation level is ~8 A/m.

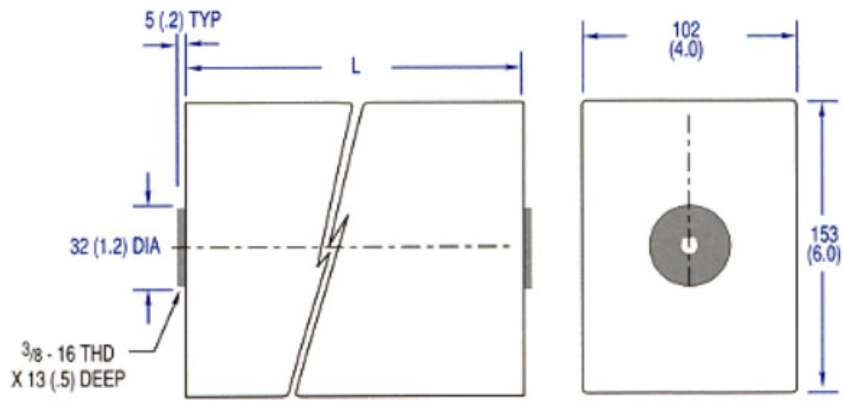

Figure 52: Part #37218 of General Atomic.

**Step-up transformer**

The voltage pulse at the input of the step-up transformer is described by the expression:

$$U_{C1} = U_{01}/2 \cdot (1+\cos(\omega t)),$$

where $U_{01}$ is the charging voltage of the primary storage capacitor. The effective volt-seconds of the driving voltage can be derived by integrating this expression over one half of the oscillation period; at this moment, the capacitor voltage reaches zero and the thyratron switches off:

$$\int U_{C1} dt = U_{01}/4f.$$

Here $k$ is the step-up ratio of the transformer (see above). With $k = 5$, $U_{01} = 30$ kV one obtains $\int U_{C1} dt = 0.03$ V·s.

Assuming amorphous magnetic materials or silicon steel as the material for the transformer with useful magnetic field span of 2 T, the required surface area of the transformer's core cross-section is

$$S\,[\text{m}^2] = \int U_{C1} dt\,/(\Delta B \cdot N) = 0.015/N,$$

where $N$ is the number of turns in the primary winding of the transformer.

As was mentioned earlier, the maximum current in the primary winding is ~2500 A, and in the secondary it is ~500 A. The average power which is handled by the transformer is ~10 kW. These parameters set the scale of this piece of equipment. Getting low fringe inductance is not a goal for this element, so higher number of turns in the primary can be used to save on the cross-section of the core. With the available range of the flux inductance change $\Delta B = 2.0$ T, using 5 tuns in the primary one obtains the required cross-section of the core of $S_{core} \approx 30$ cm$^2$. The number of turns in the secondary $N_2 = 25$. Possible design approach for the transformer is shown in Figure 53.

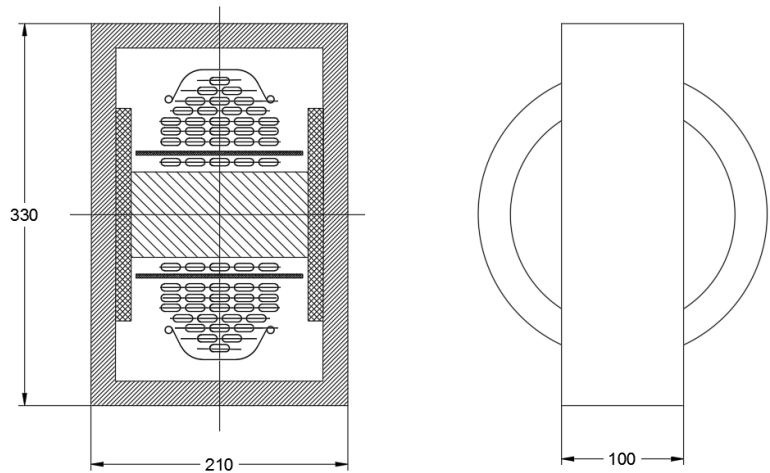

Figure 53: Step-up transformer. Dimensions are in millimeters.

### Choke

As it was shown earlier, expected inductance of the choke is ~0.2 mH. As it is in the secondary circuit of the pulse transformer, the maximum voltage across this element at the start of the PFL charging cycle is 150 kV. When the PFL is fully charged, the whole choke is at the maximum potential of 150 kV. As in the case of the pulse transformer, the effective volt-seconds applied to the choke are defined by the following expression:

$$\int U_{PFL} dt = U_{02}/4f.$$

Assume the maximum voltage in the secondary circuit is 150 kV, and the effective volt-seconds $\int U_{PFL} dt = 0.15$ V·s. For amorphous magnetic materials or silicon steel, as the material for the transformer with useful magnetic field span of 2 T, the required surface area of the transformer's core cross-section is:

$$S\,[\text{m}^2] = \int U_{C1} dt\,/(\Delta B \cdot N) = 0.075/N.$$

Assuming also $N = 25$ we obtain the cross-section area $S = 0.003$ m$^2$.

The required inductance of the choke is determined by a gap in the core. In this case the inductance can be evaluated using the following expression:

$$L = \mu_0 \cdot N^2 \cdot S_{core} / g,$$

where $g$ is a gap in the core. Using expression for the cross-section, we obtain:

$$g = \mu_0 \cdot N \cdot \int U_{C1} dt / (\Delta B \cdot L).$$

To obtain $L = 0.2$ mH with $N = 25$ and $S_{core} = 0.003$ m$^2$ one needs to choose the gap of $g = \sim 12$ mm. Figure 54 shows possible design of the choke with $N = 25$, which is very similar to that of the step-up transformer.

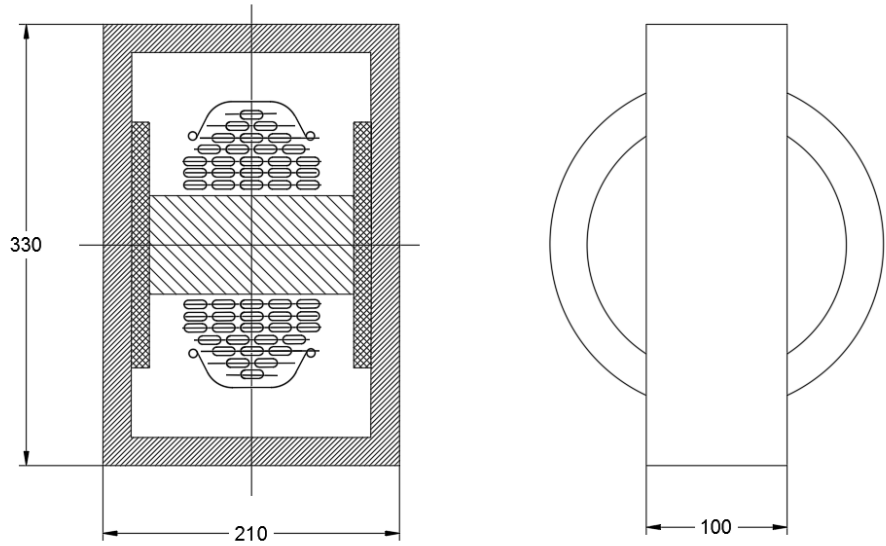

Figure 54: 150 kV choke with the inductance L = 0.2 mH. Dimensions are in millimeters.

As the choke is placed between the HV electrode of the pulse transformer and the PFL the whole assembly must be suspended in the insulating media capable to withstand ~200 kV. Dry outgassed transformer oil suits well for this purpose.

Let's evaluate losses in the core of the choke and the transformer assuming that the 2605SC Metglas material is used. The magnetization rate of the material with $\Delta B = 2$ T and $f = 250$ kHz is $\sim 3 \cdot 10^6$ T/s. According to [43], one should expect ~1100 J/m$^3$ energy loss per cycle. As the volume of the core is $\sim 1.5 \cdot 10^{-3}$ m$^3$ the losses per cycle of ~1.6 J are expected. With the 200 Hz repetition rate, ~330 W average power loss is expected in each of these two elements.

As was shown earlier, the maximum value of the charging current is ~500 A.

The skin layer at 250 kHz is ~0.1 mm, so we can evaluate the resistance of the winding in the choke and in the secondary of the transformer, which is ~0.06 Ohm. Then the power loss in each of these windings during the charging cycle is:

$$P_{wind} \approx 0.5 \cdot I_m^2 \cdot R_{wind} = 7500 \text{ W}.$$

Assuming the 1 μs charging time and the 200 Hz repetition rate, one obtains the average power loss in both windings of ~3 W.

The obtained relatively modest total power losses in the choke and in the transformer enable a consideration of more aggressive design. Even a usage of silicon steel as the core material looks possible.

**Saturable switch of the PFN**

Saturable switch of the PFN (S2 in Figure 51) is one of key elements of the circuit that requires a separate study. In this section, an attempt is made to put together a workflow needed to develop a pulse sharpening circuit of the last stage of the PFN for the LIA. A simplified schematic of the latest version of the PFN is shown in Figure 51. According to this schematic, the switch S1 connects the storage capacitor C1' to a step-up transformer (presented as magnetization inductance $L_M$). A pulsed forming line (50 Ohm PFL) is charged by the current in the secondary of the transformer through the choke CH. The required charging voltage of the PFL is $U_0 = 150$ kV to get 75 kV at the input of the accelerating cell of the LIA. The following shape of the charging voltage is expected:

$$U_{PFL} = U_0(1 - \cos(\omega t))/2.$$

With the required pulse duration of the accelerating voltage of ~380 ns, in order to have reasonably uniform distribution of the voltage along the PFL, the frequency $f = \omega/2\pi \approx 250$ kHz of the charging wave was chosen to get the charging time $t_{charge} = T/2 = 1/(2f) = \pi/\omega \approx 2$ μs.

Switch (saturable choke) S2 must open when the voltage across the PFL gets close to the maximum. Transition from the "closed" stage to the "open" stage will define the rise time of the voltage pulse on the load. The efficiency of the circuit will be defined by the ratio of the inductances of the choke **S2** in the unsaturated and in the saturated stages and of the inductor of the accelerating cell in the unsaturated stage $L_{cell}$. The inductance of the choke in the unsaturated step must be much higher than that of the cell inductor: $L_{S2\_unsat} \gg L_{cell}$. In the saturated stage, this inductance must be much lower: $L_{S2\_sat} \ll L_{cell}$.

Both the transition time and the efficiency of the switching strongly depend on the properties of material used for fabrication of the saturable core of the choke and the inductor of the accelerating cell. Figure 47 and Figure 48 show DC magnetization loops for two most probable candidate materials to be used to fabricate the core of the saturable choke: Metglas® 2605SC [43] and Metglas® 2605CO [45-48]. Table 9 compares relevant properties of the two materials [48], [49].

To find the unsaturated inductance of the choke, one needs to know permeability of the material. It is clear from Figure 47 and Figure 48 that both materials have rectangular DC magnetization curve, so the initial permeability is not well defined. With the expected relatively high magnetic flux density rise rate, eddy current effect is sizable, and deformation of the magnetization loop during pulse excitation needs to be taken into account. High frequency unsaturated permeability will be lower than the DC one. The saturation magnetization, on the other hand, is independent on the frequency.

**Table 9: Comparison of properties of 2606CO and 2605SC**

|  | 2605CO | 2605SC |
| --- | --- | --- |
| Composition | $Fe_{67}Co_{18}B_{14}Si_1$ | $Fe_{81}B_{13.5}Si_{3.5}C_2$ |
| Density (g/cm$^3$) | 7.56 | 7.32 |
| Resistivity (μΩ·m) | 1.30 | 1.25 |
| Curie temperature (°C) | 415 | 370 |
| Saturation induction $B_c$ (T) | 1.8 | 1.6 |
| Residual magnetization $B_r$ (T) | 1.7 | 1.5 |
| Flux Swing $\Delta B$ (T) | 3.5 | 3.1 |
| Coercive force $H_c$ (A/m) | 3.2 | 2.4 |

A way to evaluate effective permeability of an unsaturated core ($\mu_{unsat}$) was proposed in [45] and [46]. The authors used specific energy loss density $w$ [J/m$^3$] in the core of a saturable choke to evaluate the effective permeability.

As the initial state of the pulsed system is usually set $H = 0$ and $B = -B_r$, the flux change $\Delta B$ can be expressed as

$$\Delta B = \mu_{unsat} \cdot \mu_0 \cdot H_{sat}$$

If we assume that the effective permeability is constant during the cycle,

$$\langle H \rangle = H_{sat}/2 \ .$$

Then

$$w = 1/2 \cdot \mu_{unsat} \cdot \mu_0 \cdot H_{sat} \cdot H_{sat} = 1/2 \cdot (\Delta B)^2 / (\mu_{unsat} \cdot \mu_0),$$

and

$$\mu_{unsat} = 1/2 \cdot (\Delta B)^2 / (\mu_0 \cdot w).$$

To find the effective unsaturated permeability we need to know expected specific energy deposition in the core of a device. As significant part of the power loss in the core is due to eddy currents, it depends on the rate of the flux change in the material. In Figure 55, the energy loss in the 2605CO material is plotted versus the time to saturation, which is equivalent to the rate of the flux density change rate.

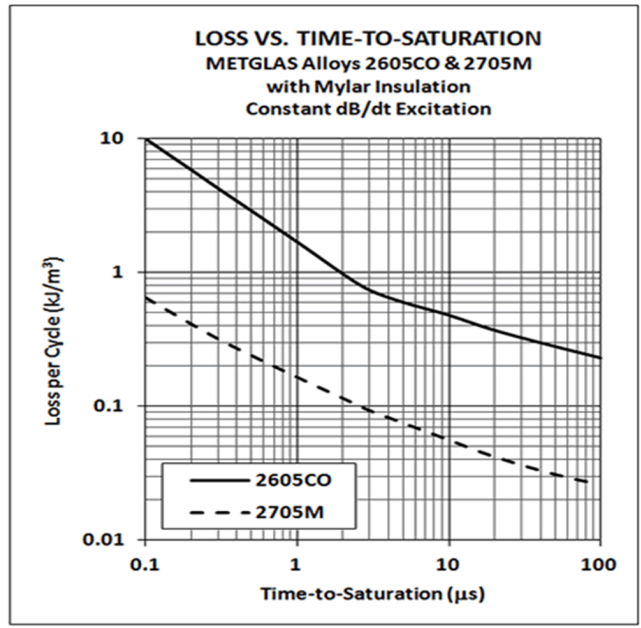

Figure 55: Energy loss per cycle as function of the saturation time [49].

Having the power loss information for the 2605CO material, we can use it to find the power loss in the core of the accelerating cell assuming the flux density rise in the core being constant during the duration of the voltage pulse. With the saturation time ~380 ns and the 3.5 T achievable flux density swing, the rate of the flux density change $dB/dt \approx 9.2 \cdot 10^6$ T/s. According to the plot in Figure 55, the expected power loss in the core of the section is ~3200 J/m³. With the volume of the core of ~0.0116 m³, the total heating energy deposited in the core during one cycle is ~37 J. This energy loss during one cycle is close to what was found earlier for the 2605SC material.

Knowing the specific energy loss for this core, we can find the effective permeability of the core of the LIA cell in the unsaturated stage:

$$\mu_{unsat} = 3.5^2/(2 \cdot \mu_0 \cdot 3200) = 1523.$$

As the input voltage across the winding of the saturable switch S2 is known, we can find the specific energy loss in this case following a method suggested in Refs. [47] and [50]. This method is based on the saturation wave model, which works well when the magnetization rate is high.

Following [47], expression for the magnetic field $H(t)$ required to magnetize a ribbon with thickness $d$ and resistivity $\rho$ is

$$H(t) = H_c + \left(\frac{d^2}{4\rho}\right)\left(\frac{\Delta B}{2B_c}\right)\left(\frac{dB}{dt}\right),$$

where $\Delta B(t) = B(t) + B_r$ is an efficient flux density swing, $B_c$ is material saturation field, $B_r$ is residual field, $H_c$ is the DC coercive force.

This expression is used to find the heating loss density in the material per one magnetization cycle for any value of $\Delta B$:

$$w(t) = H_c \Delta B + \left(\frac{d^2}{8\rho}\right)\left(\frac{\Delta B^2}{2B_c}\right)\left(\frac{dB}{dt}\right).$$

Using data from the Table 9, the power loss density per one cycle can be found: $w \approx 525$ J/m³. Then, the effective permeability of the core material in the saturable switch can be evaluated:

$$\mu_{unsat} = 3.5^2/(2 \cdot \mu_0 \cdot 525) \approx 9330.$$

Let's evaluate inductance of the accelerating cell during active stage. Taking dimensions of the cell from Figure 49, we get the effective inductance of the unsaturated inductor:

$$L_{ind} = \mu \mu_0 S/l \approx 20 \text{ μH}.$$

To find the unsaturated and saturated inductances of the saturable switch S2, we need to know the cross-section area of the core and the number of turns. They can be found by using the general expression

As the time change of the charging voltage $U_{PFL}$ is known, we can find integral in the left side of this expression:

$$\int_0^{T/2} U_{PFL} dt = \frac{U_0 T}{4}.$$

Now we can evaluate the required cross-section of the choke's core. With $\Delta B = 3$ T, and the period $T = 4$ μs, we obtain $A \cdot N = U_0 \cdot T / (4\Delta B) \approx 0.05$ m². Assuming $N = 5$ we obtain $A \approx 0.01$ m².

To evaluate the values of the unsaturated and saturated inductances of the saturable switch, let's assume that the geometry of the choke is close to that of the cell, shown in Figure 49. In the fully saturated state the inductance is:

$$L_{S2\_sat} = \mu_0 N^2 A/l \approx 0.36 \text{ μH}.$$

In the unsaturated state it is (μ = 50000):

$$L_{S2\_unsat} = \mu \mu_0 N^2 A/l \approx 3.3 \text{ mH}.$$

As the inductance of the cell is ~20 μH, the condition $L_{S2\_sat} \ll L_{cell} \ll L_{S2\_unsat}$, which defines the effectiveness of the saturable switch, is satisfied.

In real life, the full saturation can never be achieved, so the inductance of the saturated choke is going to be higher. To take this increase into account, studies of material properties must be performed using prototypes of all elements in the compression circuit.

In conclusion to this section, it is safe to say that there are no prohibitive stoppers for an implementation of the scheme of the PFN shown in Figure 49. Consequently, the concept of the PFN looks feasible as well as a simple one-stage scheme. The mass of the required tape looks quite large, and an optimization of the scheme, as well as a search for a new material, are required in the next step of the design.

### 10.5. Summary

The concept of the induction linac looks feasible at this stage. The linac contains:

- A 50 A, 300 kV electron gun with the beam area compression 6:1. The gun has the dispenser cathode 50 mm in diameter, operating at 1050C. The maximal current density is 12 A/cm$^2$, which provides reasonably long cathode lifetime, 20,000-30,000 hours.
- The gun is matched to the first solenoid using the concept successfully used in [38-40]. A trimming coil and a yoke are used to provide the magnetic force lines in the gun, which coincide the beam trajectories.
- The solenoidal focusing is used to keep the beam maximal radius in the linac of 10 mm. The field in the first solenoid is < 500 Gs. The heat deposition in the solenoid is well below 200 W.
- The linac contains 727 accelerating cells with lengths of 141 mm each. Each cell provides the energy gain of 75 keV.
- The total length of the linac is ~100 m for a 55-MeV beam (without an injector).
- The power dissipation in the cores of each cell is ~6 kW. The total power dissipation in the LIA is ~6.3 MW.

- A concept scheme of a pulser for one cell of the LIA is proposed. It meets the requirements of the voltage and pulse length.

# Appendix A: Cooling system concept for 270-GeV protons

## Main parameters of an induction linac-based electron cooler

| | |
|---|---|
| Proton beam energy | 270 GeV |
| Relativistic factor, $\gamma$ | 289 |
| Proton ring circumference (it is used to calculate cooling rates only) | 3834 m |
| Cooling length section | 80 m |
| Normalized rms proton beam emittances (x/y) | 3/0.5 µm |
| Proton beam rms momentum spread | $<3\times10^{-3}$ |
| $\beta$-functions of proton beam at the cooling midpoint | 80 m |
| Proton beam rms size (hor/ver) | 0.9/0.4 mm |
| Electron beam energy (50 – 150 MeV) | 147 MeV |
| Electron beam current (50 – 100 A) | 100 A |
| Cathode diameter | 25 mm |
| Cathode temperature | 1050ºC |
| Longitudinal magnetic field in cooling section, $B_0$ | 780 G |
| Electron beam rms momentum spread, initial/final | $(1.0/1.25)\cdot10^{-3}$ |
| Rms electron angles in cooling section | 4.8 µrad |
| Rms electron beam size in cooling section | 2.2 mm |
| Electron beam rms norm. mode emittances at injection, $\varepsilon_{1n}/\varepsilon_{2n}$, µm | 220/0.042 |
| Number of cooling turns in the electron storage ring | 6,000 |
| Longitudinal cooling time (emittance)[*] | 23 min |
| Transverse cooling time (emittance)[*] | 30 min |

[*] This small-amplitude cooling time accounts for the growth of emittances due to IBS in the cooling ring. It does not account for imperfections in the electron beam. For a 10-ms cycle, the averaged cooling times decrease to 31 and 38 min for the longitudinal and transverse planes, respectively.

Figure 56 shows the proposed cooler concept for 100-270 GeV protons.

A 55-MeV electron linac (suitable for a 100-GeV cooler) would consist of 725 induction cells, each 141 mm long and providing the energy gain of 75 keV. The average power dissipation per cell is about 6 kW. Increasing the electron beam energy to 150 MeV can be accomplished by increasing the number of cells in proportion to beam energy. It might be also possible to employ a multi-pass acceleration system (Figure 57). This multi-turn concept requires a separate study.

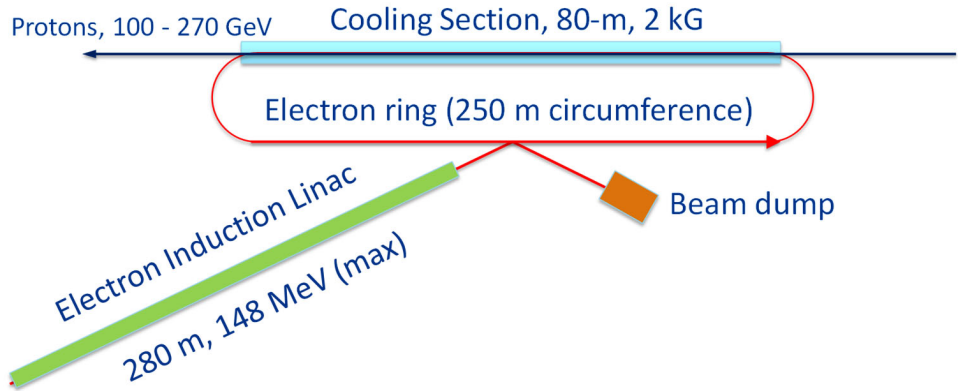

Figure 56: Schematic of a proposed induction linac-based electron cooler for 100-270 GeV protons.

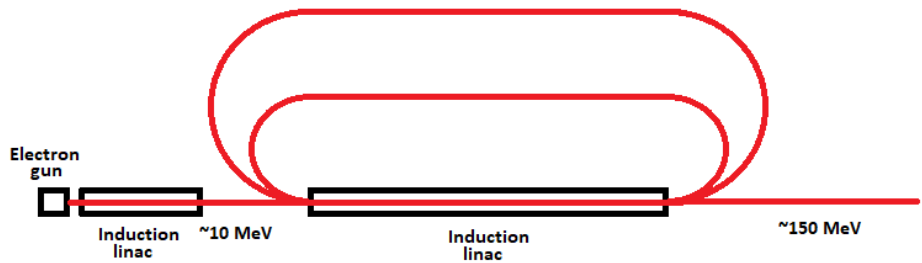

Figure 57: A multi-turn linac concept.

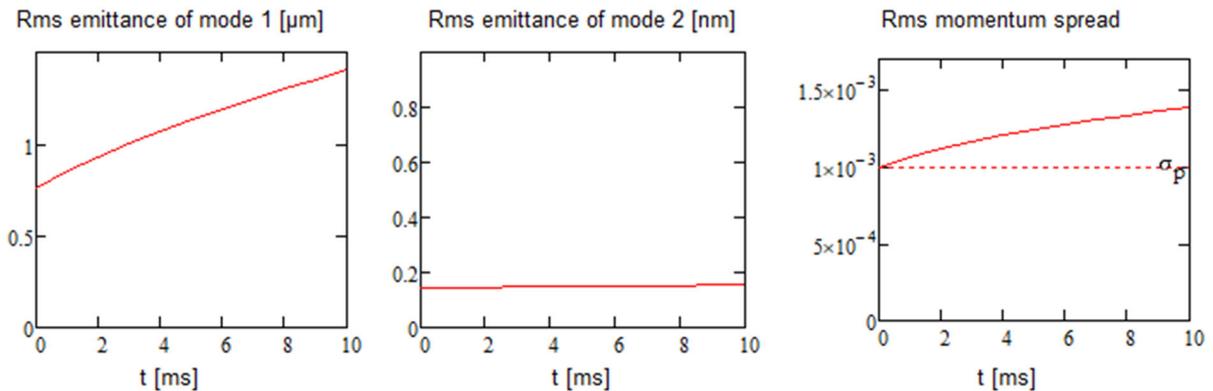

Figure 58: Dependencies of rms non-normalized mode emittances and the momentum spread on time for a 100-A, 147 MeV electron beam due to intra-beam scattering.

Figure 58 presents the dependences of beam emittances and the momentum spread on time for a 100-A electron beam at 150 MeV in a cooler ring. The rms initial momentum spread of $\sim 1 \cdot 10^{-3}$ is set by the requirement to suppress the microwave instability. The transverse emittances are set by the emittance of the gun and the magnetic field at the cathode. One can see that the beam emittances stay reasonably small during 5 ms. This determines the required repetition rate of 200 Hz. Operation at 100 Hz is also feasible. It will reduce the cooling rates by ~30% (see Figure 59 below).

Figure 59 shows the instantaneous rms emittance cooling rates for 270-GeV protons for 10 ms cycle.

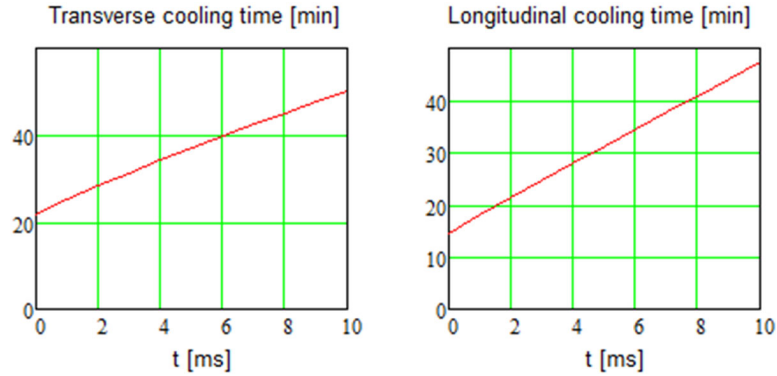

Figure 59: Instantaneous rms emittance cooling time for 270-GeV protons as a function of time.

Finally, we consider how the transverse cooling force depends on the betatron amplitude of a proton. Figure 60 shows the dependence of the cooling rate on the dimensionless particle betatron amplitude.

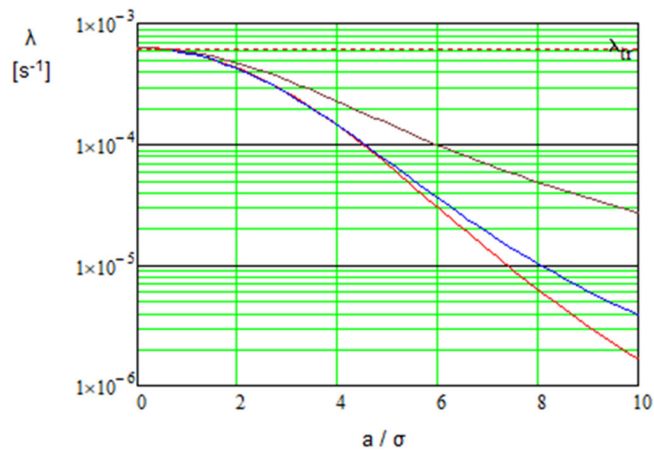

Figure 60: Dependence of cooling rate on the dimensionless proton betatron amplitude; brown (top) – the electron beam size is much larger than the amplitude of betatron motion of a proton, blue (center) – the betatron oscillations are in one plane and the electron beam has its design size and, red (bottom) line – the design rms electron beam and the betatron oscillations have equal amplitudes for the horizontal and vertical planes.

Other effects in the electron beam have been also considered in sufficient details. We have identified two critical effects, which require further studies: (1) transverse electron beam heating due to electron beam space charge and (2) longitudinal electron beam instability driven by coherent synchrotron radiation.

## Acknowledgments

This manuscript has been authored by Fermi Research Alliance, LLC under Contract No. DE-AC02-07CH11359 with the U.S. Department of Energy Office of Science, Office of High Energy Physics.

The authors acknowledge Andrey Ivanov for providing the code WinSAM and help during the work, and John Petillo for providing MICHELLE code and permanent consulting.